\documentclass[sigconf,nonacm]{aamas}

\usepackage{balance}

\usepackage{graphicx}
\usepackage{natbib}
\usepackage{caption}

\usepackage{booktabs}
\usepackage[most]{tcolorbox}
\usetikzlibrary{positioning,arrows.meta,fit}
\definecolor{promptdarkblue}{RGB}{30, 58, 138}
\definecolor{promptlightblue}{RGB}{219, 234, 254}
\definecolor{promptborder}{RGB}{147, 197, 253}
\definecolor{promptheaderbg}{RGB}{191, 219, 254}

\usepackage{multirow}
\usepackage{subcaption}

\newcommand{\interpretation}{I}

\newcommand{\satisfaction}{\models}

\newcommand{\program}{\Pi}

\newcommand{\system}{EntailLLM}

\makeatletter
\gdef\@copyrightpermission{
  \begin{minipage}{0.2\columnwidth}
   \href{https://creativecommons.org/licenses/by/4.0/}{\includegraphics[width=0.90\textwidth]{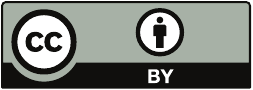}}
  \end{minipage}\hfill
  \begin{minipage}{0.8\columnwidth}
   \href{https://creativecommons.org/licenses/by/4.0/}{This work is licensed under a Creative Commons Attribution International 4.0 License.}
  \end{minipage}
  \vspace{5pt}
}
\makeatother

\setcopyright{ifaamas}
\makeatletter
\gdef\acmConference@shortname{Preprint}
\gdef\acmConference@date{under review\@gobble}
\gdef\acmConference@venue{}
\makeatother

\title[\system: Verifying Agent-Generated Vulnerability Discovery Paths with Domain Knowledge via Logic Programming]{\system: Verifying Agent-Generated Vulnerability Discovery Paths with Domain Knowledge via Logic Programming}

\author{Kaustuv Mukherji}
\email{kmukherj@syr.edu}
\author{Jaikrishna Manojkumar Patil}
\email{jpatil01@syr.edu}
\author{Colton Payne}
\email{crpayne@syr.edu}
\author{Paulo Shakarian}
\email{pashakar@syr.edu}
\affiliation{
  \institution{Syracuse University}
  \city{Syracuse}
  \state{NY}
  \country{USA}}

\author{Nigel Stepp}
\email{ndstepp@hrl.com}
\author{Dana Warmsley}
\email{dmwarmsley@hrl.com}
\author{Evelyn Kim}
\email{ekim@hrl.com}
\affiliation{
  \institution{HRL Laboratories, LLC}
  \city{Malibu}
  \state{CA}
  \country{USA}}

\begin{abstract}
Large language model agents are increasingly used to analyze potential software
vulnerabilities, but their outputs can silently violate domain knowledge,
limiting their reliability in safety-critical settings such as medical devices.
Prior work either treats that output as a prediction to be scored or constrains
it to walks within a single knowledge graph; neither checks whether reasoning
over a binary is entailed by an independent body of domain knowledge.
We present \system, a framework that includes a specialized LLM agent and a validation mechanism. The agent proposes a traversal path of the function call graph for a vulnerability, which is checked against a separate knowledge graph and a logic program by an entailment procedure under annotated logic.
We evaluate \system\ by examining six specialized agents across three CWE classes, and
eight binaries varying in size from 405 to 12{,}696 functions.  The agents improve pooled entailment from $77\%$ to $98\%$, with entailment decreasing in only 3 of 118 configurations. We found that the best-performing agent combines self-refinement and a priori domain knowledge.  \system\ is deployed end-to-end on real medical-device binaries, reaching $98\%$ pooled entailment without per-device
tuning. Our framework inherits the formal guarantees of generalized annotated logic, providing logical verification of agent output that is both explainable and grounded in well-defined semantics.
\end{abstract}


\newtcolorbox{promptbox}[1][]{
  enhanced,
  colback=promptlightblue,
  colframe=promptborder,
  boxrule=0.4pt,
  arc=2pt,
  left=0pt, right=0pt, top=0pt, bottom=0pt,
  fontupper=\small,
  #1
}
 
\newcommand{\promptsection}[1]{%
  \par\noindent
  \begin{tcolorbox}[
    width=\linewidth,
    colback=promptheaderbg,
    colframe=promptheaderbg,
    boxrule=0pt, arc=0pt,
    left=6pt, right=6pt, top=2pt, bottom=2pt,
    boxsep=0pt,
    before skip=0pt, after skip=3pt,
  ]
  {\color{promptdarkblue}\bfseries\sffamily\small #1}
  \end{tcolorbox}%
}
 
\newenvironment{promptcontent}{%
  \par\noindent\hspace{6pt}%
  \begin{minipage}{\dimexpr\linewidth-12pt\relax}%
}{%
  \end{minipage}\par\vspace{2pt}%
}

\begin{document}

\maketitle 

\section{Introduction \label{sec:intro}}

Language model agents are increasingly applied to vulnerability discovery, from
detection and triage to exploit reasoning and patch
generation~\cite{Sheng2025}. Yet they remain unreliable at identifying and
reasoning about vulnerabilities unaided~\cite{ullah2024llms}, they produce
confident inferences that probability-based checks cannot
catch~\cite{pan-etal-2023-logic}, and they mislead analysts when their
suggestions go unchecked~\cite{basque2026decompiling}. These problems worsen for
bespoke reverse-engineering tasks in domains such as medical devices, whose
firmware is long-lived, hard to patch, and rarely shipped with
source~\cite{unit42infusion2022}. Analysis proceeds on decompiled binaries, where
the semantic cues of source code are absent, and intent must be reconstructed from
low-level structure. Yet analysts already use LLMs to decide where to look in a binary~\cite{basque2026decompiling}, and attackers increasingly do the same \cite{fang2024llm}. What such a model proposes is therefore both a prediction of how the code will be explored and something that needs checking for alignment with the domain knowledge.

\begin{figure}[tbp]
\centering
\includegraphics[width=0.99\columnwidth]{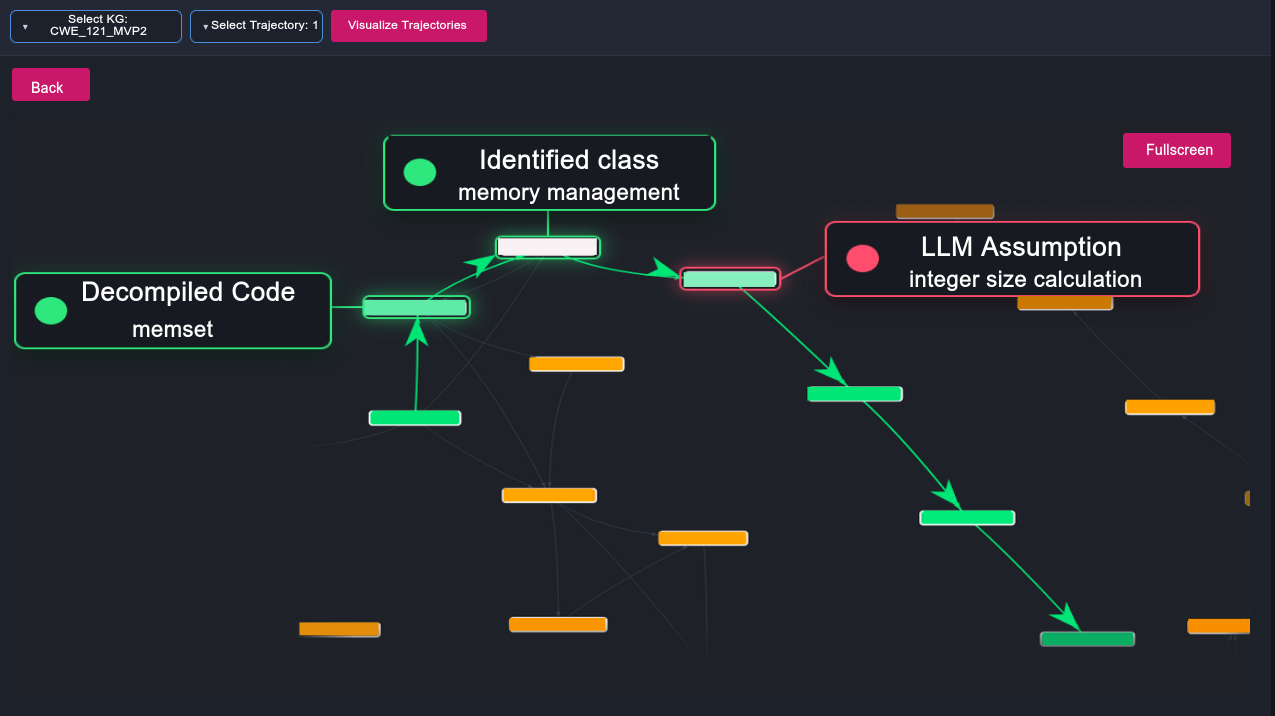}
\Description{Screenshot of the deployed interface showing part of a binary's call graph. A path proposed by the agent is drawn as green nodes joined by bold arrows; other functions appear as orange nodes with faded edges. Callouts mark three steps on the path: decompiled code calling memset, the identified class memory management, and, highlighted in red, the assumption integer size calculation, which the domain knowledge does not support.}
\caption{An agent's mistake caught by domain knowledge non-entailment. The green path (arrows) through the
codebase was selected. Orange nodes (with faded edges) denote other functions in the binary.}
\label{fig:llm-mistake}
\end{figure}

We present \system, a framework deployed at scale on real medical-device firmware, that verifies agent-proposed paths against domain knowledge. We show that when that knowledge is used to guide the agent, its proposals are entailed more often. Figure~\ref{fig:llm-mistake} is a screenshot of \system\ catching
erroneous LLM-agent output by flagging an LLM conclusion the domain knowledge does not support. The library function \texttt{memset} is correctly assigned
to the \emph{memory-management} class, but the LLM then assumes it can perform
an \emph{integer size calculation}---an inference that is structurally
unremarkable yet supported by nothing \texttt{memset} does.
The distinction that makes this possible is the difference between \emph{statistical} clean-up:
pruning candidate paths by structural signals such as the call graph, which an
LLM or a graph-based method already does well, and \emph{semantic} clean-up:
rejecting a path that violates the meaning of the domain even when it is
structurally plausible. Our logic-based reasoner supplies the latter exactly, with a trace for every decision. Table~\ref{tab:nollm} shows why the problem is hard in practice, where device binaries are often large. Without a neural model like an LLM, paths could be formed directly from the call graph under the same constraints. But enumerating them is intractable on large binaries (up to $2.1\times10^{6}$ direct six-edge call chains on \texttt{unsafelib}, growing about $2.4\times$ per added edge), and a sample of them is entailed only about half the time (No-LLM column).
\system\ comprises an LLM-based agent that proposes paths, and a reasoner that treats each proposed path as a hypothesis and admits it only if a
domain-knowledge graph (DKG) entails it. The path is a
traversal of the binary's call graph while the constraints come from a
separate DKG, so verification is an alignment between two graphs rather
than a walk within one. The domain encoding instantiates hand-crafted schemas
over the knowledge graph's triples, yielding conjunctive, co-dependent-label
rules grounded in CWE, MITRE, and device documentation. Using an LLM makes the agentic search tractable by producing a manageable number of proposals from the binary's call graph, and the paths are entailed more often than in the No-LLM case (Unguided column). An LLM, trained to respond to various user requests in language, captures human-like intuition. However, we want the agentic output to align with the domain knowledge. We use entailment as a procedure to determine this alignment. If the domain knowledge is correct, we would expect an entailed agentic output to likely lead to an actual vulnerability. However, here we focus only on how to best answer the entailment question.  We have ongoing work to evaluate the end-to-end system in terms of finding vulnerabilities. And this includes validation of the domain knowledge itself. In this work, we try to find an effective way to guide the output of a neural agent by providing semantic labels from a symbolic system. 
As shown in Table~\ref{tab:nollm}, guiding the agent with labels derived from the domain knowledge raises entailment further.
The results show that symbols from a symbolic reasoner can guide neural generators like LLMs, and the neurosymbolic system performs better than a symbolic agent or an agent with an unaided LLM. To identify the most effective way of guiding a neural agent with symbolic knowledge, we evaluate our framework for six agents across several LLM models, binaries, and three different CWEs.
We find that adding domain knowledge with a self-refinement prompting strategy raises
the fraction of paths entailed from $71\%$ to $97\%$ while increasing the
number of entailed paths by $31\%$, from $1{,}297$ to $1{,}705$, over all three Common Weakness Enumeration (CWE) classes.

Our contributions are: (i) a neurosymbolic approach to agentic binary analysis in which domain knowledge guides an agent's proposed paths through code, and a symbolic reasoner checks if those paths are entailed; (ii) a deployed, end-to-end system that localizes and explains
non-entailments, reporting the first unsupported step with the rule firings
behind it; and (iii) an evaluation of six agents across three CWE classes, four LLMs, and eight binaries, in which guidance from domain knowledge raises
pooled entailment across agents from $77\%$ to $98\%$, decreasing in only 3 of 118
configurations, and the gain persists when guidance and verification use disjoint halves of the knowledge graph.

\begin{table}[tbp]
\centering
\caption{Entailment (\%) of paths generated by the No-LLM agent (250 samples), the unguided and the guided self-refinement agent, pooled over four LLMs.}
\label{tab:nollm}
\begin{tabular}{@{}lcrrr@{}}
\toprule
\textbf{Binary} & \textbf{CWE} & \textbf{No-LLM} & \textbf{Unguided} & \textbf{Guided} \\
\midrule
unsafelib     & 121 & 40.0 & 61.2 & 95.5 \\
mupdf-x11     & 416 & 46.4 & 63.8 & 93.8 \\
mutool        & 416 & 54.8 & 78.5 & 96.5 \\
\bottomrule
\end{tabular}
\end{table}

\noindent\textbf{Related Work.}
Work that reports gains from LLMs on security tasks pairs the model with
structure rather than trusting it alone~\cite{li2024llmxcpg}. Applying LLMs
directly to stripped or decompiled binaries recovers source-like code, symbols,
and readable decompiler
output~\cite{tan2024llm4decompile,xie2024resym,degpt2024}, improving the
representation an analyst works from but not verifying that reasoning over it is
consistent with what is known about the vulnerability class. Other work gates
LLMs with structure to similar ends, translating a problem into symbolic form for
a solver~\cite{pan-etal-2023-logic} or constraining generation to paths grounded
in a knowledge graph~\cite{luo2024rog,luo2025gcr} or letting an LLM agent beam-search a knowledge graph for reasoning paths~\cite{sun2024think}. These align reasoning to a
walk within a single KG and are evaluated on natural-language QA. Our
arrangement is closest to shielding in safe reinforcement
learning~\cite{alshiekh2018safe} and to logic-based LLM guardrails~\cite{kang2025r2guard}, where the agent proposes freely, and the logic
decides what passes. But a shield strong enough to block unsafe behavior
typically blocks useful behavior with it. Here we show that enriching the
generator with the domain knowledge's own derived vocabulary raises the entailed rate more than shielding alone and the count as well for the agent following a self-refinement strategy. Classical
taint-based discovery and its recent LLM-driven
variants~\cite{Redini2020,liu2025latte} share our binary-level, source-free
setting but aim to emit vulnerability alerts; where they use the LLM to
produce the analysis, we use logic to check it. A detailed literature review
appears in the Appendix.

\section{Problem Description}
\label{sec:problem}

\noindent\textbf{Technical Preliminaries.}
\label{sec:prelims}
We reason in generalized annotated logic~\cite{ks92}, which supplies a
single formal semantics generalizing fuzzy and other real-valued logics, and we
work in a tractable fragment that admits exact polynomial-time
deduction~\cite{shakarian2022extensions} and represents time
explicitly~\cite{shakarian2012annotated}. We recall only what is needed
here and refer the reader to that work for the full development. An
\emph{annotated literal} $a{:}\mu$ pairs a literal with an annotation
$\mu=[\ell,u]\subseteq[0,1]$. The lattice
is ordered so that $[0,1]$---total uncertainty---is the bottom element, while
tight intervals such as $[1,1]$ and $[0,0]$ sit at the top. This makes
the logic open-world: an unobserved literal is not false but merely
uncertain, and inference narrows it, so the partial and noisy observations
recovered from a binary need not be forced to true or false. A \emph{temporal
annotated fact} (TAF) asserts such a literal at a time point, written
$a{:}(\mu,t)$, and rules take the form $\ell_0{:}\mu_0 \xleftarrow{\Delta t}
\ell_1{:}\mu_1 \wedge \cdots \wedge \ell_m{:}\mu_m$, where a body, defined as a conjunct in the standard way, satisfied at
$t$ makes the head hold at $t+\Delta t$. A program $\program$ is a set of TAFs
and rules. An \emph{interpretation} $I$ maps each ground literal and time point
to an annotation; it satisfies $a{:}\mu$ at $t$ when $\mu \sqsubseteq I(a,t)$,
and is a model of $\program$ when it satisfies every rule and TAF. Deduction
iterates a fixpoint operator $\Gamma$ to convergence, yielding the \emph{minimal
model} $\interpretation^\ast$. $\Gamma$ is monotonic, converges in polynomial number of applications, and decides \emph{entailment}. Entailment is therefore an exact,
decidable consequence of the program's semantics rather than a heuristic score,
and because every interpretation is produced by an identifiable rule firing, each inference
carries a trace of the facts and rules that led to it.

\noindent\textbf{Entailment Checking Agent Paths.}
\label{sec:problem-entailment}
Given a target binary and a weakness class (CWE), the agent predicts exploration
workflows a human analyst might follow through the binary's functions toward a
vulnerability of that class. Weakness classes come from the Common Weakness Enumeration (CWE) maintained by MITRE~\cite{mitre_cwe}. The NIST National Vulnerability Database (NVD)~\cite{nist_nvd} catalogues publicly disclosed vulnerabilities (CVEs) and maps them to CWE weakness types. The input is the binary's function call graph,
functions as nodes and calls as edges, with each function labeled with a set of libc
labels; the output is a sequence of call-graph nodes, each reachable from the previous one within a few call edges. Our prompts steer exploration toward the target weakness class, and entailed traversals feed a fuzzer downstream in the deployed system, so the traversal is the input to discovery rather than a substitute for it. We
evaluate CWE-121 (stack-based buffer overflow), CWE-415 (double free), and
CWE-416 (use-after-free). The agent performs a standard beam search over LLM outputs to check for paths realizable in a given binary. The logical verification step is only applied after that, to check if such a path adheres to domain knowledge.
We first fix a domain-knowledge program $\program = \program_{TAFs} \cup \program_{Rules}$, whose facts encode the DKG and whose rules are described in Section~\ref{sec:logic-program}. We denote a sequence of code blocks, that is, the binary's functions, the nodes of its call graph, as: $\tau = \langle CB_0, \ldots, CB_n \rangle$. Then a candidate trajectory generated by an agent linked by a predicate called $\mathit{stepFrom}$ is encoded by TAFs, and can be written as
\begin{equation}
\begin{aligned}
    F(\tau) = \; &\{\, \mathit{analystAt}(CB_0):(\mu_0, 0) \,\} \\
    &\cup\; \{\, \mathit{stepFrom}(CB_{i-1}, CB_i):(\top, i-1)\,\}_{i=1}^{n}
    \notag
\end{aligned}
\label{eq:traj-facts}
\end{equation}
asserting that each step is proposed at
the corresponding time point, and we reason over $\program \cup F(\tau)$, whose
minimal model we denote $\interpretation^\ast$.
Note that $\tau$ and $\program$ live in different graphs: $\tau$ is a chain in
the binary's call graph, whose connectivity is guaranteed by construction, while
$\program$'s facts encode the DKG's vulnerability concepts and typed semantic
relations. Entailment is therefore an alignment between the two, not the
existence of a path within either.
A step $CB_{i-1} \to CB_i$ is entailed when the domain knowledge verifies
that the analyst can follow $F(\tau)$ to reach $CB_i$ at time $i$:
\begin{equation}
    \program \cup F(\tau) \;\models_{\mathit{ent}}\;
    \mathit{analystAt}(CB_i):(\mu, i),
    \; \mu \succeq \mu_{\min}
\notag
\label{eq:step-ent}
\end{equation}
where $\mu_{\min}$ is the minimum admissible bound. If no chain of
rule firings can verify it, then the step is a non-entailment, meaning
the agent has proposed a move that is not supported by the domain knowledge. The entire trajectory is entailed $\program \models_{\mathit{ent}} \tau$ when every step is entailed, i.e.
\begin{equation}
\forall i \in \{1,\ldots,n\}:
    \interpretation^\ast \satisfaction_i
    \mathit{analystAt}(CB_i):\mu_i, \; \mu_i \succeq \mu_{\min}
\notag
\label{eq:traj-ent}
\end{equation}
The entailment-checking problem is to decide whether
$\program \models_{\mathit{ent}} \tau$ and, when it does not, to return the first
failing step $i^\ast = \min\{\, i : \program \cup F(\tau)
\not\models_{\mathit{ent}} \mathit{analystAt}(CB_i):(\mu,i) \,\}$. We retain
exactly the entailed trajectories. Because $\interpretation^\ast$ is computed
exactly and in polynomial time (Theorems~3.2 and~3.4
of~\cite{APTL}), this decision and the localization of $i^\ast$ are exact rather
than heuristic.

We focused development on binary analysis because of our target deployment; the
framework itself is agnostic to whether the call graph and its labels come from
a decompiler or from source, and we leave source-code analysis to future work,
studying a complete binary pipeline here. Several domain-specific reasons
motivate the choice. Device firmware ships without source, and the compiled
artifact is what runs, and optimization and field patching can leave it diverging from any source tree. We consume decompiler output rather than raw bytes because the decompiler recovers
the call graph and, from symbol and relocation tables, the imported libc calls we
use as labels---beacons that survive stripping, so no debug information is
needed. We verify all traversals rather than only paths terminating in a
vulnerability. Reverse engineers proceed by beacon-driven exploration,
recognizing API calls, strings, and constants to form and test
hypotheses~\cite{votipka2020observational,sisco2017modeling,bryant2011software},
so restricting attention to vulnerability terminating paths would discard the exploratory
structure that makes the output useful and presuppose the analyst's conclusion.
Finally, admitting only domain-entailed paths reduces the volume of unsupported
paths a human must triage and yields explainable traces an auditor can inspect. Entailment measures conformance to the encoded domain knowledge. Our problem setting is defensive in nature. We want to predict how an analyst might approach a codebase, and identify which type of agent yields the exploration most consistent with domain knowledge.

\section{System description \label{sec:system}}

\begin{figure}[tbp]
\centering
\begin{tikzpicture}[
    font=\small,
    stage/.style={draw, rounded corners=2pt, align=center, minimum height=6mm,
                  inner xsep=1.6mm, inner ysep=1mm, fill=blue!8},
    io/.style={draw, align=center, minimum height=6mm,
               inner xsep=1.6mm, inner ysep=1mm, fill=gray!12},
    ctx/.style={draw, dashed, align=center, minimum height=6mm,
                inner xsep=1.6mm, inner ysep=1mm, fill=white},
    flow/.style={-{Stealth[length=2mm]}, thick},
    bi/.style={{Stealth[length=2mm]}-{Stealth[length=2mm]}, thick},
    fb/.style={-{Stealth[length=2mm]}, thick, dashed},
    lbl/.style={inner sep=0.6mm}
]
\node[ctx]   (decomp) at (0.80, 0.00) {Decompiler};
\node[io]    (inputs) at (3.50, 0.00) {libc labels \& call graph};
\node[stage] (llm)    at (3.50,-1) {Agent (generator)};
\node[io]    (dk)     at (6.60,-1) {Domain\\Knowledge};
\node[stage] (reason) at (3.50,-2) {Reasoner (verifier)};
\node[ctx]   (fuzzer) at (6.70,-2) {Fuzzer};

\draw[flow] (decomp) -- (inputs);
\draw[flow] (inputs) -- (llm);
\draw[flow] (llm)    --  (reason);
\draw[bi]   (dk.south west) -- (reason.north east);
\draw[flow] (reason) -- (fuzzer);
\draw[fb]   (dk) -- node[lbl, above] {+DK} (llm);
\end{tikzpicture}
\Description{Pipeline diagram. A decompiler produces libc labels and a call graph, which feed the agent (generator). The agent's proposed paths go to the reasoner (verifier), which exchanges information with the domain knowledge and passes entailed paths to a fuzzer. A dashed arrow labeled +DK runs from the domain knowledge to the agent.}
\caption{The end-to-end \system\ pipeline. Solid boxes are \system\ stages,
gray boxes are its inputs, and dashed boxes place the system in the larger
cybersecurity workflow.}
\label{fig:architecture}
\end{figure}

\begin{figure}[tbp]
\begin{promptbox}
\promptsection{System role}
\begin{promptcontent}
You are a security analyst specializing in defensive binary code review\ldots\ generate realistic code-review paths --- ordered sequences of semantic function behaviors an analyst would examine when assessing a binary for a specific weakness type.
\end{promptcontent}

\promptsection{Context}
\begin{promptcontent}
\textit{CWE:} CWE-121, Stack-based Buffer Overflow \textit{(id, name, MITRE description).}\\
\textit{Label vocabulary:} groups of libc labels from Ghidra static analysis, each group co-occurring in one code region, e.g.\\
\texttt{Group 1: [gets, sprintf, strspn]}\\
\hspace*{1em}$\vdots$ \textit{(N groups per binary)}\\[3pt]
\textbf{\textit{+DK.}} The only change under domain-knowledge enrichment: each group is augmented with reasoner-derived DK labels, e.g.\ Group 1 becomes 
\texttt{[gets, sprintf, strspn, missing\_bounds\_check,}
\texttt{unchecked\_memory\_write, stack\_pointer\_overwrite, \ldots]}
\end{promptcontent}

\promptsection{Constraints}
\begin{promptcontent}
Paths must be mutually diverse; framed as analyst exploration (not every stage need relate to the CWE); use only labels from the vocabulary; one label group per stage, no cross-group mixing; minimum 6 edges per path; at most 50 paths.
\end{promptcontent}

\promptsection{Output format}
\begin{promptcontent}
A single JSON object \texttt{\{"paths": [\ldots]\}}; each path has \texttt{path\_id}, an ordered \texttt{cluster\_sequence} of label subsets, an \texttt{exploration\_score} in $[0,1]$, and a short \texttt{explanation}. No text outside the JSON.
\end{promptcontent}

\promptsection{Technique (varies for different agents)}
\begin{promptcontent}
\textit{Zero-shot:} Generate the paths directly.\\[2pt]
\textit{CoT:} First emit a \texttt{label\_analysis} triage of each group (\texttt{immediately\_suspicious} / \texttt{worth\_investigating} / \texttt{likely\_irrelevant}), then generate paths with an added per-stage \texttt{stage\_reasoning} field.\\[2pt]
\textit{Self-Refinement:} Three sequential calls:
(1)~\emph{Generate} an initial path set (as zero-shot).
(2)~\emph{Critique}: the paths are returned to the model, which reviews them for CWE relevance, diversity, ignored labels, and score calibration, replying with plain-text feedback.
(3)~\emph{Refine}: regenerate an improved path set addressing that critique.
\end{promptcontent}
\end{promptbox}
\Description{A boxed prompt template divided into five headed sections: system role, context (the CWE and label vocabulary, with the +DK enrichment), constraints, output format, and technique (zero-shot, chain-of-thought, or self-refinement).}
\caption{Structured prompt template for CWE-121. Full prompts
appear in the appendix.}
\label{fig:prompt}
\end{figure}

Figure~\ref{fig:architecture} shows the \system~ framework within an end-to-end cybersecurity pipeline. The
input to the system is the analyzed binary's call graph together with the libc
labels of its functions. From these inputs, \system\
proceeds through two core stages: agentic path generation and logical
verification, before emitting the paths that survive verification.

\noindent\textbf{Labeled call graph.} In our experiment, we use Ghidra's static analysis to recover the binary's structural representation. It
disassembles the machine code, reconstructs functions and their boundaries,
produces decompiled C-like output, and recovers the inter-procedural call graph
with cross-references. From symbol and relocation information, it resolves the
names of imported libc functions (e.g., \texttt{strcpy}, \texttt{gets},
\texttt{realloc}, \texttt{system}), yielding for each function in the call graph
the set of libc calls it references. Selecting a path from this graph without a parsimony requirement leads to arbitrary outcomes. Evaluating them all is intractable. In our agent, we leverage an LLM's documented capability for searching for potential vulnerability discovery as a parsimony requirement.

\noindent\textbf{Domain Knowledge (DK).}
\label{sec:kg}
DK is a CWE-focused, code-centric ontology together with instantiated knowledge
graphs for CWE-121, CWE-415, and CWE-416. We note that our approach is agnostic to the underlying knowledge graph, and it does not necessarily have to be associated with a CWE.  Ablation studies in section~\ref{sec:circular-arg} show the utility of our method irrespective of the domain knowledge. The ontology used defines a layered class hierarchy, from concrete
beacons, code entities, code operations, and code patterns, to fault conditions
and outcome-level faults and CWE-level vulnerability classes. They are connected by typed
relations such as \texttt{is\_a},
\texttt{can\_cause}, and \texttt{mitigates}. Each graph is populated by automated
extraction from technical corpora (reverse-engineering manuals, security
advisories, man pages), then refined by schema validation with targeted semantic
review; together they span 105--188 entities and 232--403 relations over 14--16
relation types (per-graph counts in Appendix). A central
modeling choice represents faults as unsafe variants of otherwise-neutral
operations; for example, an \texttt{out\_of\_bounds\_write} is an
\texttt{unsafe\_variant\_of} a generic \texttt{memory\_write}. Safe and
unsafe behavior coexist in one graph, and the reasoner can express how missing
validation turns an ordinary operation into a memory-corruption event. DK also
supplies the mapping rules that lift raw libc labels onto its own vocabulary during reasoning, so
the labels attached to functions are expressed in the terms against
which paths are later checked for entailment. For the guided agents (\textit{\textbf{+DK}} experimental setting), we pre-run a single inference step over the call graph, with only rules that derive high-level concept labels for each function. These labels are then appended to that function's label group in the prompt, before any path is generated. The model receives only these per-function labels, never the knowledge graph's relations or the traversal rules that decide entailment. We hypothesize that making semantically meaningful domain knowledge labels available to the agent would improve the proposed paths. Viewed another way, by adding domain knowledge in the LLM prompt, we aim to bias the conditional probability of generated tokens towards an entailed solution. As labels come from the same graph that the verifier reasons with, in Section~\ref{sec:circular-arg} we split each graph into two disjoint halves, guide generation with one and verify with the other. We find that the half that never supplied a path's labels entails it about as often as the half that did, showing no sign of circularity or leakage. Instead, if we added the whole domain knowledge in the context window, it would leak the graph relations to the generator, turning path proposal into a lookup of transitions the verifier already accepts and making the result circular.

\noindent\textbf{Step 1: Agentic Generation.}
The agent's first component is an LLM. The LLM does not propose paths directly, but rather path templates. Figure~\ref{fig:prompt} shows
the prompt structure. Every configuration receives the same request, $N_{\text{req}}$ templates under identical beam-search settings.
The model is given the target CWE and the binary's label
vocabulary, and asked for path templates: a sequence of label subsets rather than of functions. 
The
constraints ensure that the model cannot attribute behavior the binary does not exhibit, and that they are framed as analyst
exploration rather than as vulnerability reports, so not every stage needs to be related to the CWE. The prompt allows a model to return fewer templates than requested, so that it is not forced to pad its output with low-quality proposals.
The agent then processes the LLM's output using beam search, which searches the binary's call graph for a function carrying each stage's labels, with consecutive stages at most $h$ call edges apart. In real life, an analyst often skips over minor functions in an execution path and focuses on what they deem worthy of more time. We choose $h=3$ hops in the call graph to find a middle ground in terms of permissivity. Another common analyst behavior is backtracking while analyzing the codebase; the search takes that into account by allowing transitions in either call direction (down to a callee or back up to a caller). The intuition is that the template, through label subsets, ensures key functions are picked up. It then discards templates it cannot complete. From each realizable template, the agent keeps up to $N_{\text{path}}$ distinct realized paths. A realized path is a sequence of steps an analyst might take while examining the code in search of a vulnerability. Each step is a function that exists in the binary, at most three hops from its neighbor. It is not required to be an execution path (that is a question for a fuzzer tool), but rather a possible exploration path. The deduplicated realized paths are the generator agent's output and are referred to as `paths' throughout this paper.

\noindent\textbf{Step 2: Reasoner Verification.}
\label{sec:logic-program}
The reasoner takes the generated paths as input and executes the domain
knowledge as a logic program. The logic program comprises the DKG's facts and mapping rules together with
42 non-ground rules obtained by instantiating a small set of hand-written
schemas over the graph's entity--relation--entity triples. The rules fall into three families: label propagation, transitive closure, and analyst traversal with step control, of which
Table~\ref{tab:example_rules} gives one traversal rule and one
label-propagation rule. Inference is carried out with PyReason~\cite{pyreason2023}, an open-source inference engine for generalized annotated logic that reasons directly over knowledge graphs. 
Verifying a single step may require several rules. A label-propagation rule may first combine the raw libc labels at one block into
a semantic label, then a DK mapping rule lifts that label onto a DKG concept, and a
transitive-closure rule chains relations across several hops, before a traversal
rule can fire and validate the analyst's move. Because PyReason computes the
minimal model of the whole program, the inference is exact rather than a heuristic
graph walk. If no combination of rules can validate a step, that step is
non-entailed by the domain knowledge. We report the step along with the full reasoning
trace (Figure~\ref{fig:deployment-trace}). If any one step of a path is non-entailed, the entire path is non-entailed.

\section{Experiments \label{sec:experiments}}
\subsection{Experimental Setup}
\label{sec:exp-setup}

We evaluate across three weakness classes---CWE-121 (stack-based buffer
overflow), CWE-415 (double free), and CWE-416 (use-after-free). Standard target binaries (\texttt{std}) are drawn from BinPool~\cite{binpool2025}, a
dataset of Debian binaries curated from historical CVEs; it spans 603 CVEs across 89 CWE
classes, so the CWEs we target are represented by real compiled programs whose
call graphs and libc usage resemble production firmware. Real-world targets (\texttt{rw}) are firmware extracted
from deployed medical devices, confidentially provided, and are the targets \system\ is required to analyze in our production environment.
Table~\ref{tab:callgraphs} lists the eight binaries behind the results reported
here. \texttt{Ezurio} and \texttt{htmldoc} binaries are evaluated for multiple CWEs, making 11 binary--CWE pairs. Sizes span two orders of magnitude, from \texttt{gpac}
(405 nodes, 571 edges) to \texttt{unsafelib} (12{,}696 nodes, 40{,}881 edges),
exercising the pipeline on both small and large call graphs.

\begin{table*}[tbp]
\centering
\caption{Two example rules from the logic program and their description in natural language. $\mu_{\mathrm{pair}}$ is a pairwise-minimum bound function over annotations in the rule body, applied to the head atom if the rule is fired. Annotations in the body are bounds that must be contained by any possible grounding during the reasoning process.  The full program appears in Appendix.}
\label{tab:example_rules}
\begin{tabular}{p{0.45\textwidth}p{0.5\textwidth}}
\toprule
{\bf Rule} & {\bf English Description} \\
\midrule

\(analystAt(CB_2):\mu_{\mathrm{pair}} \xleftarrow\ analystAt(CB_1):[0.25,1] \wedge hasLabel(CB_1, L_{c}):[0.1,1] \wedge hasLabel(CB_2, L_{e}):[0.1,1] \wedge can\_cause(L_{c}, L_{e}):[0.1,1] \wedge stepFrom(CB_1, CB_2):[1,1]\) &
If the analyst is at code block $CB_1$, which is labeled with $L_c$, and code block $CB_2$ is labeled with $L_e$, and the domain knowledge asserts $L_c$ \texttt{can\_cause} $L_e$, then the analyst may advance to $CB_2$.\\
\cmidrule{1-2}
\(hasLabel(X, copy\_operation):[0.9,1] \xleftarrow\ hasLabel(X, sprintf):[0.6,1] \wedge hasLabel(X, strcpy):[0.6,1]\) &
If code block $X$ is observed to use both \texttt{sprintf} and \texttt{strcpy}, it is inferred with at least 90\% confidence to perform a \texttt{copy\_operation}.\\
\bottomrule
\end{tabular}

\end{table*}

\begin{table}[!tbp]
\centering
\caption{The binaries used across the target
CWEs.}
\label{tab:callgraphs}
\begin{tabular}{@{}lllcc@{}}
\toprule
\textbf{CWE} & \textbf{Binary} & \textbf{Type} & \textbf{Nodes} & \textbf{Edges} \\
\midrule
\multirow{3}{*}{CWE-416} & gpac          & std &    405 &    571 \\
                         & mupdf-x11     & std & 3{,}341 & 11{,}883 \\
                         & mutool        & std & 3{,}569 & 14{,}051 \\
\cmidrule{1-5}
\multirow{3}{*}{CWE-121} 
                         & htmldoc       & std & 1{,}086 &  2{,}815 \\
                         & insulininject & rw & 1{,}010 &  2{,}774 \\
                         & unsafelib    & rw & 12{,}696 & 40{,}881 \\
\cmidrule{1-5}
\multirow{2}{*}{CWE-415} & dropbear      & std &    816 &  2{,}416 \\
                         & htmldoc       & std & 1{,}086 &  2{,}815 \\
\cmidrule{1-5}
All & ezurio   & rw & 1{,}155 &  3{,}322 \\
\bottomrule
\end{tabular}

\end{table}
\noindent\textbf{Agents and Models.} The agents are classified by the prompting strategy (\textit{zero-shot, chain-of-thought, self-refinement}) used. Self-refinement~\cite{madaan2023self} is a metacognitive strategy~\cite{wang2024metacognitive} in which the model critiques and then revises its own proposals. Then we also separate agents based on whether or not domain knowledge was used to guide the agent a priori.
All agents are evaluated with four LLMs in the paper: Opus-4.8, GPT-5.5, Llama~4 Scout, and
Kimi~K2.5, spanning proprietary and open-weight families.
This gives us six agents and a total of 24 configurations
per binary-CWE pair. Our initial selection of LLMs also included Claude Fable~5, but its built-in
safeguards, which redirect queries on sensitive topics including
cybersecurity to a less specialized model, were triggered by our
vulnerability-analysis prompts and prevented its use in our agents.

\noindent\textbf{Hyperparameters.}
All LLMs use greedy decoding ($t{=}0.0$), since we evaluate each
model's most likely output rather than a sample. We request
the LLM for $N_{\text{req}}=50$ templates per call, but it had to be lowered to $20$ for Llama~4
Scout, whose $8{,}192$-token output cap truncates a 50-template JSON. Paths must
span at least six call-graph edges, lowered to four for \texttt{gpac}, whose
405-node graph is too shallow to admit many six-edge paths. Consecutive steps of a path may be up to $h=3$ call edges apart (maximum hop distance). The parameter $N_{\text{path}}$, the
number of realized paths kept per realizable template, was tuned---over
$\{1,2,5,10\}$: smaller values degraded results and larger ones raised runtime, 
and LLM cost for no gain, so we use $5$. Full settings and rationale are in
Appendix.

\noindent\textbf{Metrics.}
\label{sec:exp-metrics}
\texttt{\#Paths} is the number of unique paths an agent outputs. \texttt{\#Entailed} gives how many of those the domain knowledge entails, with 
    $\text{Entailment\%} = \#\text{Entailed}\,/\,\#\text{Paths} \times 100$.
\texttt{\#Paths} directly depends on how many templates a model returns and how many of those are realizable. \texttt{\#Entailed} can therefore never exceed \texttt{\#Paths}, and a low or high number on its own does not reflect the performance. On the other hand, a high entailment\% signifies a good result, and a higher \texttt{\#Entailed} is desirable alongside it. We tracked template returns for CWE-121 and CWE-416: across these two classes, models returned $53\%$ of the $8{,}160$ templates requested, and $62\%$ of the returned templates were realizable (Appendix). Figure~\ref{fig:arrows} plots \texttt{\#Entailed} against \texttt{Entailment\%}; The full table for every configuration is in the appendix. An agent-proposed path for which no chain of rule firings can validate every step is a non-entailed trajectory, as the domain knowledge does not support it. When
aggregating over configurations we pool \texttt{\#Entailed} and \texttt{\#Paths} rather than
averaging per-configuration percentages, so that generations are weighted by the
evidence they carry.

\noindent\textbf{Compute.}
All experiments ran on a single server (AMD EPYC 9755, 128 cores, 754\,GiB RAM,
Ubuntu 24.04, Python 3.10), with every model accessed through a hosted API
(Anthropic, OpenAI, AWS Bedrock). Peak resident memory was $\sim$21\,MB for the
generation stage and $\sim$931\,MB for the reasoner, about $0.1\%$ of available
RAM, so the pipeline reproduces on commodity hardware.

\subsection{Results and Discussion}
\label{sec:exp-results}

\begin{figure*}[!tbp]
\centering
\includegraphics[width=\textwidth]{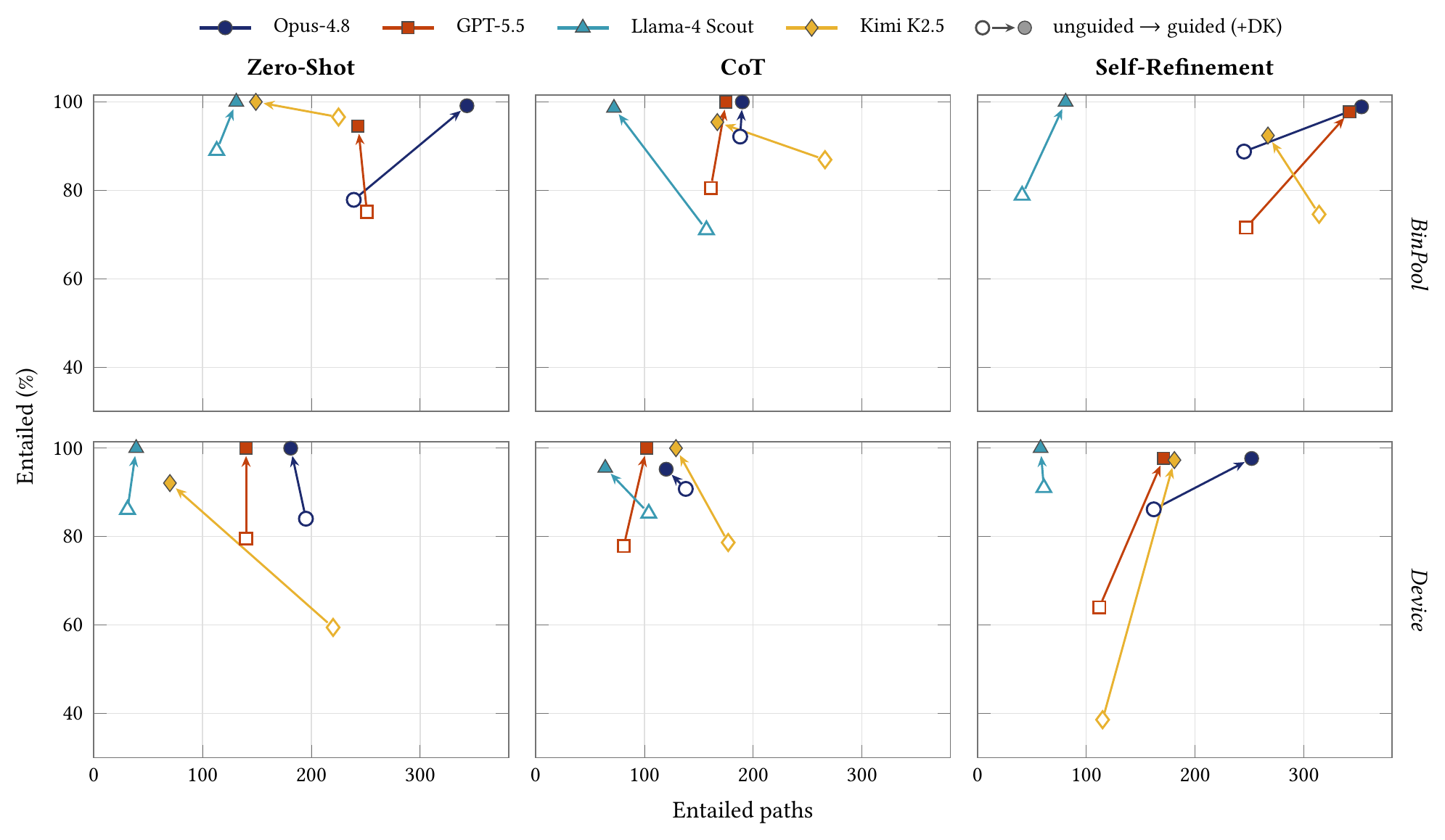}
\Description{Six scatter panels in two rows and three columns. Columns are the prompting strategies (Zero-Shot, CoT, Self-Refinement); the top row, labeled BinPool, pools the standard binaries and the bottom row, labeled Device, the real-world medical-device binaries, each over all three CWEs. The x-axis is the number of entailed paths and the y-axis the percentage of paths entailed. Each of the four LLMs has its own colour and marker shape; an arrow runs from a hollow marker (unguided agent) to a filled marker (guided agent, +DK).}
\caption{Effect of domain knowledge, pooled over all three CWEs. (Top) the standard BinPool binaries (gpac, mupdf-x11, mutool, htmldoc, dropbear) and (bottom) the real-world medical-device binaries (insulininject, unsafelib, ezurio). Columns are different agents, and panels share both axes. For an agent, each arrow is for a different LLM, going from entailment rate without domain knowledge (hollow) to that with domain knowledge (+DK, filled). Better results are up and to the right.}
\label{fig:arrows}
\end{figure*}
\begin{figure}[!tbp]
\centering
\includegraphics[width=\columnwidth]{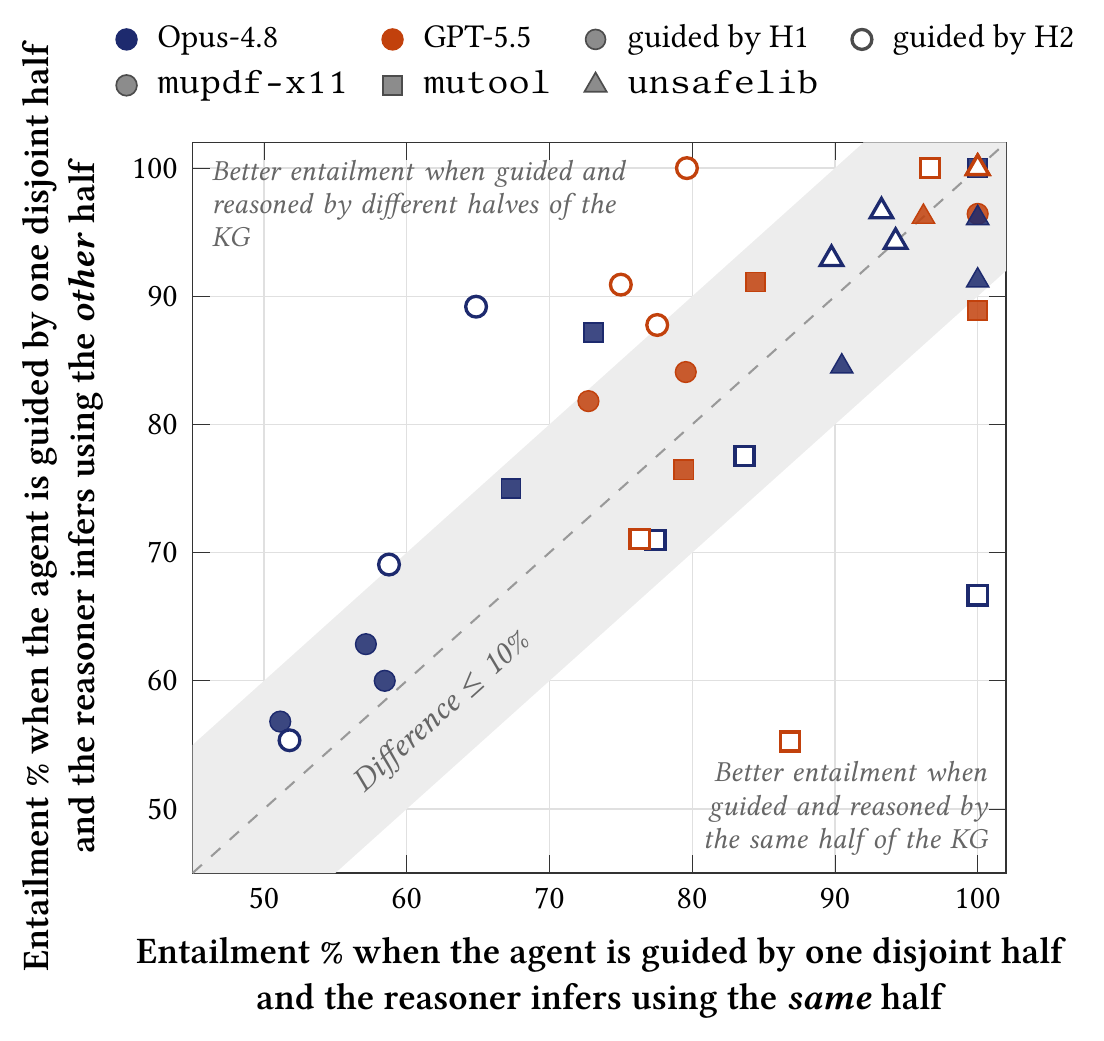}
\Description{A scatter plot with 36 points, one per binary, model, prompting strategy and guiding half. The x-axis is the percentage of paths entailed when the agent is guided by one disjoint half of the knowledge graph and the reasoner uses the same half; the y-axis is the percentage when the reasoner uses the other half. Colour marks the model, shape the binary, and filled or hollow markers the guiding half. A shaded band marks differences of at most 10 points around the diagonal, with the corners labeled for better entailment under different halves (top left) and under the same half (bottom right); most points lie inside the band, on both sides of the diagonal.}
\caption{Disjoint halves of the DK are used to guide the agent separately. Then each output is verified against both halves. Entailment rates are plotted for three binaries (shape), two LLMs (colour) for all guided agents. Filled and unfilled markers denote the two halves of the KG. The x-axis gives entailment under the guiding half and the y-axis under the other half. Most points lie in the shaded band where the difference is at most $10\%$, showing no leakage or circularity.}
\label{fig:kgsplit-all}
\end{figure}

Figure~\ref{fig:arrows} shows the effect of domain knowledge on the standard and real-world binaries (top and bottom rows), pooled over all three CWEs. Per-configuration counts are in the Appendix.

\noindent\textbf{Domain knowledge improves entailment for both binary types.}
Pooled entailment increases from $80.9\%$ to $97.7\%$ for standard BinPool binaries and from $71.6\%$ to $98.0\%$ for real-world medical device binaries, and the number of configurations at $100\%$ entailment rises from 22 to 55 and from 17 to 43. We see this in figure~\ref{fig:arrows} as all arrows point upward. Across the 118 configurations where both the unguided and guided agent produced at least one path, entailment improves in 77, is unchanged in 38, and
only decreases in 3. A one-sided Wilcoxon signed-rank test over those 118 paired configurations
confirms the shift ($W^{+}=3180$, $W^{-}=60$, $p<10^{-13}$; median $+8.8$
points); the 38 unchanged configurations were already at $100\%$ entailment and
are excluded as ties. The result holds for each binary type separately
($\text{std } p=7.1\times10^{-14}$; $\text{rw } p=1.3\times10^{-6}$). We exercise caution with this evidence as these configurations are not fully independent: configurations that share a model, a binary--CWE pair, or a strategy are correlated, so we note that the p-value above likely overstates the evidence. However, treating each binary--CWE pair as a single observation, pooled entailment still improves on all 11 pairs (exact one-sided sign test, $p=4.9\times10^{-4}$), and it improves for every agent with all models. This shows the modularity of our approach of being able to switch out the LLM in the overall framework.

\noindent\textbf{Guided Self-Refinement Agent yields the most entailed paths.}
It produces more entailed paths than any other agent: 1{,}043 on
the standard binaries against 866 for zero-shot and 604 for chain-of-thought (CoT),
and 662 on the device binaries against 430 and 415. We see for most LLMs, arrows for self-refinement in figure~\ref{fig:arrows} point up and to the right. It is also the only agent
for which guidance raises the entailed count along with the
rate on both binary types. For some configurations, the agent returns no paths: either the LLM returns no templates, or the agent finds none of the proposed templates realizable. Excluding those configurations, self-refinement agents add 196 paths on standard binaries and 212 on device binaries, and
do so in 29 of their 39 individual configurations against 22 of 40 for
zero-shot and 16 of 39 for CoT agents. We note the unguided self-refinement agent is actually the least reliable---$61.8\%$ entailment on
device binaries. So while it has the most to gain from guidance, the logical layer utilizes that exploration capability of the self-refinement agent and produces improved verified output. One possible explanation is the metacognitive nature of self-refinement. Its critique step potentially allows the model to use the domain knowledge to review its own proposals before revising them.
Figure~\ref{fig:arrows} also shows the cost in output volume. For zero-shot and CoT agents, several arrows go up and left, meaning that while domain knowledge raises the rate, the agent also returns fewer entailed paths (most sharply Kimi~K2.5 zero-shot on the device binaries, from 220 to 70 entailed paths). Whereas for the self-refinement agent, the arrows point up and to the right in six of eight model--binary-type pairs. Counted per model, binary and CWE, guided self-refinement raises or maintains the entailment rate in 38 of 39 configurations, and both the rate and the entailed count in 29 (CWE-121: 14 of 16; CWE-415: 5 of 8; CWE-416: 10 of 15). Results for individual configurations are in the Appendix.

\noindent\textbf{Results translate to real-world binaries.}
The real-world binaries used in these experiments are not publicly available. They were obtained through partnerships with medical device vendors. Entailment for unguided agents on the device binaries is lower than the BinPool targets ($71.6\%$ vs $80.9\%$), yet with domain knowledge the two converge to
$98.0\%$ and $97.7\%$. This shows that the approach generalizes to uncommon binaries.

\noindent\textbf{Separating guidance from verification.} \label{sec:circular-arg} A natural concern could be that the guidance labels come from the same graph that the symbolic system uses for reasoning, and whether the \textbf{\textit{+DK}} setting simply rewards the agent for matching its evaluator. In this neurosymbolic framework, the neural LLM of the agent never trains on the domain knowledge, and the symbols are presented to it only during runtime through the prompt. It has no visibility of the logical rules or the relations in the knowledge graph, which drive the entailment check. Critically, no model is fine-tuned on these labels or on entailment outcomes. That the labels help is therefore an empirical finding, and this work attempts to find an effective way of guiding the neural system.
To rule out circularity, we split each knowledge graph into two disjoint halves. Each half keeps every node and edge class. Classes are coarse-grained. For example, labels denoting different types of functions all belong to the same generic class `\texttt{functions}'. Each half receives only half of the edges of each class. So the two halves share no edges. The split is made on the graph before the reasoner applies the transitive-closure rules, so a half can only infer what its own edges imply. On the three binaries with more than $3{,}000$ call-graph nodes (\texttt{mupdf-x11}, \texttt{mutool}, \texttt{unsafelib}), we run the agents with Opus-4.8 and GPT-5.5, using labels from one half only, and verify each set of paths against the same half and against the other half (tables in the Appendix). If the gain came from the agent seeing its evaluator's facts, paths would be entailed more often by the half that guided them, and the points in Figure~\ref{fig:kgsplit-all} would be below and to the right of the diagonal. However, across the $36$ comparisons, the mean difference between other-half and same-half entailment is only $+0.8$ points, and $27$ of the $36$ points lie within $10\%$ points of the diagonal. This shows that guidance does not unfairly inflate the experimental results, rather, the semantic concepts are a useful tool that can be used to guide the neural component of a neurosymbolic framework.

\noindent\textbf{Stability across runs.}
To check that these gains are not an artifact of a single run, we repeated the
full pipeline ten times and measured the
spread in entailment across repetitions. As expected, LLM API responses varied, but domain knowledge both raised entailment and sharply narrowed its spread across
runs under all three prompting strategies. Results were analogous to a single run. Plots are included in the Appendix.

\section{\system~Deployment}
\label{sec:deploy}

\begin{figure}[tbp]
    \centering
    \includegraphics[width=0.99\columnwidth]{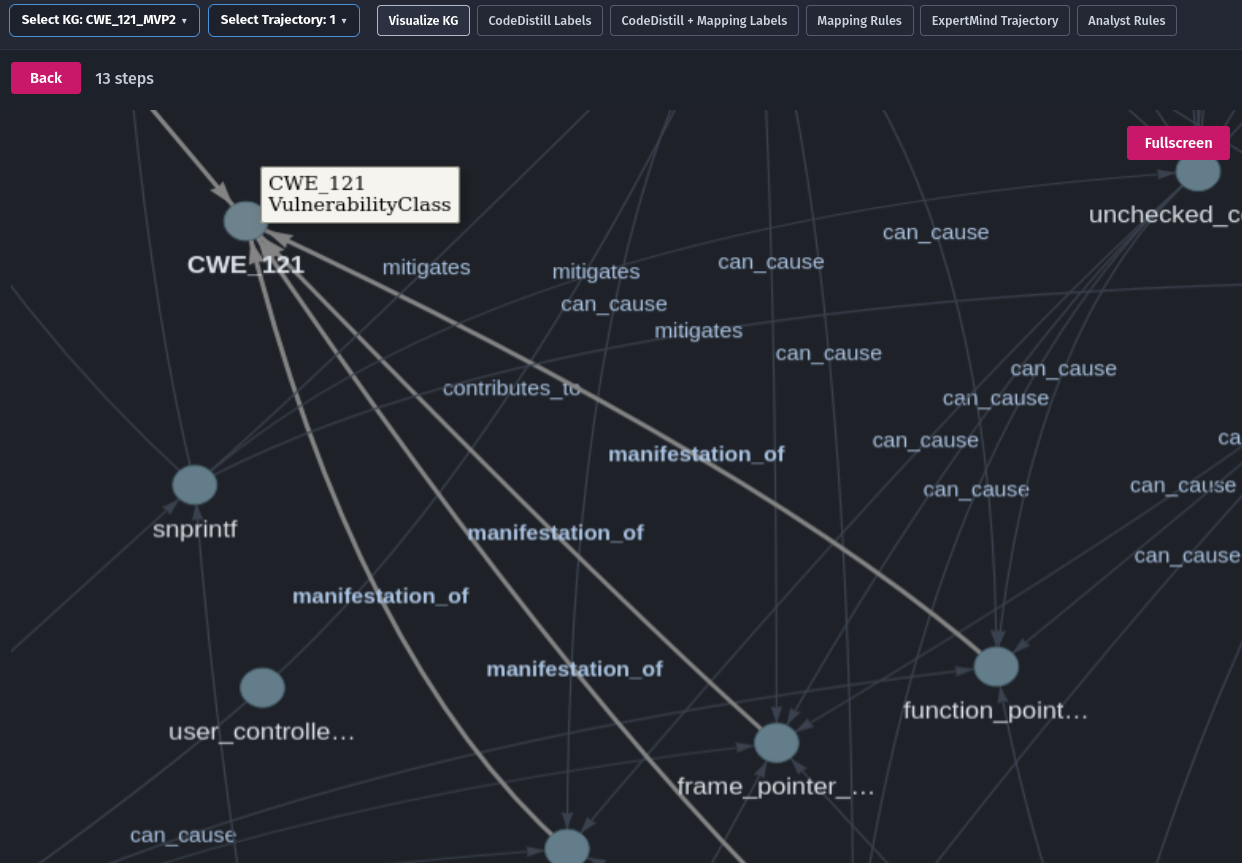}
    \Description{Screenshot of part of the CWE-121 domain-knowledge graph in the deployed interface, with highlighted entities such as function pointer linked to the CWE-121 vulnerability class through relations such as manifestation of.}
    \caption{An excerpt of the \texttt{CWE\_121} DK graph in the deployed system. Highlighted portions show how entities like \texttt{function\_pointer} relate to the \texttt{CWE\_121} vulnerability class through KG relations like \texttt{manifestation\_of}.}
    \label{fig:deployment-kg}
\end{figure}

\system\ is deployed end-to-end inside a vulnerability-discovery pipeline,
occupying the solid boxes of Figure~\ref{fig:architecture}. A decompiler
upstream supplies the labeled call graph; \system\ proposes candidate analyst
paths over it, checks each against the domain knowledge, and passes only
the entailed ones to downstream tooling such as a fuzzer. Results on the three binaries extracted from deployed medical devices are
reported in Figure~\ref{fig:arrows} (bottom row).
\system~runs as an interactive web application served from a REST API. The interface mirrors the system's data flow. Figure~\ref{fig:llm-mistake} is a screenshot of the UI showing the graph of an analyzed binary with an agent-proposed path highlighted. Figure~\ref{fig:deployment-kg} shows a snippet of the analyst knowledge graph on the UI.

\begin{figure}[tbp]
    \centering
    \includegraphics[width=\columnwidth]{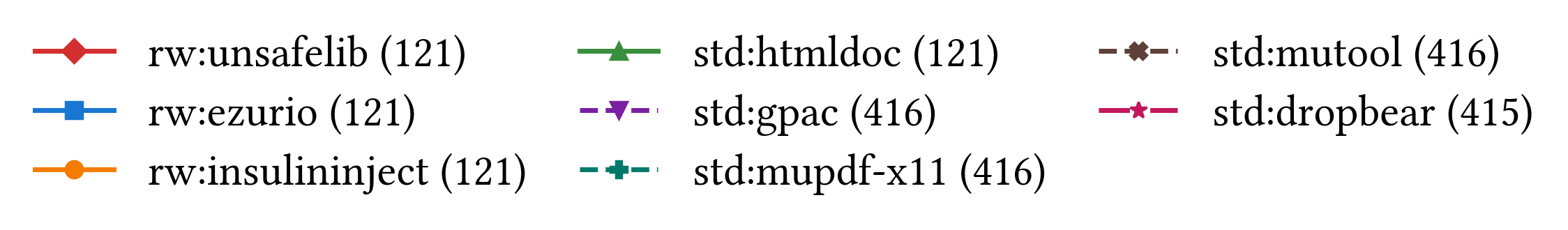}\\[2pt]
    \includegraphics[width=0.49\columnwidth]{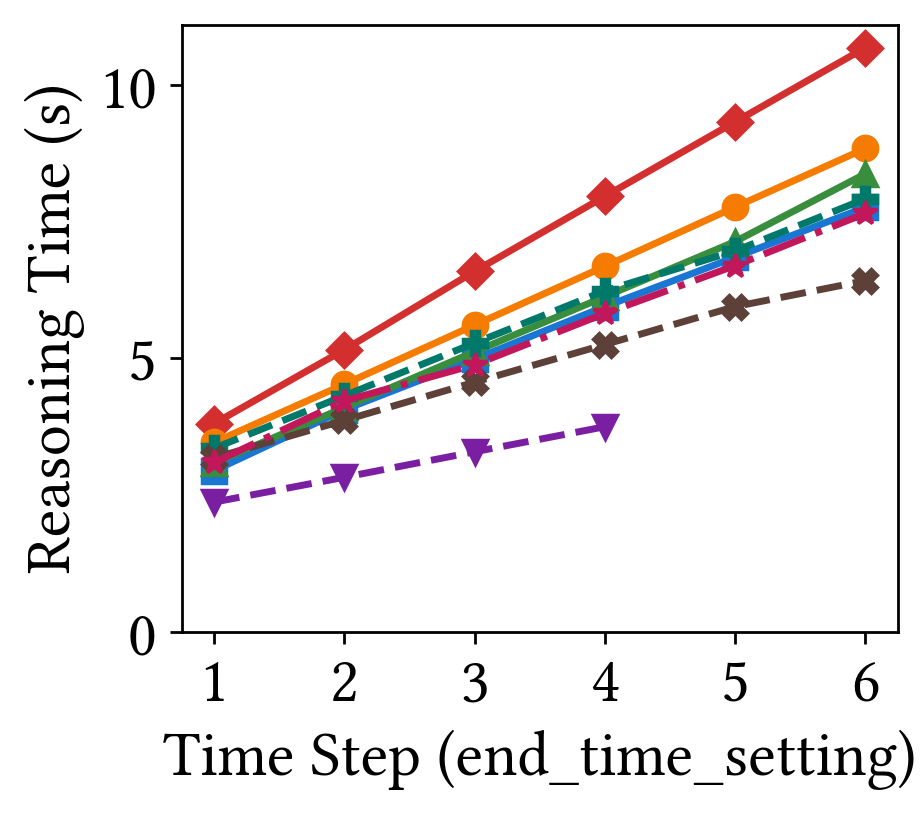}\hfill
    \includegraphics[width=0.49\columnwidth]{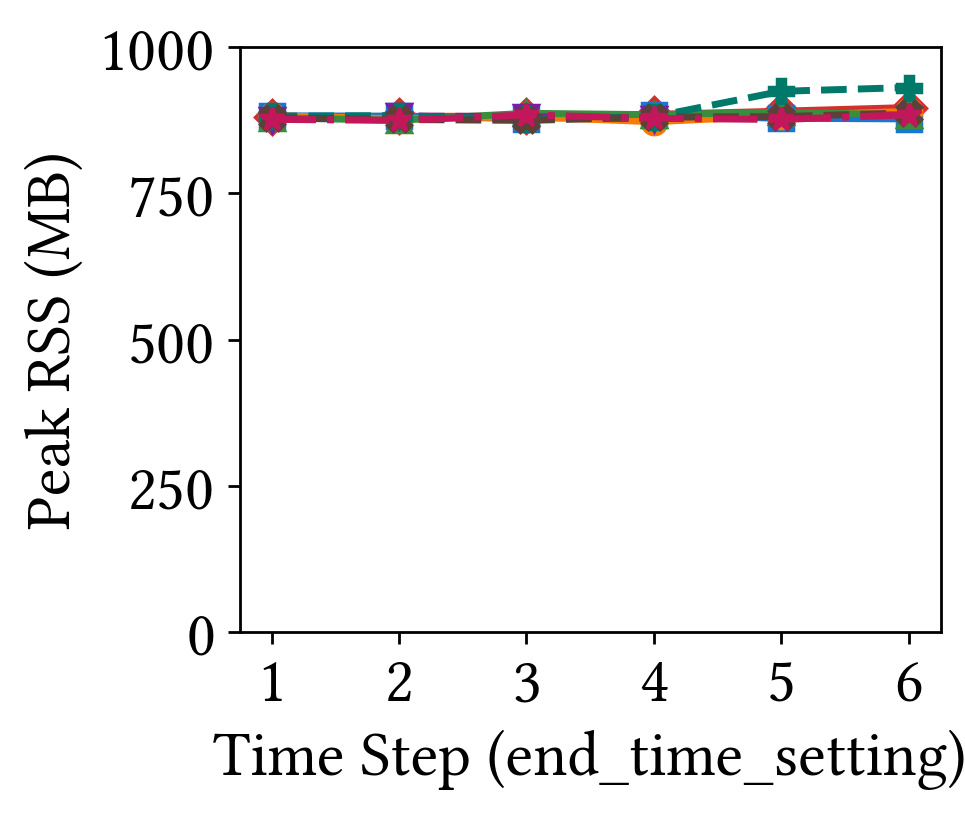}
    
    \Description{A shared legend above two side-by-side plots for eight binaries. The left plot shows reasoning time growing roughly linearly with the number of reasoning steps; the right plot shows peak memory staying mostly constant.}
    \caption{Plots showing how
    (left) reasoning time and (right) peak memory of the reasoner scale for eight binaries.}
    \label{fig:scalability}
\end{figure}

\noindent\textbf{Scalability.}
Figure~\ref{fig:scalability} shows the scaling capability of the reasoner. Memory remains mostly constant, while runtime scales linearly with steps of reasoning.

\begin{figure}[tbp]
    \centering
    \includegraphics[width=0.99\columnwidth]{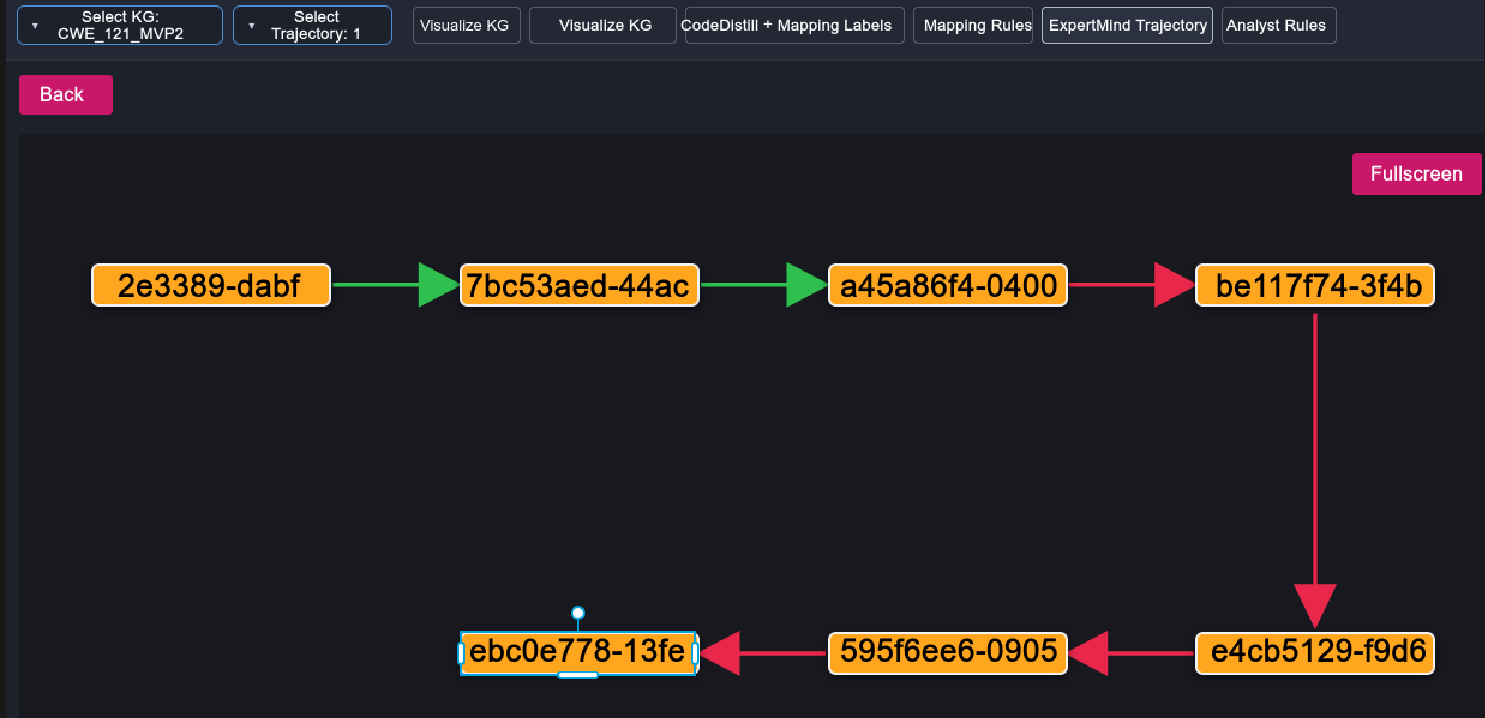}
    \Description{Screenshot of the deployed interface showing one analyst path as a chain of seven function nodes, each labeled with a short hexadecimal identifier and joined by arrows. The first two arrows are green, marking entailed steps; the remaining four arrows are red, marking non-entailed steps.}
    \caption{The explainable trace showing entailment (first two, in green) and non-entailment (rest, in red) for the analyst path selected by the agent in Figure~\ref{fig:llm-mistake}.}
    \label{fig:deployment-trace}
\end{figure}

\noindent\textbf{Trace.} For every non-entailment, a reasoning trace, shown in Figure~\ref{fig:deployment-trace}, is produced. This shows the exact step that was non-entailed by domain knowledge. On click on the UI, a more detailed trace showing the complete reasoning path leading to the non-entailment conclusion is available, and an example is provided in the Appendix. 

\noindent\textbf{Compute.}
The system is deployed on an AWS EC2 \texttt{r5ad.2xlarge} instance (8 cores, AMD EPYC 7571, 64 GB RAM)---an order of magnitude smaller than the experiment
server of Section~\ref{sec:exp-setup}.

\noindent\textbf{Downstream vulnerability discovery.}
During deployment, \system's trajectory-generation and symbolic-verification approach was integrated into an operational vulnerability-analysis platform for medical-device software and firmware. The pipeline supplies structured reasoning results from \system~to downstream vulnerability discovery and exploit validation agentic cybersecurity tool, which is a proprietary product of an independent external company. In the deployed system, this tool used the trajectories from \system~to guide exploration through a binary in order to generate executable proofs of vulnerability. The cybersecurity tool validated that during deployment, these paths led to actual exploits. We have found
validated vulnerability labels for real device firmware to be scarce, and we are
working to obtain them through coordinated disclosure, vendor collaboration, and
expert red-teaming. Once obtained, we plan to run further evaluations and release a full report on the deployed system.

\section{Conclusion and Future Work \label{sec:conclusion}}

We presented \system, a neurosymbolic framework that pairs agentic generation of analyst
traversal paths through a software codebase with logic-programming verification
against domain knowledge. An output path is not merely a plausible
traversal of the function call graph according to an LLM-based agent, but provably entailed
by the encoded domain knowledge. We examined six agents with four LLMs, for eight binaries and domain
knowledge raises pooled entailment from $77.0\%$ to $97.8\%$, decreasing for only 3 of 118 configurations. This shows the potential of guiding a neural agent with symbolic knowledge. The strongest agent follows a self-refinement prompting strategy guided by domain knowledge, which raises entailment from $71.2\%$ to $97.2\%$ while increasing the number of entailed paths from $1{,}297$ to $1{,}705$. 
In real-world medical-device firmware, in comparison to standard dataset binaries, the entailment rate drops for unguided agents. However, with domain knowledge, rates for both classes of binaries converge. This shows that the system can generalize and the improvements transfer to deployed firmware.

Other than a full report on the deployed system, we plan to extend our work in two directions. First, we plan to include a neurosymbolic corrector module that takes a non-entailed path as an input and attempts to produce a logically entailed alternative with minimal edits. Second, we plan to study the utility of incorporating a cognitive model~\cite{anderson1997act} in our framework.

\begin{acks}
This research was, in part, funded by the Advanced Research Projects Agency for Health (ARPA-H). The views and conclusions contained in this document are those of the authors and should not be interpreted as representing the official policies, either expressed or implied, of the U.S. Government.
Research in this paper is related to the invention described in U.S. provisional patent application 64/077,998: Automatic Vulnerability Analysis of Software Consistent with Domain Knowledge.
\end{acks}

\bibliographystyle{ACM-Reference-Format} 
\bibliography{master}

\end{document}


\maketitle
\typeout{SUPP textwidth = \the\textwidth}

\appendix

\begin{multicols}{2}
\small
\tableofcontents
\end{multicols}
\vspace{2pt}

\section{Stability and the No-LLM Baseline}
\label{app:stability}

\begin{figure}[!htbp]
\centering
\includegraphics[width=0.9\columnwidth]{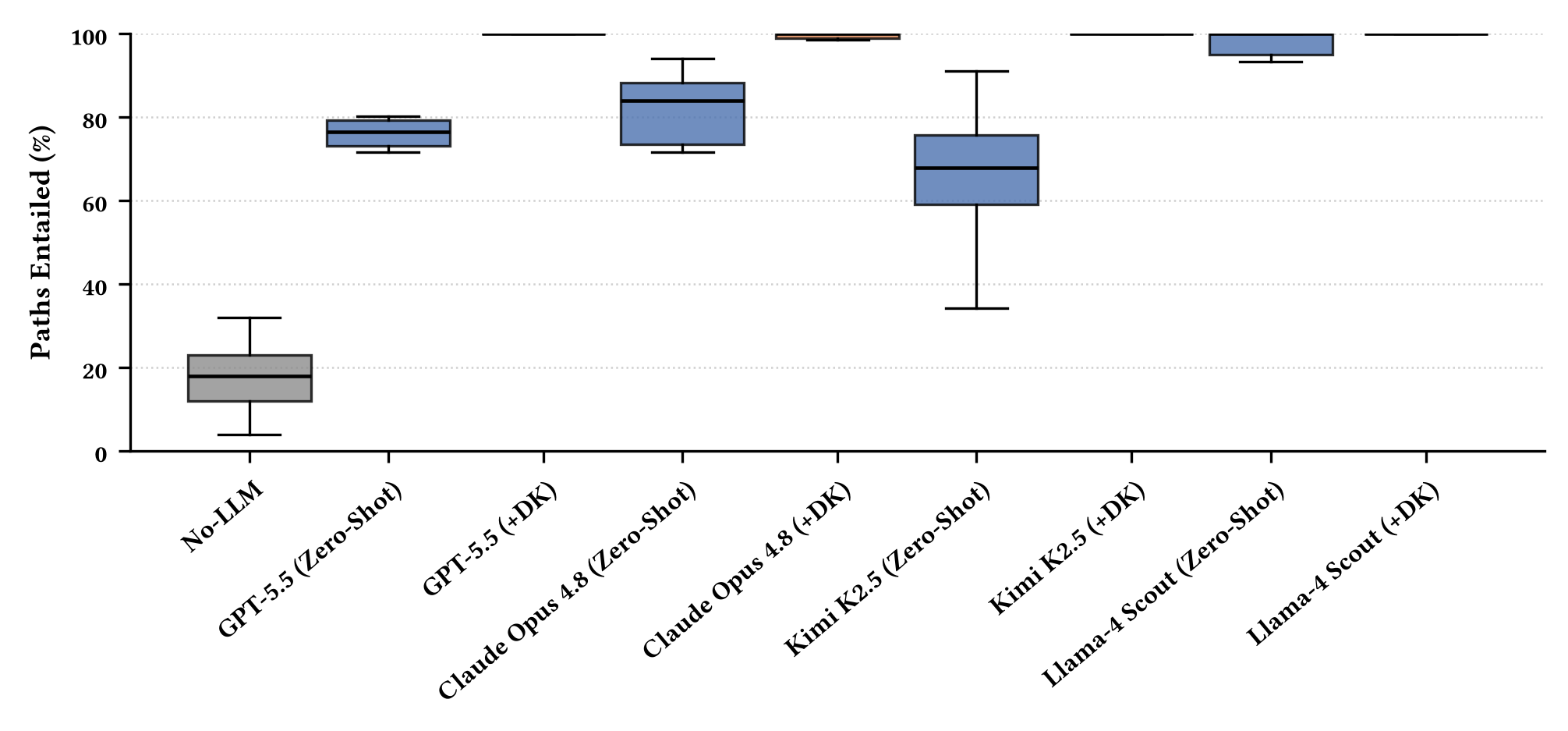}
\caption{Zero-shot on \texttt{insulininject} (CWE-121). Each box is ten
independent runs: grey is No-LLM, blue the unguided agent, orange the
guided agent (+DK) with the same LLM. No-LLM attains
$18.0\pm8.5\%$ entailment. A box collapsed to a single line indicates zero
variance across runs.}
\label{fig:stab-ins-ct1}
\end{figure}

\begin{figure}[!htbp]
\centering
\includegraphics[width=0.9\columnwidth]{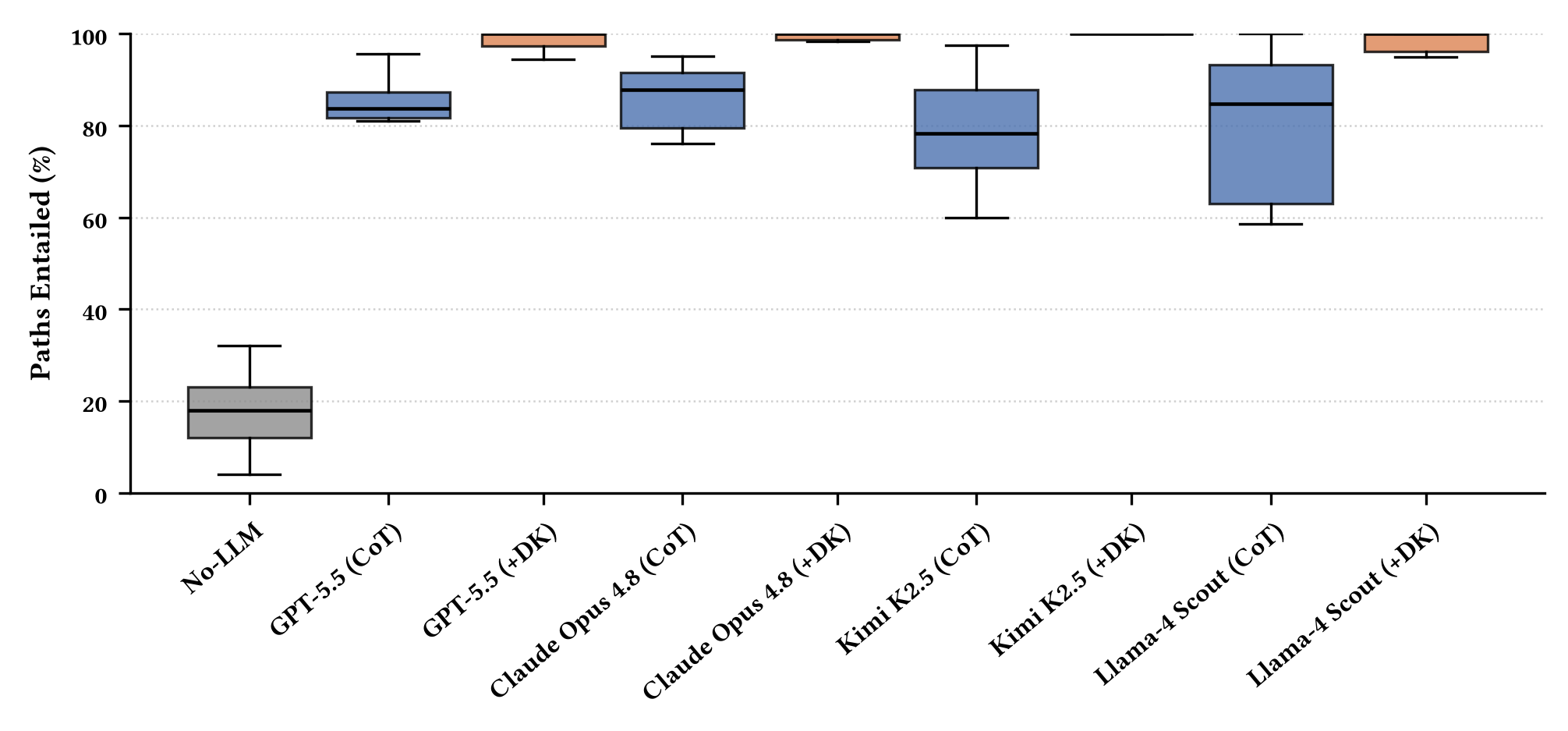}
\caption{Chain-of-thought on \texttt{insulininject} (CWE-121). Conventions as in
Figure~\ref{fig:stab-ins-ct1}.}
\label{fig:stab-ins-ct2}
\end{figure}

\begin{figure}[!htbp]
\centering
\includegraphics[width=0.9\columnwidth]{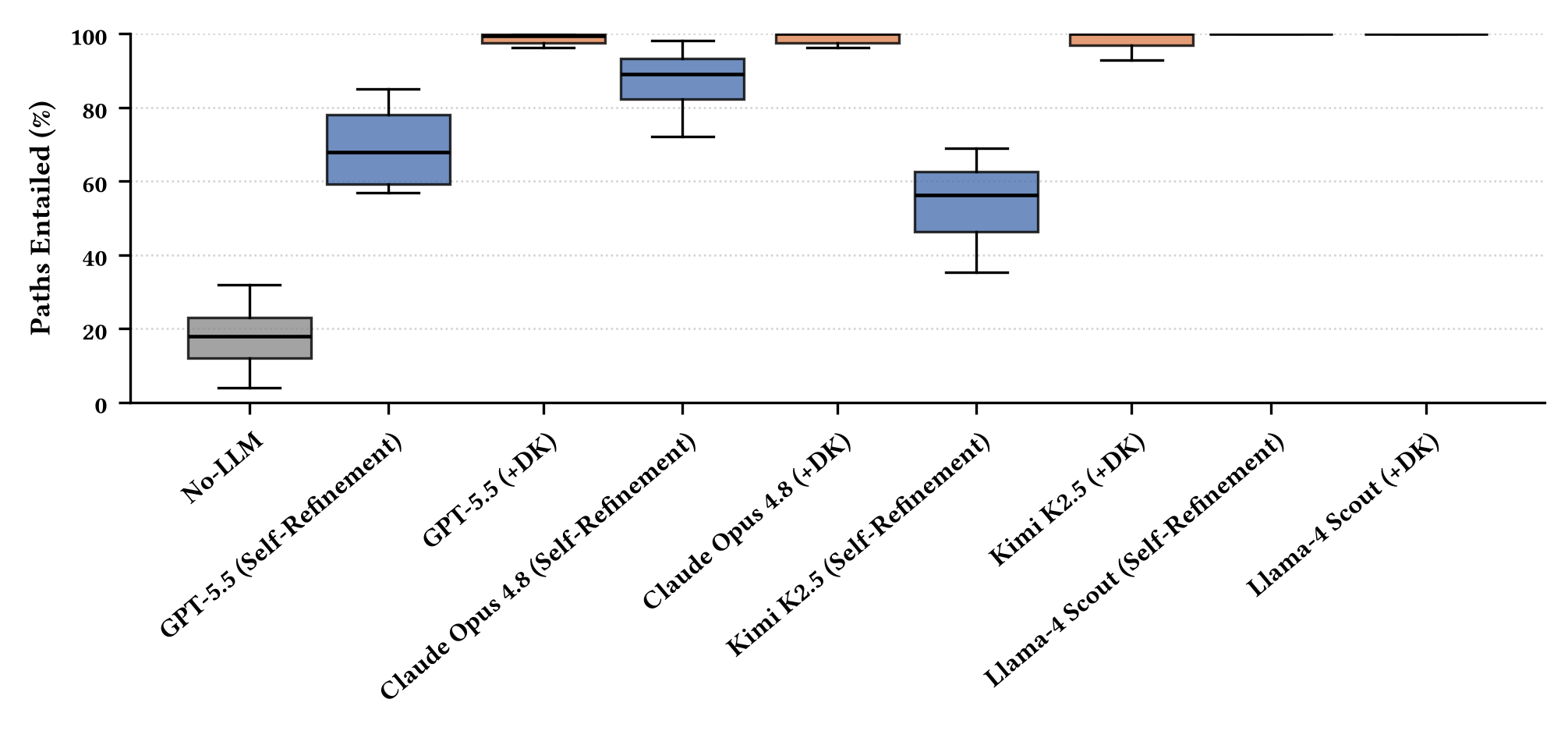}
\caption{Self-Refinement on \texttt{insulininject} (CWE-121).}
\label{fig:stab-ins-ct3}
\end{figure}

\begin{figure}[!htbp]
\centering
\includegraphics[width=0.9\columnwidth]{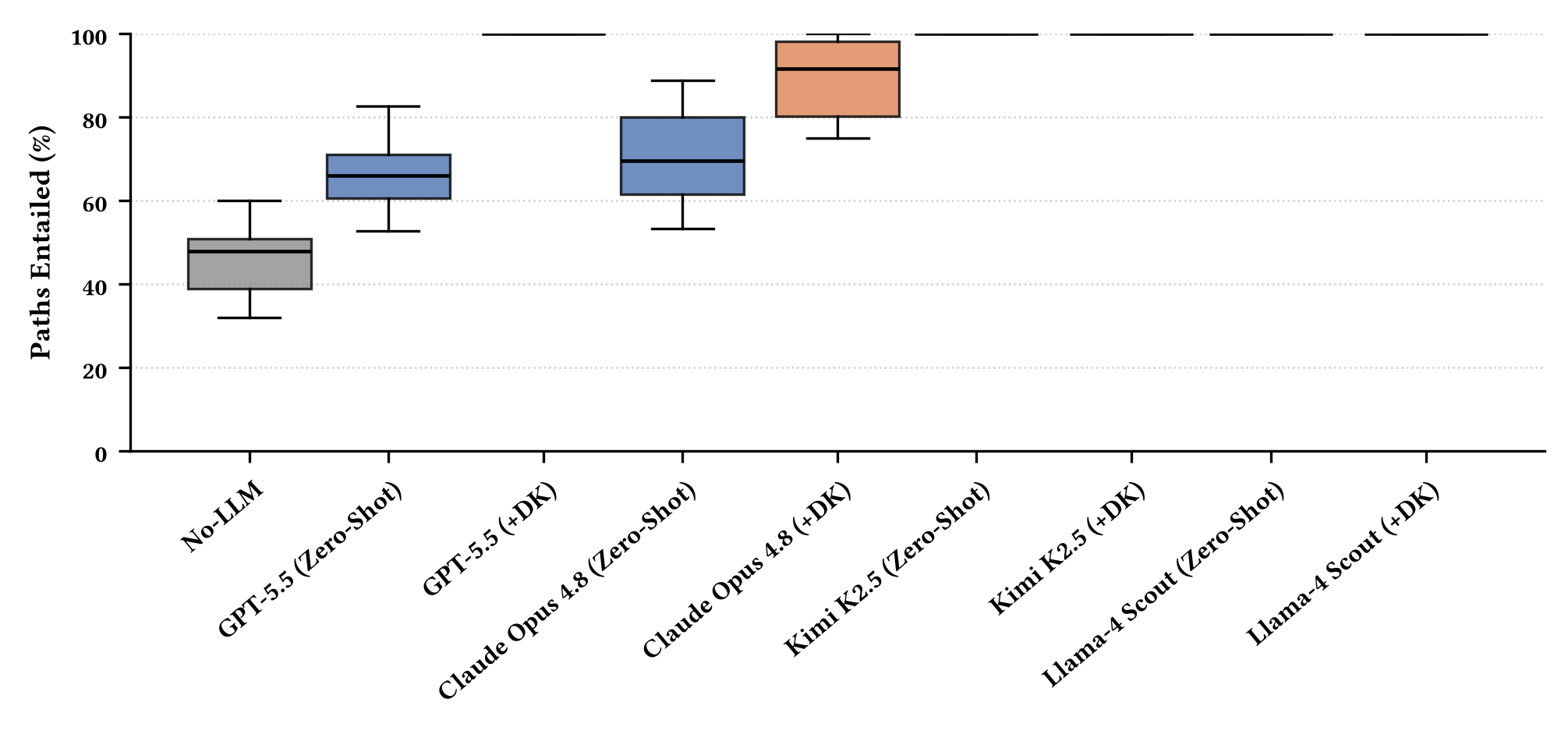}
\caption{Zero-shot on \texttt{mupdf-x11} (CWE-416). Conventions as in
Figure~\ref{fig:stab-ins-ct1}; No-LLM attains $46.4\pm9.1\%$, so the
margin over No-LLM is smaller on this target than on
\texttt{insulininject}.}
\label{fig:stab-mup-ct1}
\end{figure}

\begin{figure}[!htbp]
\centering
\includegraphics[width=0.9\columnwidth]{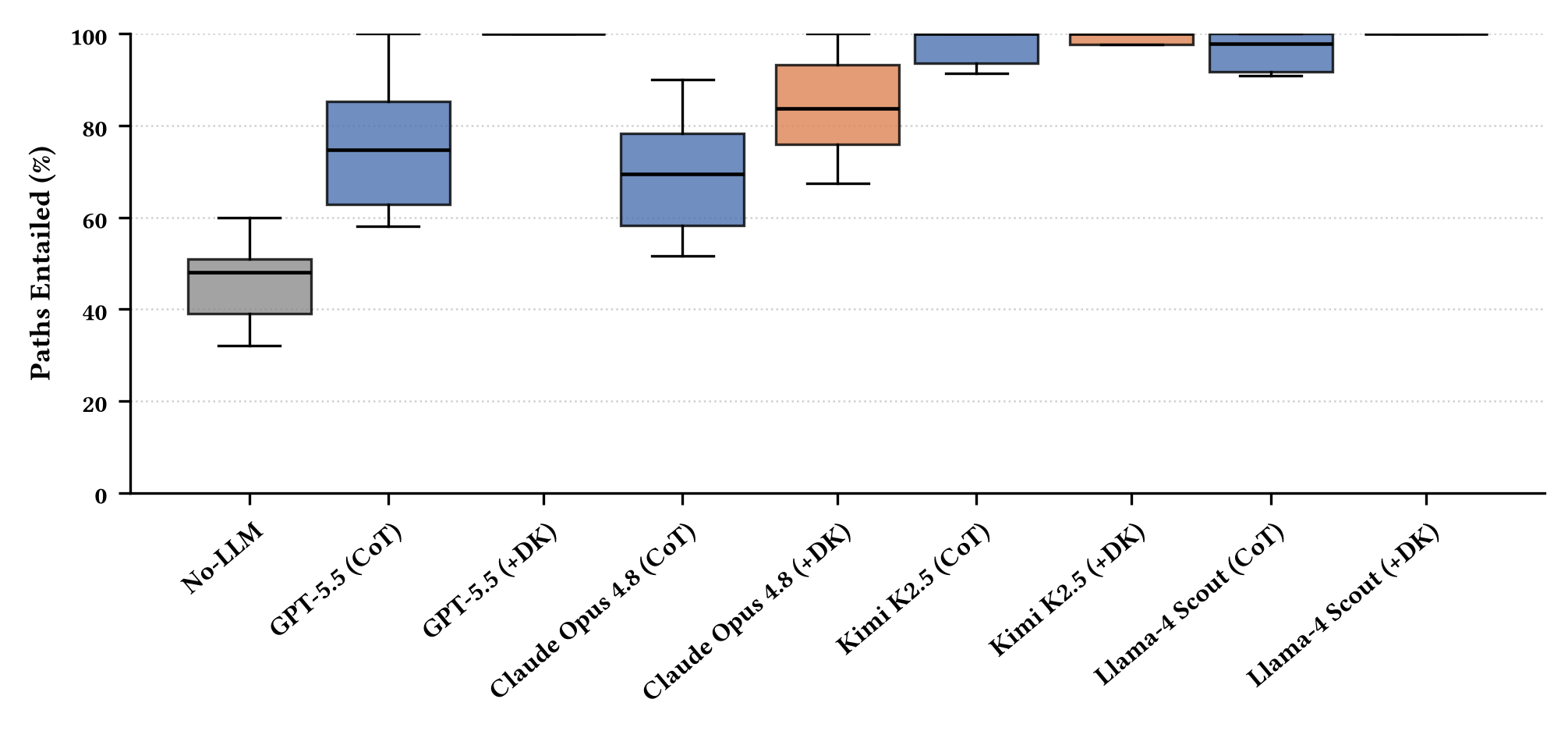}
\caption{Chain-of-thought on \texttt{mupdf-x11} (CWE-416). Conventions as in
Figure~\ref{fig:stab-ins-ct1}.}
\label{fig:stab-mup-ct2}
\end{figure}

\begin{figure}[!htbp]
\centering
\includegraphics[width=0.9\columnwidth]{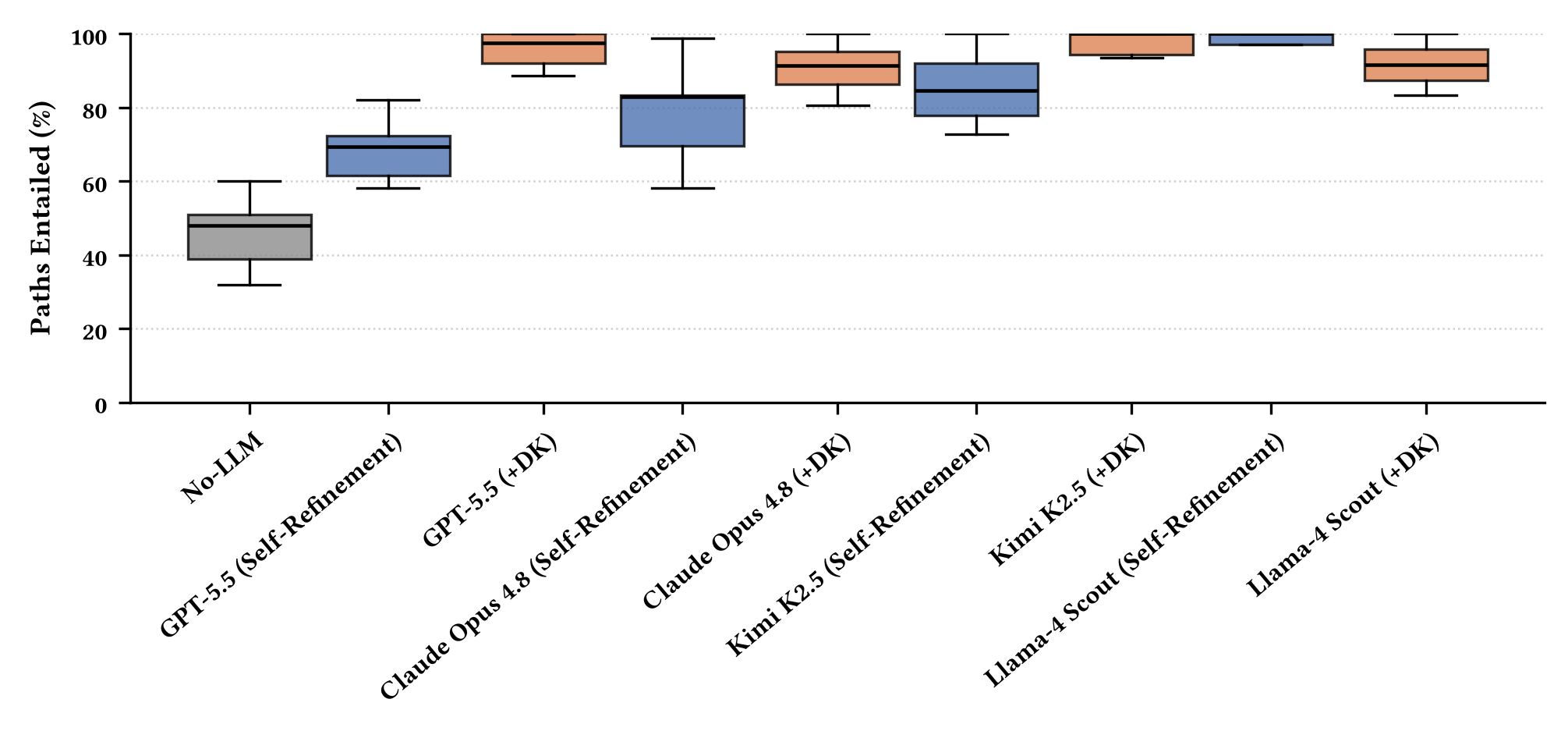}
\caption{Self-Refinement on \texttt{mupdf-x11} (CWE-416). Conventions as in
Figure~\ref{fig:stab-ins-ct1}.}
\label{fig:stab-mup-ct3}
\end{figure}

To check that the reported experimental results are not an artifact of a single run, and to
establish what entailment rate an uninformed generator would achieve, we
repeated the full pipeline ten times for each of the six agents (three prompting strategies, unguided and guided) with each of the four LLMs on two cases: \texttt{insulininject}
under CWE-121 and \texttt{mupdf-x11} under CWE-416. Alongside these we ran ten
independent rounds of a \emph{No-LLM} baseline, in which 25 analyst paths are drawn
directly from the binary's call graph, matched to the LLM conditions on path
length and on the label-group constraint, so that the only difference is which
labels a path visits. No-LLM replaces the LLM's templates with direct path generation over the call graph; no model is involved. Figures~\ref{fig:stab-ins-ct1}--\ref{fig:stab-mup-ct3}
show the resulting distributions; all settings are otherwise those of the
experimental setup in the main paper.

No-LLM attains $18.0\pm8.5\%$ entailment on \texttt{insulininject} and
$46.4\pm9.1\%$ on \texttt{mupdf-x11}. Every model--strategy--%
combination runs on these two targets exceeds No-LLM under a one-sided
Mann--Whitney test against the ten No-LLM rounds ($p<0.05$; most at
$p<10^{-3}$). Pooled over models, the unguided agents sit $57$--$64$ points
above No-LLM on \texttt{insulininject} and $32$--$38$ points above on
\texttt{mupdf-x11}; with domain knowledge (guided agents) the margins widen to $79$--$81$ and
$47$--$51$ points. We make two observations. First, every agent that uses an LLM generates better paths than the No-LLM agent, which forms paths directly from the call graph. Second, Entailment is further improved by the addition of domain knowledge across several runs.

Averaged over the twelve model--strategy combinations per target, the standard
deviation of entailment across runs falls from $10.7$ to $4.2$ on
\texttt{insulininject} and from $9.7$ to $5.8$ on \texttt{mupdf-x11}. The effect
is largest where unaided behaviour is least reliable: Kimi~K2.5 under
self-refinement moves from $54.1\pm11.3\%$ to $98.3\pm2.8\%$ on
\texttt{insulininject}, and GPT-5.5 under zero-shot from $66.3\pm9.9\%$ to
$99.6\pm0.8\%$ on \texttt{mupdf-x11}. Several guided agents entail every path in all ten runs, so their spread is zero. Although generation uses
greedy decoding, hosted inference APIs are not bit-reproducible; variable request
batching perturbs logits enough to flip near-tied argmax decisions, which is the
source of the spread reported here.

A run contributes no measurement when the model returns no usable path set. This
occurs once in 240 runs on \texttt{insulininject} but in $10.4\%$ of runs on
\texttt{mupdf-x11}, and is concentrated in Llama-4~Scout, whose $8{,}192$-token
output cap truncates longer responses. Aggregated over all conditions,
generation failure accounts for $0\%$ of GPT-5.5 runs, $3.9\%$ of Kimi~K2.5,
$6.7\%$ of Opus-4.8, and $24.4\%$ of Llama-4~Scout.

\section{Per-Configuration Results}
\label{app:perconfig}

Tables~\ref{tab:std} and~\ref{tab:rw} list every configuration behind the
plots in the main paper for CWE-121 and CWE-416; the CWE-415 configurations are
in Table~\ref{tab:cwe415}. \#Paths is the number of unique paths an agent outputs and
\#Entailed the number the domain knowledge entails, with the entailment
percentage in parentheses. Bold marks, for each model and strategy, the larger value of the two conditions (without and with +DK), separately for \#Paths, \#Entailed and the entailment percentage.

\begin{table*}[!htbp]
\centering
\caption{Unique paths realized (\#Paths) and entailed (\#Entailed) for the four standard BinPool binaries, each validated against the DK graph of the CWE shown in parentheses, without and with domain knowledge (+DK).}
\label{tab:std}
\setlength{\tabcolsep}{3pt}
\begin{tabular}{llcccccccc}
\toprule
 &  & \multicolumn{2}{c}{std:gpac (416)} & \multicolumn{2}{c}{std:mupdf-x11 (416)} & \multicolumn{2}{c}{std:mutool (416)} & \multicolumn{2}{c}{std:htmldoc (121)} \\ \cmidrule(lr){3-4}\cmidrule(lr){5-6}\cmidrule(lr){7-8}\cmidrule(lr){9-10}
\textbf{Model} & \textbf{Strategy} & Paths & Entailed & Paths & Entailed & Paths & Entailed & Paths & Entailed \\ \midrule
 \multirow{6}{*}{Opus-4.8} & Zero-Shot & 3 & 3(100) & 69 & 40(57.97) & 52 & 38(73.08) & 70 & 69(98.57) \\
  & \quad + DK & \textbf{8} & \textbf{8(100)} & \textbf{101} & \textbf{101(100)} & \textbf{62} & \textbf{59(95.16)} & \textbf{98} & \textbf{98(100)} \\ \cmidrule{2-10}
  & CoT & \textbf{17} & 16(94.12) & 47 & 38(80.85) & 5 & 5(100) & \textbf{87} & \textbf{87(100)} \\
  & \quad + DK & 16 & 16\textbf{(100)} & \textbf{58} & \textbf{58(100)} & \textbf{13} & \textbf{13(100)} & 55 & 55(100) \\ \cmidrule{2-10}
  & Self-Refinement & 8 & 8(100) & 63 & 50(79.37) & 65 & 64\textbf{(98.46)} & 62 & 60(96.77) \\
  & \quad + DK & \textbf{15} & \textbf{15(100)} & \textbf{71} & \textbf{70(98.59)} & \textbf{84} & \textbf{81}(96.43) & \textbf{86} & \textbf{86(100)} \\
\cmidrule{1-10}
 \multirow{6}{*}{GPT-5.5} & Zero-Shot & \textbf{12} & \textbf{5}(41.67) & 65 & 45(69.23) & \textbf{76} & \textbf{60}(78.95) & \textbf{72} & \textbf{71}(98.61) \\
  & \quad + DK & 1 & 1\textbf{(100)} & \textbf{66} & \textbf{61(92.42)} & 44 & 42\textbf{(95.45)} & 41 & 41\textbf{(100)} \\ \cmidrule{2-10}
  & CoT & \textbf{2} & \textbf{2(100)} & \textbf{48} & 29(60.42) & \textbf{30} & \textbf{30(100)} & 45 & 44(97.78) \\
  & \quad + DK & 1 & 1(100) & 30 & \textbf{30(100)} & 27 & 27(100) & \textbf{46} & \textbf{46(100)} \\ \cmidrule{2-10}
  & Self-Refinement & 11 & 9(81.82) & 61 & 35(57.38) & \textbf{77} & 48(62.34) & 77 & 73(94.81) \\
  & \quad + DK & \textbf{33} & \textbf{33(100)} & \textbf{64} & \textbf{64(100)} & 59 & \textbf{57(96.61)} & \textbf{86} & \textbf{86(100)} \\
\cmidrule{1-10}
 \multirow{6}{*}{\shortstack[l]{Llama-4\\Scout}} & Zero-Shot & \textbf{6} & \textbf{6(100)} & 14 & 9(64.29) & \textbf{5} & 1(20) & \textbf{78} & \textbf{73}(93.59) \\
  & \quad + DK & 1 & 1(100) & \textbf{30} & \textbf{30(100)} & 4 & \textbf{4(100)} & 36 & 36\textbf{(100)} \\ \cmidrule{2-10}
  & CoT & \textbf{29} & \textbf{19}(65.52) & \textbf{73} & \textbf{54}(73.97) & 10 & 8(80) & \textbf{65} & \textbf{65(100)} \\
  & \quad + DK & 7 & 7\textbf{(100)} & 18 & 18\textbf{(100)} & \textbf{19} & \textbf{18(94.74)} & 22 & 22(100) \\ \cmidrule{2-10}
  & Self-Refinement & 5 & 5(100) & 15 & 10(66.67) & 4 & 0(0) & 28 & 26(92.86) \\
  & \quad + DK & \textbf{13} & \textbf{13(100)} & \textbf{33} & \textbf{33(100)} & 4 & \textbf{4(100)} & \textbf{31} & \textbf{31(100)} \\
\cmidrule{1-10}
 \multirow{6}{*}{Kimi K2.5} & Zero-Shot & 13 & 13(100) & 14 & 14(100) & \textbf{33} & \textbf{33(100)} & \textbf{167} & \textbf{159}(95.21) \\
  & \quad + DK & \textbf{29} & \textbf{29(100)} & \textbf{30} & \textbf{30(100)} & 23 & 23(100) & 20 & 20\textbf{(100)} \\ \cmidrule{2-10}
  & CoT & \textbf{6} & \textbf{6(100)} & \textbf{79} & \textbf{79(100)} & \textbf{102} & \textbf{83}(81.37) & 32 & 32(100) \\
  & \quad + DK & 5 & 5(100) & 44 & 44(100) & 72 & 64\textbf{(88.89)} & \textbf{37} & \textbf{37(100)} \\ \cmidrule{2-10}
  & Self-Refinement & \textbf{11} & \textbf{11(100)} & \textbf{90} & 51(56.67) & \textbf{142} & \textbf{114}(80.28) & 60 & 60(100) \\
  & \quad + DK & 2 & 2(100) & 72 & \textbf{58(80.56)} & 82 & 79\textbf{(96.34)} & \textbf{94} & \textbf{94(100)} \\
\bottomrule
\end{tabular}

\end{table*}

\begin{table*}[!htbp]
\centering
\caption{Unique paths realized (\#Paths) and entailed (\#Entailed) for the three real-world medical-device binaries, each validated against the DK graph of the CWE shown in parentheses, without and with domain knowledge (+DK).}
\label{tab:rw}
\setlength{\tabcolsep}{3pt}
\begin{tabular}{llcccccccc}
\toprule
 &  & \multicolumn{2}{c}{rw:insulininject (121)} & \multicolumn{2}{c}{rw:unsafelib (121)} & \multicolumn{2}{c}{rw:ezurio (121)} & \multicolumn{2}{c}{rw:ezurio (416)} \\ \cmidrule(lr){3-4}\cmidrule(lr){5-6}\cmidrule(lr){7-8}\cmidrule(lr){9-10}
\textbf{Model} & \textbf{Strategy} & Paths & Entailed & Paths & Entailed & Paths & Entailed & Paths & Entailed \\ \midrule
 \multirow{6}{*}{Opus-4.8} & Zero-Shot & \textbf{124} & \textbf{106}(85.48) & 47 & 37(78.72) & 29 & 26(89.66) & 6 & 6(100) \\
  & \quad + DK & 33 & 33\textbf{(100)} & \textbf{54} & \textbf{54(100)} & \textbf{40} & \textbf{40(100)} & \textbf{27} & \textbf{27(100)} \\ \cmidrule{2-10}
  & CoT & \textbf{51} & \textbf{48(94.12)} & \textbf{67} & \textbf{56}(83.58) & 28 & 28(100) & 4 & 4(100) \\
  & \quad + DK & 31 & 26(83.87) & 42 & 41\textbf{(97.62)} & \textbf{47} & \textbf{47(100)} & 4 & 4(100) \\ \cmidrule{2-10}
  & Self-Refinement & \textbf{73} & \textbf{70}(95.89) & 69 & 47(68.12) & 31 & 31(100) & 2 & 2(100) \\
  & \quad + DK & 29 & 29\textbf{(100)} & \textbf{87} & \textbf{81(93.1)} & \textbf{97} & \textbf{97(100)} & \textbf{16} & \textbf{16(100)} \\
\cmidrule{1-10}
 \multirow{6}{*}{GPT-5.5} & Zero-Shot & \textbf{69} & \textbf{60}(86.96) & \textbf{53} & \textbf{41}(77.36) & 24 & 21(87.5) & \textbf{23} & \textbf{14}(60.9) \\
  & \quad + DK & 37 & 37\textbf{(100)} & 35 & 35\textbf{(100)} & \textbf{61} & \textbf{61(100)} & 7 & 7\textbf{(100)} \\ \cmidrule{2-10}
  & CoT & \textbf{45} & \textbf{34}(75.56) & 35 & 24(68.57) & 15 & 14(93.33) & \textbf{9} & \textbf{9(100)} \\
  & \quad + DK & 26 & 26\textbf{(100)} & \textbf{42} & \textbf{42(100)} & \textbf{28} & \textbf{28(100)} & 6 & 6(100) \\ \cmidrule{2-10}
  & Self-Refinement & 73 & 40(54.79) & \textbf{66} & 42(63.64) & 16 & 14(87.5) & \textbf{11} & \textbf{8}(72.7) \\
  & \quad + DK & \textbf{87} & \textbf{83(95.4)} & 63 & \textbf{63(100)} & 16 & \textbf{16(100)} & 5 & 5\textbf{(100)} \\
\cmidrule{1-10}
 \multirow{6}{*}{\shortstack[l]{Llama-4\\Scout}} & Zero-Shot & 12 & 7(58.33) & 4 & 4(100) & \textbf{17} & \textbf{17(100)} & 3 & 3(100) \\
  & \quad + DK & \textbf{15} & \textbf{15(100)} & \textbf{7} & \textbf{7(100)} & 12 & 12(100) & \textbf{5} & \textbf{5(100)} \\ \cmidrule{2-10}
  & CoT & \textbf{70} & \textbf{58(82.86)} & 30 & 24(80) & \textbf{17} & \textbf{17(100)} & \textbf{5} & \textbf{5(100)} \\
  & \quad + DK & 9 & 6(66.67) & \textbf{42} & \textbf{42(100)} & 16 & 16(100) & 0 & NA \\ \cmidrule{2-10}
  & Self-Refinement & 22 & 22(100) & \textbf{15} & 9(60) & \textbf{18} & \textbf{18(100)} & 12 & 12(100) \\
  & \quad + DK & \textbf{30} & \textbf{30(100)} & 13 & \textbf{13(100)} & 9 & 9(100) & \textbf{6} & \textbf{6(100)} \\
\cmidrule{1-10}
 \multirow{6}{*}{Kimi K2.5} & Zero-Shot & \textbf{158} & \textbf{57}(36.08) & \textbf{136} & \textbf{88}(64.71) & \textbf{62} & \textbf{61}(98.39) & \textbf{14} & \textbf{14(100)} \\
  & \quad + DK & 11 & 11\textbf{(100)} & 31 & 25\textbf{(80.65)} & 23 & 23\textbf{(100)} & 11 & 11(100) \\ \cmidrule{2-10}
  & CoT & \textbf{134} & \textbf{108}(80.6) & \textbf{69} & \textbf{48}(69.57) & 18 & 17(94.44) & 4 & 4(100) \\
  & \quad + DK & 41 & 41\textbf{(100)} & 33 & 33\textbf{(100)} & \textbf{41} & \textbf{41(100)} & \textbf{14} & \textbf{14(100)} \\ \cmidrule{2-10}
  & Self-Refinement & \textbf{140} & 8(5.71) & \textbf{108} & 60(55.56) & 50 & 47(94) & 0 & NA \\
  & \quad + DK & 41 & \textbf{41(100)} & 83 & \textbf{78(93.98)} & \textbf{62} & \textbf{62(100)} & 0 & NA \\
\bottomrule
\end{tabular}

\end{table*}

\label{app:cwe415}%

We note that
\texttt{ezurio} yields output in only 9 of its 24 configurations---all of
Opus-4.8 and three configurations of GPT-5.5. Llama-4~Scout produces nothing at all under
self-refinement on any binary, and nothing on \texttt{htmldoc} under
chain-of-thought; Kimi~K2.5 produces nothing on \texttt{dropbear} under
zero-shot. These are the generation failures characterised in
Section~\ref{app:stability}, and the reason CWE-415 contributes fewer configurations (24 of 118) to the pooled figures in the main paper. These results go to show how LLM results can vary significantly between models, and how larger state-of-the-art models often outperform other models.

\begin{table}[!htbp]
\centering
\caption{CWE-415 (double free): unique paths realized (\#Paths) and entailed
(\#Entailed), without and with domain knowledge (+DK). \texttt{NA} marks a configuration that produced no realized paths. Bold as in Table~\ref{tab:std}.}
\label{tab:cwe415}
\setlength{\tabcolsep}{5pt}
\begin{tabular}{llcccccc}
\toprule
 &  & \multicolumn{2}{c}{std:htmldoc} & \multicolumn{2}{c}{std:dropbear} & \multicolumn{2}{c}{rw:ezurio} \\ \cline{3-8}
\textbf{Model} & \textbf{Strategy} & Paths & Entailed (\%) & Paths & Entailed (\%) & Paths & Entailed (\%) \\ \midrule
 \multirow{6}{*}{Opus-4.8} & Zero-Shot & \textbf{80} & \textbf{70}(87.5) & \textbf{33} & 19(57.58) & 26 & 20(76.92) \\
  & \quad + DK & 58 & 58\textbf{(100)} & 19 & 19\textbf{(100)} & \textbf{27} & \textbf{27(100)} \\\cline{2-8}
  & CoT & \textbf{45} & 41(91.11) & 3 & 1(33.33) & 2 & 2(100) \\
  & \quad + DK & 42 & \textbf{42(100)} & \textbf{6} & \textbf{6(100)} & 2 & 2(100) \\\cline{2-8}
  & Self-Refinement & 57 & 52(91.23) & \textbf{21} & 11(52.38) & 13 & 12(92.31) \\
  & \quad + DK & \textbf{84} & \textbf{84(100)} & 17 & \textbf{17(100)} & \textbf{29} & \textbf{29(100)} \\
\midrule
 \multirow{6}{*}{GPT-5.5} & Zero-Shot & 57 & 54(94.74) & \textbf{52} & 16(30.77) & 7 & 4(57.14) \\
  & \quad + DK & \textbf{66} & \textbf{66(100)} & 39 & \textbf{32(82.05)} & 0 & NA \\\cline{2-8}
  & CoT & \textbf{49} & 42(85.71) & \textbf{26} & 14(53.85) & 0 & NA \\
  & \quad + DK & 48 & \textbf{48(100)} & 23 & \textbf{23(100)} & 0 & NA \\\cline{2-8}
  & Self-Refinement & \textbf{80} & 71(88.75) & \textbf{39} & 11(28.21) & \textbf{9} & \textbf{8}(88.89) \\
  & \quad + DK & 71 & 71\textbf{(100)} & 37 & \textbf{31(83.78)} & 4 & 4\textbf{(100)} \\
\midrule
 \multirow{6}{*}{\shortstack[l]{Llama-4\\Scout}} & Zero-Shot & 19 & 19(100) & 5 & 5(100) & 0 & NA \\
  & \quad + DK & \textbf{36} & \textbf{36(100)} & \textbf{24} & \textbf{24(100)} & 0 & NA \\\cline{2-8}
  & CoT & 0 & NA & \textbf{44} & \textbf{11}(25) & 0 & NA \\
  & \quad + DK & 0 & NA & 7 & 7\textbf{(100)} & 0 & NA \\\cline{2-8}
  & Self-Refinement & 0 & NA & 0 & NA & 0 & NA \\
  & \quad + DK & 0 & NA & 0 & NA & 0 & NA \\
\midrule
 \multirow{6}{*}{Kimi K2.5} & Zero-Shot & 6 & 6(100) & 0 & NA & 0 & NA \\
  & \quad + DK & \textbf{47} & \textbf{47(100)} & 0 & NA & 0 & NA \\\cline{2-8}
  & CoT & \textbf{56} & \textbf{56(100)} & \textbf{31} & \textbf{10}(32.26) & 0 & NA \\
  & \quad + DK & 10 & 10(100) & 7 & 7\textbf{(100)} & 0 & NA \\\cline{2-8}
  & Self-Refinement & \textbf{44} & \textbf{44(100)} & \textbf{74} & \textbf{34}(45.95) & 0 & NA \\
  & \quad + DK & 11 & 11(100) & 28 & 23\textbf{(82.14)} & 0 & NA \\
\bottomrule
\end{tabular}

\end{table}

\section{Per-Binary Effect of Domain Knowledge}
\label{app:perbinary-arrows}

\begin{figure}[!htbp]
\centering
\includegraphics[width=\textwidth]{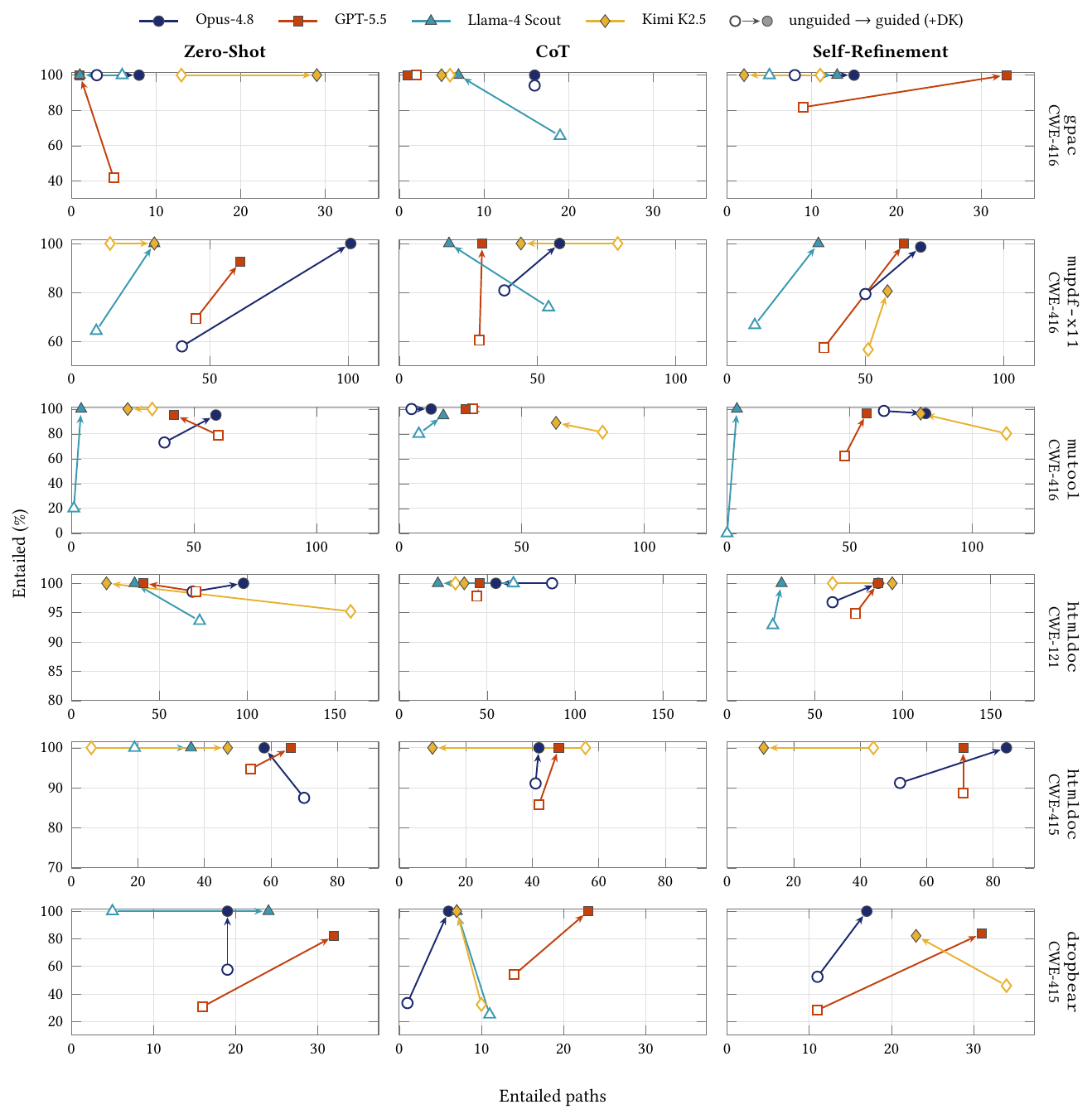}
\caption{Effect of domain knowledge on each standard BinPool binary--CWE pair. Rows are binary--CWE pairs and columns prompting strategies; each arrow is one LLM, from the unguided agent (hollow) to the guided agent (+DK, filled).}
\label{fig:arrows-std}
\end{figure}

\begin{figure}[!htbp]
\centering
\includegraphics[width=\textwidth]{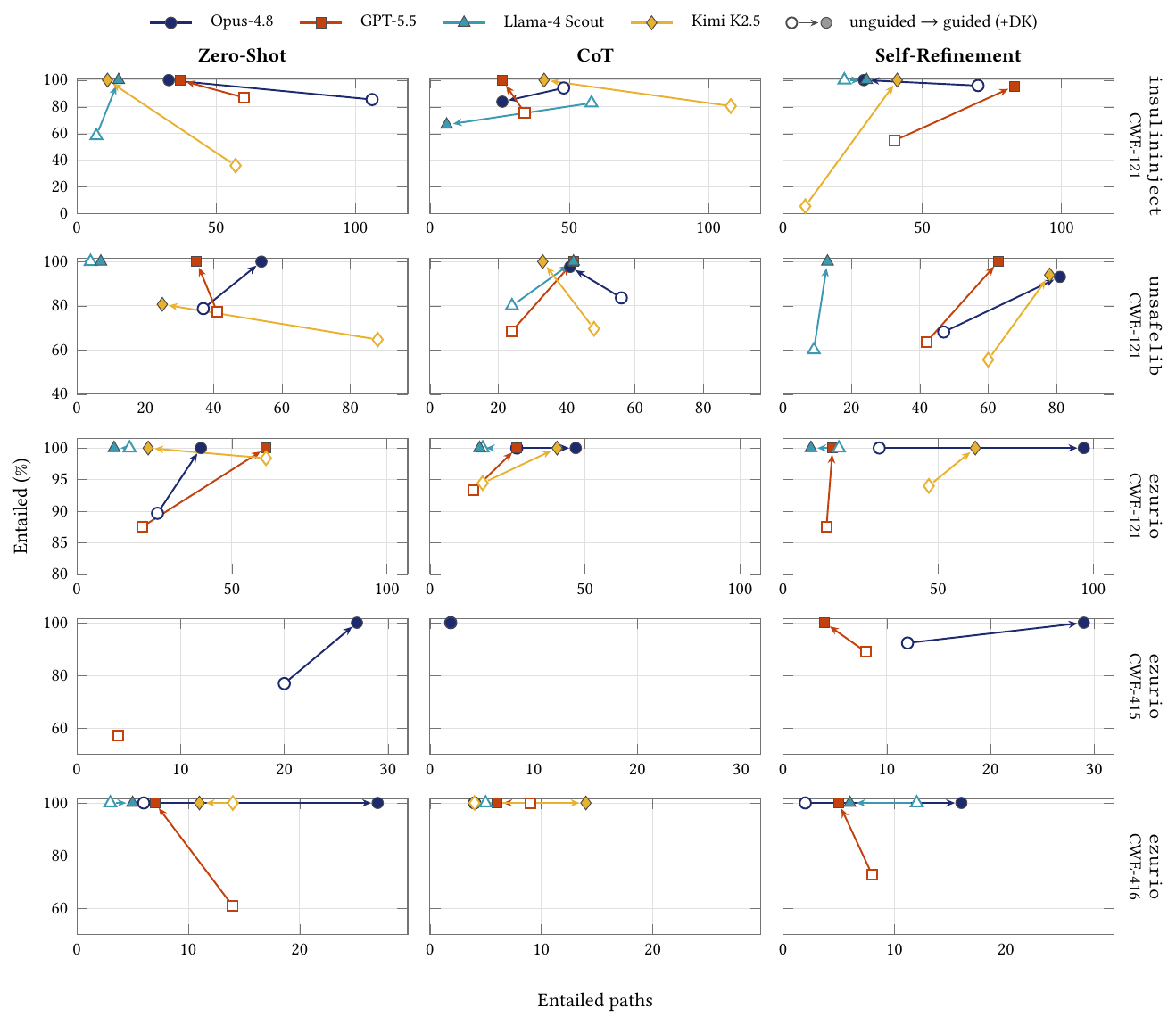}
\caption{Effect of domain knowledge on each real-world medical-device binary--CWE pair. Conventions as in Figure~\ref{fig:arrows-std}.}
\label{fig:arrows-rw}
\end{figure}

Figure~4 of the main paper pools the binaries of each type. Figures~\ref{fig:arrows-std} and~\ref{fig:arrows-rw} show the same comparison for each binary--CWE pair separately, from the per-configuration counts of Section~\ref{app:perconfig}. Each row is one binary--CWE pair and each column one prompting strategy; each arrow goes from the unguided to the guided agent for one LLM. Axes are shared within a row, since the number of paths differs widely between binaries. A missing marker means that agent produced no paths for that binary.

\section{Aggregate Analysis}
\label{app:aggregate}

This section aggregates the per-configuration results of Section~\ref{app:perconfig} along three axes: by binary, by absolute entailed-path
count, and by statistical significance. Unless stated otherwise, figures pool
\#Entailed and \#Paths over configurations rather than averaging per-configuration
percentages, so that each configuration is weighted by the evidence it carries.

\subsection{Per-Binary Results}
\label{app:perbinary}

Table~\ref{tab:app-perbinary} pools every model and strategy for each binary--CWE pair. Entailment of the unguided agents varies widely across pairs, from $40.2\%$ to $97.2\%$ with no consistent \texttt{std}/\texttt{rw} pattern, but under domain knowledge every pair reaches $91.3$--$100\%$.

\begin{table}[!htbp]
\centering\small
\caption{Pooled entailment and entailed-path counts per binary--CWE pair, over all model$\times$strategy configurations, ordered by entailment of the unguided agents within each type. Unguided entailment spans $40.2$--$97.2\%$ with no consistent \texttt{std}/\texttt{rw} pattern; under domain knowledge every pair reaches $91.3$--$100\%$.}
\label{tab:app-perbinary}
\setlength{\tabcolsep}{4pt}
\begin{tabular}{@{}llcrrrr@{}}
\toprule
& & & \multicolumn{2}{c}{\textbf{Entail\%}} & \multicolumn{2}{c}{\textbf{\#Entailed}} \\
\cmidrule(lr){4-5}\cmidrule(lr){6-7}
\textbf{Binary} & \textbf{Type} & \textbf{CWE} & Unguided & Guided & Unguided & Guided \\
\midrule
\texttt{dropbear}      & std & 415 & 40.2 &  91.3 & 132 & 189 \\
\texttt{mupdf-x11}     & std & 416 & 71.2 &  96.8 & 454 & 597 \\
\texttt{mutool}        & std & 416 & 80.5 &  95.5 & 484 & 471 \\
\texttt{gpac}   & std & 416 & 83.7 & 100.0 & 103 & 131 \\
\texttt{htmldoc}       & std & 415 & 92.3 & 100.0 & 455 & 473 \\
\texttt{htmldoc}       & std & 121 & 97.2 & 100.0 & 819 & 652 \\
\midrule
\texttt{insulininject} & rw  & 121 & 63.6 &  96.9 & 618 & 378 \\
\texttt{unsafelib}     & rw  & 121 & 68.7 &  96.6 & 480 & 514 \\
\texttt{ezurio}        & rw  & 415 & 80.7 & 100.0 &  46 &  62 \\
\texttt{ezurio}        & rw  & 416 & 87.1 & 100.0 &  81 & 101 \\
\texttt{ezurio}        & rw  & 121 & 95.7 & 100.0 & 311 & 452 \\
\bottomrule
\end{tabular}
\end{table}

Table~\ref{tab:app-nollm} extends the No-LLM comparison of the main paper,
which reports self-refinement, to every prompting strategy. Every unguided agent beats the No-LLM agent on all three binaries, and domain knowledge
raises entailment to $93.1$--$100\%$ in every case.

\begin{table}[!htbp]
\centering\small
\caption{Entailment (\%) of paths from the No-LLM agent, the unguided agent (Unguided) and the guided agent (Guided, +DK) for the three binaries with more than
$3{,}000$ call-graph nodes, pooled over the four models for each prompting
strategy and over all strategies (\emph{All}).}
\label{tab:app-nollm}
\setlength{\tabcolsep}{4pt}
\begin{tabular}{@{}lcrrrrrrrrr@{}}
\toprule
& & & \multicolumn{2}{c}{\textbf{Zero-Shot}} & \multicolumn{2}{c}{\textbf{CoT}} & \multicolumn{2}{c}{\textbf{Self-Refinement}} & \multicolumn{2}{c}{\textbf{All}} \\
\cmidrule(lr){4-5}\cmidrule(lr){6-7}\cmidrule(lr){8-9}\cmidrule(lr){10-11}
\textbf{Binary} & \textbf{CWE} & \textbf{No-LLM} & Unguided & Guided & Unguided & Guided & Unguided & Guided & Unguided & Guided \\
\midrule
\texttt{mupdf-x11} & 416 & 46.4 & 66.7 & 97.8 & 81.0 & 100.0 & 63.8 & 93.8 & 71.2 & 96.8 \\
\texttt{mutool}    & 416 & 54.8 & 79.5 & 96.2 & 85.7 &  93.1 & 78.5 & 96.5 & 80.5 & 95.5 \\
\texttt{unsafelib} & 121 & 40.0 & 70.8 & 95.3 & 75.6 &  99.4 & 61.2 & 95.5 & 68.7 & 96.6 \\
\bottomrule
\end{tabular}
\end{table}

\subsection{Entailed-Path Counts}
\label{app:counts}

The main paper plots entailed paths against entailment rate. Table~\ref{tab:app-counts} gives the
underlying counts by binary type and strategy. Guidance raises the pooled entailment rate for every agent type, but the number of entailed paths moves in both directions. The self-refinement agent shows the best improvements both in rate and number of entailments.

\begin{table}[!htbp]
\centering\small
\caption{Entailment rate and absolute entailed-path count by binary type and
strategy, pooled over all three weakness classes. \emph{Count up} is the number
of configurations in which domain knowledge raised the entailed count, out of
those producing output in both conditions. The self-refinement agent is the only one
for which the count rises on both target types; chain-of-thought reduces it in
both.}
\label{tab:app-counts}
\begin{tabular}{@{}llrrrrr@{}}
\toprule
& & \multicolumn{2}{c}{\textbf{Entail\%}} & \multicolumn{2}{c}{\textbf{\#Entailed}} & \\
\cmidrule(lr){3-4}\cmidrule(lr){5-6}
\textbf{Type} & \textbf{Strategy} & Unguided & Guided & Unguided & Guided & \textbf{count up} \\
\midrule
std & Zero-Shot       & 82.7 & 98.1 & 828 &  866 & 14/23 \\
std & CoT             & 82.9 & 98.5 & 772 &  604 & 10/23 \\
std & Self-Refinement & 77.4 & 96.8 & 847 & 1043 & 17/22 \\
\midrule
rw  & Zero-Shot       & 72.0 & 98.6 & 586 &  430 &  8/17 \\
rw  & CoT             & 82.9 & 97.9 & 500 &  415 &  6/16 \\
rw  & Self-Refinement & 61.8 & 97.8 & 450 &  662 & 12/17 \\
\bottomrule
\end{tabular}

\end{table}

\subsection{Statistical Significance}
\label{app:stats}

\begin{table}[!htbp]
\centering\small
\caption{One-sided Wilcoxon signed-rank tests on paired (baseline, +DK)
entailment percentages. $n$ is the number of configurations with output in both
conditions, $n'$ the number after excluding ties; every tie is a configuration
already at $100\%$ entailment without domain knowledge. Median is the change in
percentage points; $r$ is the matched-pairs rank-biserial effect size. The
second row is CWE-121 and CWE-416 only; the main paper now reports the first row, all three CWEs.}
\label{tab:app-wilcoxon}
\setlength{\tabcolsep}{4pt}
\begin{tabular}{@{}lrrrrrrr@{}}
\toprule
\textbf{Subset} & $n$ & $n'$ & $W^{+}$ & $W^{-}$ & $p$ & median & $r$ \\
\midrule
All three CWEs  & 118 & 80 & 3180 & 60 & $3.7\times10^{-14}$ & $+8.83$  & $0.963$ \\
CWE-121 \& 416  &  94 & 62 & 1905 & 48 & $3.8\times10^{-11}$ & $+6.54$  & $0.951$ \\
\midrule
\texttt{std}    &  68 & 46 & 1078 &  3 & $7.1\times10^{-14}$ & $+8.14$  & $0.994$ \\
\texttt{rw}     &  50 & 34 &  572 & 23 & $1.3\times10^{-6}$  & $+10.73$ & $0.923$ \\
\midrule
CWE-416         &  46 & 25 &  324 &  1 & $6.0\times10^{-8}$  & $+6.70$  & $0.994$ \\
CWE-121         &  48 & 37 &  667 & 36 & $9.7\times10^{-7}$  & $+6.54$  & $0.898$ \\
CWE-415         &  24 & 18 &  171 &  0 & $3.8\times10^{-6}$  & $+11.88$ & $1.000$ \\
\bottomrule
\end{tabular}

\end{table}

Across all 118 paired configurations, entailment improves in 77, is unchanged in
38, and decreases in 3; over the 94 configurations of CWE-121 and CWE-416 the counts
are 59, 32 and 3. Every subset is significant at $p<10^{-5}$. On CWE-415 no
configuration decreases at all, giving $W^{-}=0$ and a maximal effect size.

Note that, configurations are not fully independent---each model appears in 30 and each
binary--CWE pair in 12---so we treat the per-subset tests as descriptive rather
than as independent confirmatory comparisons, and apply no multiplicity
correction. At the level of these clusters the effect holds without exception. Pooled entailment improves with domain knowledge on all 11 binary--CWE pairs (exact one-sided sign test, $p=4.9\times10^{-4}$), for all four models and for all three strategies, which is the most a sign test can show with four or three observations ($p=0.06$ and $p=0.13$). It also improves on 35 of the 42 model--pair combinations and falls on none; the remaining 7 are already at $100\%$ without domain knowledge.
\section{Circularity and Leakage experiment: Disjoint Knowledge-Graph Halves}
\label{app:kgsplit}

Tables~\ref{tab:exp2-zs}--\ref{tab:exp2-sr} give the full results behind the split-graph experiment of the main paper (Section~4.2). Rows A1--A3 re-verify existing path sets against the full graph and each half: A1 the unguided agent, A2 the guided agent (labels from the full graph), and A3 the No-LLM agent. Rows B1 and B2 are guided agents whose labels come from one half only, verified against the same half and the other half. Table~\ref{tab:exp2-P1-full} gives the full A and B rows for the \texttt{unsafelib} partition.

\begin{table*}[!htbp]
\centering
\setlength{\tabcolsep}{3pt}
\caption{Partial-knowledge-graph ablation, \textbf{Zero-Shot} agents. Models Opus-4.8 and GPT-5.5; binaries mupdf-x11 (CWE-416), mutool (CWE-416), unsafelib (CWE-121). Each binary's CWE domain-knowledge (DK) graph is split into two disjoint halves (H1/H2): every half keeps all node and edge \emph{types} but only half of the \emph{instances} of each edge type. Each entry is \#Entailed/\#Paths\,(\%): \#Paths unique paths, \#Entailed of them entailed by the reasoner over the Full graph or one half. \textbf{A1}: unguided agent; \textbf{A2}: guided agent, labels from the full graph; \textbf{A3}: No-LLM agent; \textbf{B1}/\textbf{B2}: guided agent with labels from one half only (B1: DK$=$H1, B2: DK$=$H2), validated on both halves---\emph{same} when the validation half equals the DK half, \emph{cross} otherwise. A3 re-verifies a single 25-path No-LLM sample; Table~1 of the main paper reports the mean over ten such samples.}
\label{tab:exp2-zs}
\begin{tabular}{llcccccc}
\toprule
 & & \multicolumn{3}{c}{\textbf{Opus-4.8}} & \multicolumn{3}{c}{\textbf{GPT-5.5}} \\
\cmidrule(lr){3-5}\cmidrule(lr){6-8}
\textbf{Agent} & \textbf{KG} & mupdf-x11 & mutool & unsafelib & mupdf-x11 & mutool & unsafelib \\
\midrule
\multirow{3}{*}{A1 (unguided)} & Full & 40/69 (58.0) & 38/52 (73.1) & 37/47 (78.7) & 45/65 (69.2) & 60/76 (78.9) & 41/53 (77.4) \\
 & H1 & 24/69 (34.8) & 33/52 (63.5) & 35/47 (74.5) & 29/65 (44.6) & 41/76 (53.9) & 37/53 (69.8) \\
 & H2 & 23/69 (33.3) & 27/52 (51.9) & 36/47 (76.6) & 31/65 (47.7) & 52/76 (68.4) & 37/53 (69.8) \\
\cmidrule(lr){1-8}
\multirow{3}{*}{A2 (guided)} & Full & 101/101 (100.0) & 59/62 (95.2) & 54/54 (100.0) & 61/66 (92.4) & 42/44 (95.5) & 35/35 (100.0) \\
 & H1 & 63/101 (62.4) & 34/62 (54.8) & 54/54 (100.0) & 46/66 (69.7) & 21/44 (47.7) & 35/35 (100.0) \\
 & H2 & 63/101 (62.4) & 50/62 (80.6) & 54/54 (100.0) & 48/66 (72.7) & 24/44 (54.5) & 35/35 (100.0) \\
\cmidrule(lr){1-8}
\multirow{3}{*}{A3 (No-LLM)} & Full & 12/25 (48.0) & 17/25 (68.0) & 7/25 (28.0) & 12/25 (48.0) & 17/25 (68.0) & 7/25 (28.0) \\
 & H1 & 5/25 (20.0) & 10/25 (40.0) & 5/25 (20.0) & 5/25 (20.0) & 10/25 (40.0) & 5/25 (20.0) \\
 & H2 & 6/25 (24.0) & 16/25 (64.0) & 7/25 (28.0) & 6/25 (24.0) & 16/25 (64.0) & 7/25 (28.0) \\
\cmidrule(lr){1-8}
\multirow{2}{*}{B1 (H1-guided)} & H1 (same) & 38/65 (58.5) & 35/52 (67.3) & 76/84 (90.5) & 35/44 (79.5) & 27/34 (79.4) & 51/51 (100.0) \\
 & H2 (cross) & 39/65 (60.0) & 39/52 (75.0) & 71/84 (84.5) & 37/44 (84.1) & 26/34 (76.5) & 51/51 (100.0) \\
\cmidrule(lr){1-8}
\multirow{2}{*}{B2 (H2-guided)} & H1 (cross) & 31/56 (55.4) & 38/49 (77.6) & 82/87 (94.3) & 43/49 (87.8) & 21/38 (55.3) & 43/43 (100.0) \\
 & H2 (same) & 29/56 (51.8) & 41/49 (83.7) & 82/87 (94.3) & 38/49 (77.6) & 33/38 (86.8) & 43/43 (100.0) \\
\bottomrule
\end{tabular}
\end{table*}

\begin{table*}[!htbp]
\centering
\setlength{\tabcolsep}{3pt}
\caption{Partial-knowledge-graph ablation, \textbf{CoT} agents. Notation as in Table~\ref{tab:exp2-zs}.}
\label{tab:exp2-cot}
\begin{tabular}{llcccccc}
\toprule
 & & \multicolumn{3}{c}{\textbf{Opus-4.8}} & \multicolumn{3}{c}{\textbf{GPT-5.5}} \\
\cmidrule(lr){3-5}\cmidrule(lr){6-8}
\textbf{Agent} & \textbf{KG} & mupdf-x11 & mutool & unsafelib & mupdf-x11 & mutool & unsafelib \\
\midrule
\multirow{3}{*}{A1 (unguided)} & Full & 38/47 (80.9) & 5/5 (100.0) & 56/67 (83.6) & 29/48 (60.4) & 30/30 (100.0) & 24/35 (68.6) \\
 & H1 & 20/47 (42.6) & 0/5 (0.0) & 53/67 (79.1) & 11/48 (22.9) & 28/30 (93.3) & 22/35 (62.9) \\
 & H2 & 23/47 (48.9) & 5/5 (100.0) & 52/67 (77.6) & 27/48 (56.2) & 28/30 (93.3) & 21/35 (60.0) \\
\cmidrule(lr){1-8}
\multirow{3}{*}{A2 (guided)} & Full & 58/58 (100.0) & 13/13 (100.0) & 41/42 (97.6) & 30/30 (100.0) & 27/27 (100.0) & 42/42 (100.0) \\
 & H1 & 46/58 (79.3) & 13/13 (100.0) & 41/42 (97.6) & 30/30 (100.0) & 27/27 (100.0) & 42/42 (100.0) \\
 & H2 & 44/58 (75.9) & 13/13 (100.0) & 41/42 (97.6) & 27/30 (90.0) & 25/27 (92.6) & 42/42 (100.0) \\
\cmidrule(lr){1-8}
\multirow{3}{*}{A3 (No-LLM)} & Full & 12/25 (48.0) & 17/25 (68.0) & 7/25 (28.0) & 12/25 (48.0) & 17/25 (68.0) & 7/25 (28.0) \\
 & H1 & 5/25 (20.0) & 10/25 (40.0) & 5/25 (20.0) & 5/25 (20.0) & 10/25 (40.0) & 5/25 (20.0) \\
 & H2 & 6/25 (24.0) & 16/25 (64.0) & 7/25 (28.0) & 6/25 (24.0) & 16/25 (64.0) & 7/25 (28.0) \\
\cmidrule(lr){1-8}
\multirow{2}{*}{B1 (H1-guided)} & H1 (same) & 20/35 (57.1) & 14/14 (100.0) & 57/57 (100.0) & 28/28 (100.0) & 18/18 (100.0) & 27/27 (100.0) \\
 & H2 (cross) & 22/35 (62.9) & 14/14 (100.0) & 52/57 (91.2) & 27/28 (96.4) & 16/18 (88.9) & 27/27 (100.0) \\
\cmidrule(lr){1-8}
\multirow{2}{*}{B2 (H2-guided)} & H1 (cross) & 33/37 (89.2) & 20/30 (66.7) & 86/89 (96.6) & 40/44 (90.9) & 30/30 (100.0) & 24/24 (100.0) \\
 & H2 (same) & 24/37 (64.9) & 30/30 (100.0) & 83/89 (93.3) & 33/44 (75.0) & 29/30 (96.7) & 24/24 (100.0) \\
\bottomrule
\end{tabular}
\end{table*}

\begin{table*}[!htbp]
\centering
\setlength{\tabcolsep}{3pt}
\caption{Partial-knowledge-graph ablation, \textbf{Self-Refinement} agents. Notation as in Table~\ref{tab:exp2-zs}.}
\label{tab:exp2-sr}
\begin{tabular}{llcccccc}
\toprule
 & & \multicolumn{3}{c}{\textbf{Opus-4.8}} & \multicolumn{3}{c}{\textbf{GPT-5.5}} \\
\cmidrule(lr){3-5}\cmidrule(lr){6-8}
\textbf{Agent} & \textbf{KG} & mupdf-x11 & mutool & unsafelib & mupdf-x11 & mutool & unsafelib \\
\midrule
\multirow{3}{*}{A1 (unguided)} & Full & 50/63 (79.4) & 64/65 (98.5) & 47/69 (68.1) & 35/61 (57.4) & 48/77 (62.3) & 42/66 (63.6) \\
 & H1 & 40/63 (63.5) & 46/65 (70.8) & 46/69 (66.7) & 10/61 (16.4) & 32/77 (41.6) & 42/66 (63.6) \\
 & H2 & 36/63 (57.1) & 40/65 (61.5) & 45/69 (65.2) & 30/61 (49.2) & 37/77 (48.1) & 38/66 (57.6) \\
\cmidrule(lr){1-8}
\multirow{3}{*}{A2 (guided)} & Full & 70/71 (98.6) & 81/84 (96.4) & 81/87 (93.1) & 64/64 (100.0) & 57/59 (96.6) & 63/63 (100.0) \\
 & H1 & 45/71 (63.4) & 56/84 (66.7) & 81/87 (93.1) & 41/64 (64.1) & 30/59 (50.8) & 63/63 (100.0) \\
 & H2 & 41/71 (57.7) & 59/84 (70.2) & 81/87 (93.1) & 46/64 (71.9) & 49/59 (83.1) & 63/63 (100.0) \\
\cmidrule(lr){1-8}
\multirow{3}{*}{A3 (No-LLM)} & Full & 12/25 (48.0) & 17/25 (68.0) & 7/25 (28.0) & 12/25 (48.0) & 17/25 (68.0) & 7/25 (28.0) \\
 & H1 & 5/25 (20.0) & 10/25 (40.0) & 5/25 (20.0) & 5/25 (20.0) & 10/25 (40.0) & 5/25 (20.0) \\
 & H2 & 6/25 (24.0) & 16/25 (64.0) & 7/25 (28.0) & 6/25 (24.0) & 16/25 (64.0) & 7/25 (28.0) \\
\cmidrule(lr){1-8}
\multirow{2}{*}{B1 (H1-guided)} & H1 (same) & 45/88 (51.1) & 57/78 (73.1) & 77/77 (100.0) & 32/44 (72.7) & 38/45 (84.4) & 76/79 (96.2) \\
 & H2 (cross) & 50/88 (56.8) & 68/78 (87.2) & 74/77 (96.1) & 36/44 (81.8) & 41/45 (91.1) & 76/79 (96.2) \\
\cmidrule(lr){1-8}
\multirow{2}{*}{B2 (H2-guided)} & H1 (cross) & 67/97 (69.1) & 44/62 (71.0) & 118/127 (92.9) & 54/54 (100.0) & 27/38 (71.1) & 54/54 (100.0) \\
 & H2 (same) & 57/97 (58.8) & 48/62 (77.4) & 114/127 (89.8) & 43/54 (79.6) & 29/38 (76.3) & 54/54 (100.0) \\
\bottomrule
\end{tabular}
\end{table*}

\begin{table}[!htbp]
\centering
\caption{\emph{Partition 1} (an alternative pair of disjoint halves of the
CWE-121 knowledge graph). Binary \texttt{unsafelib}, model Opus-4.8. The full KG is
split into four disjoint quarters, each with a quarter of the edges of every type, $Q_1\!-\!Q_4$; Partition~1 pairs
them as \textbf{Half-A}$=Q_1\!\cup\!Q_2$ and \textbf{Half-B}$=Q_3\!\cup\!Q_4$.
Rows \textbf{A1} (no domain knowledge), \textbf{A2} (+DK, full-enriched) and
\textbf{A3} (No-LLM agent) verify the existing paths again over the
Full graph and each half (no new generations). Rows \textbf{B1}/\textbf{B2}
generate paths with the domain knowledge of one half (B1: DK$=$Half-A;
B2: DK$=$Half-B) and validate on the \emph{same} half and the \emph{cross} half.
Each cell is \#Entailed$/$\#Paths~(\% entailed).}
\label{tab:exp2-P1-full}
\small
\setlength{\tabcolsep}{5pt}
\begin{tabular}{llccc}
\toprule
Agent & KG & Zero-Shot & CoT & Self-Refinement \\
\midrule
\multirow{3}{*}{A1 (unguided)}
 & Full   & $37/47$ (78.7) & $56/67$ (83.6) & $47/69$ (68.1) \\
 & Half-A & $36/47$ (76.6) & $53/67$ (79.1) & $47/69$ (68.1) \\
 & Half-B & $31/47$ (66.0) & $41/67$ (61.2) & $43/69$ (62.3) \\
\midrule
\multirow{3}{*}{A2 (guided)}
 & Full   & $54/54$ (100.0) & $41/42$ (97.6) & $81/87$ (93.1) \\
 & Half-A & $54/54$ (100.0) & $41/42$ (97.6) & $81/87$ (93.1) \\
 & Half-B & $54/54$ (100.0) & $41/42$ (97.6) & $81/87$ (93.1) \\
\midrule
\multirow{3}{*}{A3 (No-LLM)}
 & Full   & $7/25$ (28.0) & $7/25$ (28.0) & $7/25$ (28.0) \\
 & Half-A & $6/25$ (24.0) & $6/25$ (24.0) & $6/25$ (24.0) \\
 & Half-B & $4/25$ (16.0) & $4/25$ (16.0) & $4/25$ (16.0) \\
\midrule
\multirow{2}{*}{B1 (Half-A-guided)}
 & Half-A (same)  & $76/84$ (90.5) & $57/57$ (100.0) & $77/77$ (100.0) \\
 & Half-B (cross) & $71/84$ (84.5) & $52/57$ (91.2)  & $74/77$ (96.1) \\
\midrule
\multirow{2}{*}{B2 (Half-B-guided)}
 & Half-A (cross) & $82/87$ (94.3) & $86/89$ (96.6) & $118/127$ (92.9) \\
 & Half-B (same)  & $82/87$ (94.3) & $83/89$ (93.3) & $114/127$ (89.8) \\
\bottomrule
\end{tabular}
\end{table}

\section{Scalability of the Reasoner}
\label{app:scalability}

\begin{figure}[!htbp]
    \centering
    \includegraphics[width=0.75\columnwidth]{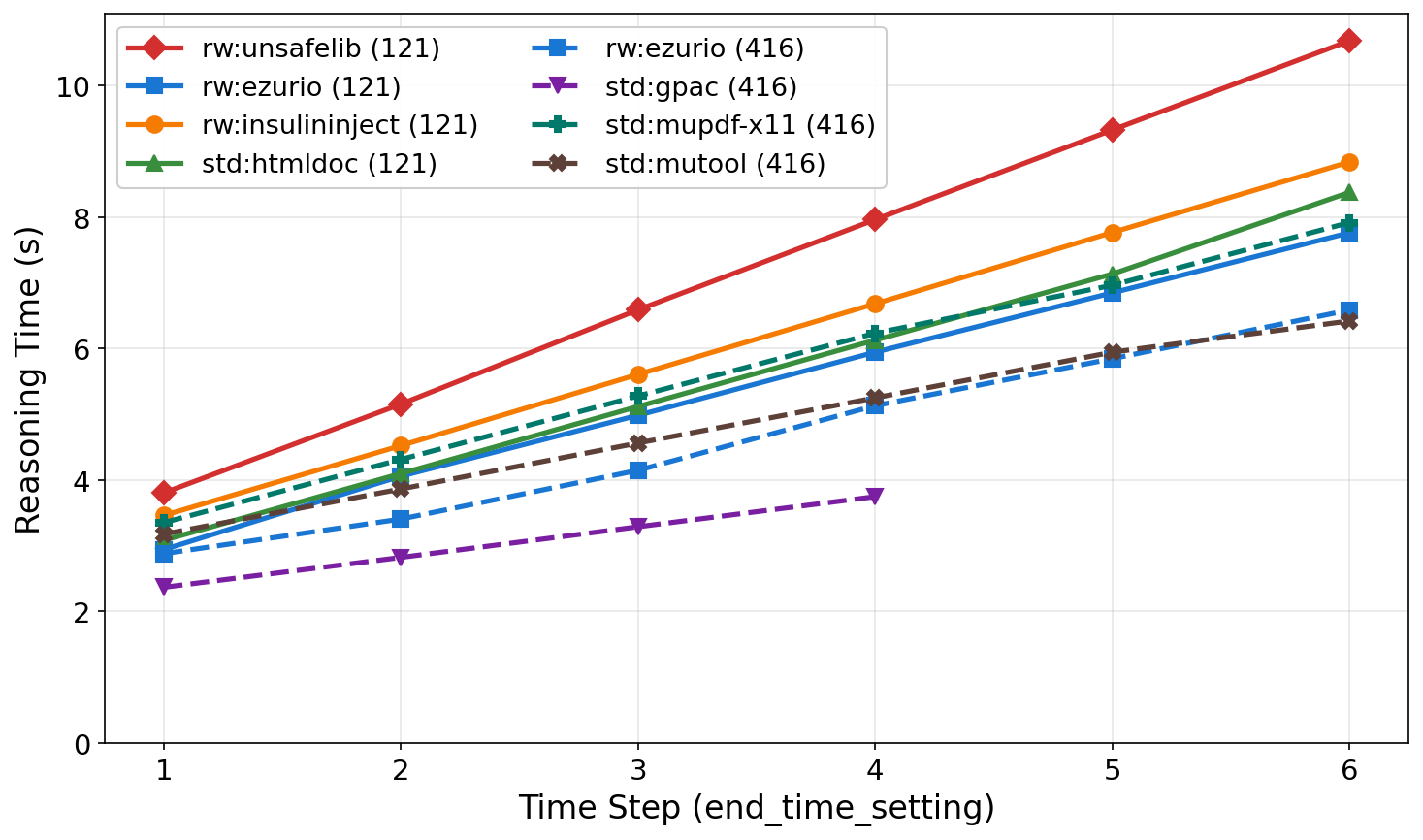}
    \caption{Reasoning runtime versus number of reasoning time steps for all
    CWE-121 (solid) and CWE-416 (dashed) binaries.}
    \label{fig:appendix-runtime-full}
\end{figure}

\begin{figure}[!htbp]
    \centering
    \includegraphics[width=0.75\columnwidth]{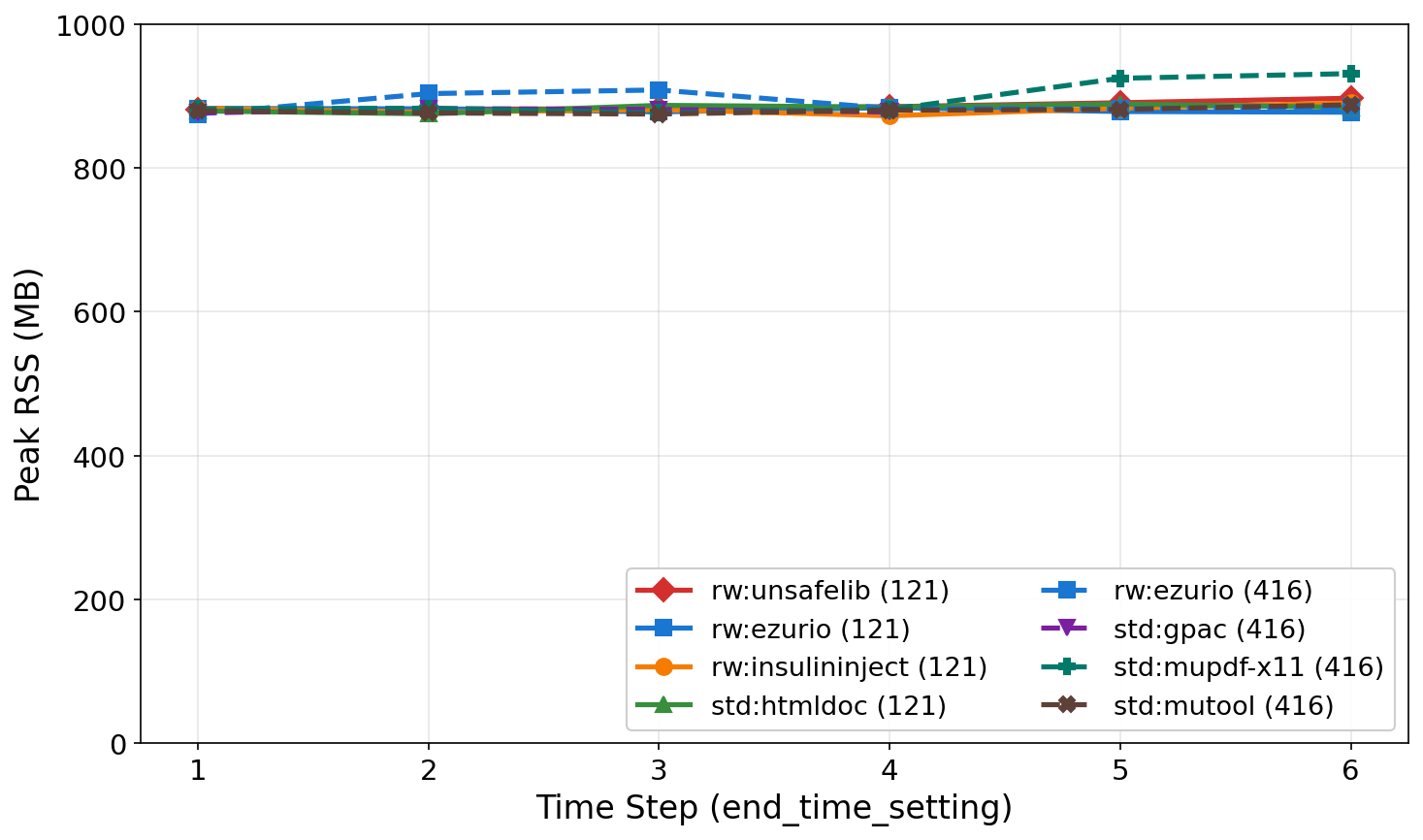}
    \caption{Peak resident set size (RSS) after reasoning versus number of
    reasoning time steps for all CWE-121 (solid) and CWE-416 (dashed) binaries.
    The footprint is dominated by fixed startup cost.}
    \label{fig:appendix-memory-full}
\end{figure}

Figures~\ref{fig:appendix-runtime-full} and~\ref{fig:appendix-memory-full}
report reasoner runtime and peak memory as the number of reasoning time steps
grows, for all CWE-121 (solid) and CWE-416 (dashed) binaries, including
\texttt{rw:ezurio} under CWE-416. The horizontal axis is PyReason's
\texttt{time} setting, which we fix to the number of steps in an analyst path:
verifying a trajectory of $n$ steps requires the reasoner to run for $n$ time
points, so this axis is the length of the trajectory being checked. Each point
is an independent run in a fresh reasoner process with \texttt{time} fixed to
the value shown.

Runtime grows linearly with the number of time steps for every binary, so the
slope of each line is the cost of one additional reasoning step. This per-step
cost is found to be proportional to how many groundings the reasoner produces at each step. The slopes actually order the binaries by
grounding volume: \texttt{rw:unsafelib} produces the most groundings per step
(${\sim}1{,}760$) and has the steepest slope (${\sim}1.4$\,s per step), while
\texttt{std:gpac} produces the fewest and has the shallowest.

Peak memory is flat at roughly $880$\,MB for every binary because the footprint
is dominated by fixed startup costs---the Python interpreter, the JIT-compiled
reasoner, and the loaded rules and knowledge graph---all done before any
reasoning occurs. The additional memory per time step of reasoning is small by comparison: running
six steps instead of one raises peak RSS by at most about $50$\,MB, about $5\%$ of the baseline, on \texttt{mupdf-x11}. The spread across binaries is small ($878$--$931$\,MB).
For deployment, we note that verification cost scales with trajectory
length and with the grounding volume of the binary--DKG pairing, not with the
number of trajectories checked, since each is verified independently and the
program is restarted after every run. As the memory footprint is dominated by
startup, a reasoner process sized for the largest binary is then sized
for all of them. The deployment instance of
$64$\,GB (Section~5 of the main paper) has roughly $70\times$ the peak
requirement of any single run.

\section{Evaluation Binaries}
\label{app:binaries}

Table~\ref{tab:app-callgraphs} lists function call graphs for every binary used in the study. Graphs
are recovered by Ghidra and restricted to the libc subset: functions that
reference at least one imported libc routine, together with the calls among
them, since these are the nodes over which entailment is defined. All eight contribute to the pooled results in the main paper; \texttt{dropbear} is evaluated only under CWE-415 (Section~\ref{app:cwe415}).

\begin{table}[!htbp]
\centering
\caption{All binary call graphs, grouped by target type and
ordered by size. Sizes span two orders of
magnitude, from 405 to 12{,}696 nodes.}
\label{tab:app-callgraphs}
\begin{tabular}{@{}llcrr@{}}
\toprule
\textbf{Binary} & \textbf{Type} & \textbf{CWE} & \textbf{Nodes} & \textbf{Edges} \\
\midrule
\multicolumn{5}{@{}l}{\emph{Standard (BinPool)}}\\
gpac          & std & 416           &    405 &    571 \\
dropbear      & std & 415           &    816 &  2{,}416 \\
htmldoc       & std & 121, 415           & 1{,}086 &  2{,}815 \\
mupdf-x11     & std & 416           & 3{,}341 & 11{,}883 \\
mutool        & std & 416           & 3{,}569 & 14{,}051 \\
\midrule
\multicolumn{5}{@{}l}{\emph{Real-world (medical device)}}\\
insulininject & rw  & 121           & 1{,}010 &  2{,}774 \\
ezurio        & rw  & 121, 415, 416 & 1{,}155 &  3{,}322 \\
unsafelib     & rw  & 121           & 12{,}696 & 40{,}881 \\
\bottomrule
\end{tabular}

\end{table}

\section{Language Models}
\label{app:llm}

Table~\ref{tab:models} lists the four LLMs used inside the agents, together
with the routing and generation settings under which each was called. All calls
go through \texttt{litellm}, which selects the API key by model-id prefix, and
all use greedy decoding ($t{=}0.0$).

\begin{table}[!htbp]
\centering\small
\caption{Language models used in our experiments. \emph{Context} and
\emph{Max out} are the model's own limits; \texttt{max-tok} and
\texttt{n-paths} are the values we request. For Kimi~K2.5 and Llama-4~Scout the
requested budget equals the model's output cap, so the cap binds. Prices are
per million input\,/\,output tokens at the time of the experiments. Temperature
is $0.0$ throughout.}
\label{tab:models}
\setlength{\tabcolsep}{4pt}
\begin{tabular}{@{}llrrrrr@{}}
\toprule
& & \multicolumn{2}{c}{\textbf{Model limits}} & \multicolumn{2}{c}{\textbf{Our settings}} & \\
\cmidrule(lr){3-4}\cmidrule(lr){5-6}
\textbf{Model} & \textbf{Route} & \textbf{Context} & \textbf{Max out}
& \texttt{max-tok} & \texttt{n-paths} & \textbf{\$/1M} \\
\midrule
Opus-4.8      & Anthropic        & 1{,}000{,}000 & 128{,}000 & 85{,}536 & 50 & 5.00\,/\,25.00 \\
GPT-5.5       & OpenAI           & 1{,}050{,}000 & 128{,}000 & 85{,}536 & 50 & 5.00\,/\,30.00 \\
Kimi K2.5     & AWS Bedrock      &   256{,}000 &  16{,}384 & 16{,}384 & 50 & 0.60\,/\,3.00 \\
Llama-4 Scout & AWS Bedrock (US) &   128{,}000 &   8{,}192 &  8{,}192 & 20 & 0.17\,/\,0.66 \\
\bottomrule
\end{tabular}

\end{table}

\section{Pipeline Configuration}
\label{app:config}

This section gives the full parameter settings, then reports the two
measurements that justify the one parameter we tuned: how many templates a model
returns and how many of those the beam search can realize
(Section~\ref{app:realization}), and what changes when the number of paths kept per template, $N_{\text{path}}$, is doubled (Section~\ref{app:nsol}).

\subsection{Hyperparameters}
\label{app:hyperparams}
The full hyperparameter values and rationale are shown in Table~\ref{tab:hyperparams}.

\begin{table}[!htbp]
\centering
\caption{Hyperparameter values and rationale.}
\label{tab:hyperparams}
\renewcommand{\arraystretch}{1.2}
\begin{tabular}{@{}lcp{0.5\textwidth}@{}}
\toprule
\textbf{Parameter} & \textbf{Value} & \textbf{Rationale} \\
\midrule
\multicolumn{3}{@{}l}{\emph{Generation (LLM)}}\\
\midrule
\texttt{temperature} & $0.0$ & Greedy decoding, to evaluate each model's most likely output rather than a sample. \\
\texttt{n-paths} ($N_{\text{req}}$) & $50$ (Llama-4: $20$) & Templates requested per call. Llama-4~Scout is reduced to $20$ because its $8{,}192$-token output cap truncates a 50-template JSON, most often under chain-of-thought. \\
\texttt{max-tokens} & $85{,}536$ & Opus-4.8 and GPT-5.5 \\
                    & $16{,}384$ & Kimi~K2.5 (equals the model's output cap) \\
                    & $8{,}192$  & Llama-4~Scout (equals the model's output cap) \\
\midrule
\multicolumn{3}{@{}l}{\emph{Beam-search realization}}\\
\midrule
$h$ (max hop distance)      & $3$   & Not tuned; chosen as a middle ground between permissive and strict realization (Section~3 of the main paper). \\
\texttt{beam-width}         & $100$ & Default, not tuned. \\
\texttt{max-back-hops}      & $3$   & Default, not tuned. \\
\texttt{hop-weight}         & $0.1$ & Default, not tuned; score penalty per hop beyond the first. \\
\texttt{bh-weight}          & $0.2$ & Default, not tuned; score penalty per back-hop. \\
\texttt{extra-label-weight} & $1.0$ & Default, not tuned; penalty per extra label on a matched node, favouring tight matches. \\
$N_{\text{path}}$ & $5$   & Realized paths kept per realizable template. Varied over $\{1,2,5,10\}$; Smaller values do not generate enough results. $N=10$ raised runtime, enrichment context size, and API cost without providing comparable entailment improvements (Section~\ref{app:nsol}). \\
\midrule
\multicolumn{3}{@{}l}{\emph{Path constraint}}\\
\midrule
\texttt{MIN\_PATH\_EDGES} & $6$ (\texttt{gpac}: $4$) & Minimum path length in edges. Lowered to $4$ for \texttt{gpac}, whose 405-node call graph is too shallow to admit many six-edge paths. \\
\bottomrule
\end{tabular}
\end{table}

\subsection{Template Generation and Realization}
\label{app:realization}

This section opens up Step~1 of the main paper, the agent's internal stages; the agent's output is the realized paths. A model does not return paths directly: it returns \emph{templates}, ordered
sequences of label subsets, which the agent's beam search must instantiate against the
actual call graph. Tables~\ref{tab:counts121} and~\ref{tab:counts416} report
these counts per configuration for CWE-121 and CWE-416. Of $8{,}160$ templates
requested across the two classes, models returned $4{,}291$ ($52.6\%$), of which
$2{,}683$ ($62.5\%$) realized as at least one connected chain, ultimately
yielding $7{,}661$ unique paths---a mean of $2.86$ per realizable template, below
the cap of $N_{\text{path}}=5$ because deduplication removes repeats. Roughly a third of the templates requested, and $62.5\%$ of those a model actually returns, therefore reach the reasoner. The rest
is discarded by beam search before any semantic check is applied.
Only the request is held fixed across configurations: every configuration asks for the same number of templates under the same beam-search settings. The number of templates returned, the number realizable, and the number of realized paths (\#Paths) are all outcomes of the experiment, which is why they differ between configurations. A realized path is a sequence of steps an analyst might take while examining the code, made of functions that exist in the binary and are statically connected in its call graph; it is not an execution path.

Domain knowledge is provided before generation rather than on realization. The return rate
rises from $46.9\%$ to $58.3\%$ under +DK. So, enrichment makes models propose more.

\begin{table}[!htbp]
\centering
\caption{Template generation and realization counts for CWE-121. All three columns count templates: Requested, the number asked for; Returned, the number the model returned; and Realizable, the number of returned templates that beam search realized as at least one path. The realized paths that result appear as Headers: Req.\ = Requested, Ret.\ = Returned, Rlz.\ = Realizable.
\#Paths in the per-configuration tables (Tables~\ref{tab:std}, \ref{tab:rw} and~\ref{tab:cwe415}). Only Requested is fixed by the experimenter; Returned, Realizable and \#Paths are outcomes.}
\label{tab:counts121}
\setlength{\tabcolsep}{4pt}
\begin{tabular}{llcccccccccccc}
\toprule
 &  & \multicolumn{3}{c}{rw:insulininject} & \multicolumn{3}{c}{rw:unsafelib} & \multicolumn{3}{c}{rw:ezurio} & \multicolumn{3}{c}{std:htmldoc} \\ \cline{3-14}
\textbf{Model} & \textbf{Strategy} & Req. & Ret. & Rlz. & Req. & Ret. & Rlz. & Req. & Ret. & Rlz. & Req. & Ret. & Rlz. \\ \midrule
 \multirow{6}{*}{Opus-4.8} & Zero-Shot & 50 & 25 & 25 & 50 & 20 & 15 & 50 & 15 & 9 & 50 & 15 & 15 \\
  & \quad + DK & 50 & 30 & 17 & 50 & 40 & 36 & 50 & 30 & 25 & 50 & 50 & 44 \\\cline{2-14}
  & CoT & 50 & 12 & 12 & 50 & 15 & 15 & 50 & 12 & 9 & 50 & 20 & 19 \\
  & \quad + DK & 50 & 20 & 12 & 50 & 20 & 20 & 50 & 20 & 16 & 50 & 25 & 23 \\\cline{2-14}
  & Self-Refinement & 50 & 18 & 18 & 50 & 18 & 17 & 50 & 15 & 9 & 50 & 17 & 16 \\
  & \quad + DK & 50 & 32 & 17 & 50 & 50 & 46 & 50 & 50 & 43 & 50 & 40 & 40 \\
\midrule
 \multirow{6}{*}{GPT-5.5} & Zero-Shot & 50 & 15 & 15 & 50 & 15 & 13 & 50 & 15 & 10 & 50 & 18 & 15 \\
  & \quad + DK & 50 & 15 & 12 & 50 & 20 & 17 & 50 & 18 & 16 & 50 & 15 & 15 \\\cline{2-14}
  & CoT & 50 & 10 & 9 & 50 & 12 & 9 & 50 & 12 & 5 & 50 & 10 & 10 \\
  & \quad + DK & 50 & 12 & 9 & 50 & 12 & 12 & 50 & 10 & 8 & 50 & 12 & 12 \\\cline{2-14}
  & Self-Refinement & 50 & 16 & 16 & 50 & 20 & 19 & 50 & 15 & 8 & 50 & 20 & 19 \\
  & \quad + DK & 50 & 20 & 20 & 50 & 20 & 20 & 50 & 20 & 4 & 50 & 20 & 20 \\
\midrule
 \multirow{6}{*}{\shortstack[l]{Llama-4\\Scout}} & Zero-Shot & 20 & 20 & 5 & 20 & 20 & 4 & 20 & 20 & 7 & 20 & 20 & 18 \\
  & \quad + DK & 20 & 20 & 9 & 20 & 20 & 2 & 20 & 20 & 4 & 20 & 20 & 18 \\\cline{2-14}
  & CoT & 20 & 20 & 16 & 20 & 20 & 15 & 20 & 20 & 4 & 20 & 15 & 15 \\
  & \quad + DK & 20 & 3 & 3 & 20 & 20 & 18 & 20 & 20 & 10 & 20 & 15 & 15 \\\cline{2-14}
  & Self-Refinement & 20 & 20 & 6 & 20 & 20 & 4 & 20 & 20 & 5 & 20 & 20 & 9 \\
  & \quad + DK & 20 & 20 & 17 & 20 & 20 & 9 & 20 & 20 & 3 & 20 & 20 & 11 \\
\midrule
 \multirow{6}{*}{Kimi K2.5} & Zero-Shot & 50 & 50 & 49 & 50 & 40 & 35 & 50 & 20 & 18 & 50 & 50 & 39 \\
  & \quad + DK & 50 & 50 & 5 & 50 & 30 & 15 & 50 & 50 & 5 & 50 & 50 & 5 \\\cline{2-14}
  & CoT & 50 & 30 & 30 & 50 & 20 & 14 & 50 & 15 & 7 & 50 & 30 & 8 \\
  & \quad + DK & 50 & 15 & 11 & 50 & 50 & 9 & 50 & 25 & 14 & 50 & 20 & 10 \\\cline{2-14}
  & Self-Refinement & 50 & 40 & 39 & 50 & 30 & 30 & 50 & 50 & 28 & 50 & 25 & 14 \\
  & \quad + DK & 50 & 50 & 19 & 50 & 40 & 39 & 50 & 50 & 18 & 50 & 50 & 28 \\
\bottomrule
\end{tabular}
\end{table}

\begin{table}[!htbp]
\centering
\caption{Template generation and realization counts for CWE-416. Columns as in
Table~\ref{tab:counts121}.}
\label{tab:counts416}
\setlength{\tabcolsep}{4pt}
\begin{tabular}{llcccccccccccc}
\toprule
 &  & \multicolumn{3}{c}{std:gpac} & \multicolumn{3}{c}{std:mupdf-x11} & \multicolumn{3}{c}{std:mutool} & \multicolumn{3}{c}{rw:ezurio} \\ \cline{3-14}
\textbf{Model} & \textbf{Strategy} & Req. & Ret. & Rlz. & Req. & Ret. & Rlz. & Req. & Ret. & Rlz. & Req. & Ret. & Rlz. \\ \midrule
 \multirow{6}{*}{Opus-4.8} & Zero-Shot & 50 & 10 & 2 & 50 & 15 & 15 & 50 & 12 & 11 & 50 & 15 & 4 \\
  & \quad + DK & 50 & 30 & 19 & 50 & 40 & 38 & 50 & 30 & 30 & 50 & 50 & 23 \\\cline{2-14}
  & CoT & 50 & 10 & 5 & 50 & 10 & 10 & 50 & 12 & 1 & 50 & 12 & 3 \\
  & \quad + DK & 50 & 20 & 16 & 50 & 20 & 20 & 50 & 20 & 16 & 50 & 20 & 14 \\\cline{2-14}
  & Self-Refinement & 50 & 14 & 5 & 50 & 15 & 15 & 50 & 15 & 15 & 50 & 12 & 2 \\
  & \quad + DK & 50 & 21 & 12 & 50 & 25 & 25 & 50 & 50 & 50 & 50 & 50 & 30 \\
\midrule
 \multirow{6}{*}{GPT-5.5} & Zero-Shot & 50 & 15 & 4 & 50 & 15 & 13 & 50 & 18 & 17 & 50 & 16 & 6 \\
  & \quad + DK & 50 & 20 & 1 & 50 & 20 & 19 & 50 & 12 & 12 & 50 & 15 & 2 \\\cline{2-14}
  & CoT & 50 & 12 & 1 & 50 & 12 & 11 & 50 & 10 & 8 & 50 & 12 & 3 \\
  & \quad + DK & 50 & 15 & 1 & 50 & 12 & 11 & 50 & 12 & 12 & 50 & 12 & 2 \\\cline{2-14}
  & Self-Refinement & 50 & 16 & 3 & 50 & 20 & 20 & 50 & 20 & 19 & 50 & 20 & 5 \\
  & \quad + DK & 50 & 26 & 15 & 50 & 20 & 20 & 50 & 22 & 22 & 50 & 20 & 1 \\
\midrule
 \multirow{6}{*}{\shortstack[l]{Llama-4\\Scout}} & Zero-Shot & 20 & 20 & 2 & 20 & 15 & 4 & 20 & 15 & 2 & 20 & 10 & 2 \\
  & \quad + DK & 20 & 20 & 5 & 20 & 10 & 10 & 20 & 10 & 6 & 20 & 10 & 2 \\\cline{2-14}
  & CoT & 20 & 20 & 10 & 20 & 20 & 19 & 20 & 4 & 2 & 20 & 5 & 1 \\
  & \quad + DK & 20 & 3 & 3 & 20 & 20 & 20 & 20 & 20 & 18 & 20 & 3 & 0 \\\cline{2-14}
  & Self-Refinement & 20 & 20 & 5 & 20 & 20 & 5 & 20 & 20 & 1 & 20 & 20 & 5 \\
  & \quad + DK & 20 & 20 & 12 & 20 & 20 & 18 & 20 & 20 & 2 & 20 & 20 & 20 \\
\midrule
 \multirow{6}{*}{Kimi K2.5} & Zero-Shot & 50 & 50 & 5 & 50 & 50 & 4 & 50 & 50 & 7 & 50 & 20 & 8 \\
  & \quad + DK & 50 & 45 & 29 & 50 & 30 & 11 & 50 & 20 & 10 & 50 & 15 & 6 \\\cline{2-14}
  & CoT & 50 & 15 & 3 & 50 & 20 & 19 & 50 & 30 & 30 & 50 & 15 & 1 \\
  & \quad + DK & 50 & 15 & 6 & 50 & 20 & 17 & 50 & 20 & 20 & 50 & 50 & 23 \\\cline{2-14}
  & Self-Refinement & 50 & 12 & 5 & 50 & 40 & 38 & 50 & 50 & 50 & 50 & 20 & 0 \\
  & \quad + DK & 50 & 25 & 6 & 50 & 30 & 28 & 50 & 30 & 27 & 50 & 20 & 0 \\
\bottomrule
\end{tabular}
\end{table}

\subsection{Sensitivity to the Beam-Search Budget}
\label{app:nsol}

$N_{\text{path}}$, the number of realized paths kept per realizable template, was
the only beam-search parameter tuned. Table~\ref{tab:cwe121_n10} repeats the
entire CWE-121 evaluation at $N_{\text{path}}=10$, with every other setting
unchanged.

Doubling the budget realizes roughly $1.7\times$ as many paths---$4{,}884$
against $2{,}838$ without domain knowledge, and $3{,}413$ against $2{,}026$ with
it. But, pooled
entailment under +DK is $97.2\%$ at $N_{\text{path}}=10$ against $98.5\%$ at
$N_{\text{path}}=5$. The effect of domain knowledge is therefore not dependent on how many realizations the beam search retains, and the smaller budget is
preferred on runtime and API cost.

\begin{table}[!htbp]
\centering
\caption{CWE-121 with $N_{\text{path}}=10$: unique paths realized (\#Paths) and
entailed (\#Entailed), without and with domain knowledge (+DK). Pooled entailment rises
from $82.3\%$ to $97.2\%$, against $78.5\%$ to $98.5\%$ at $N_{\text{path}}=5$. Bold as in Table~\ref{tab:std}.}
\label{tab:cwe121_n10}
\setlength{\tabcolsep}{3pt}
\begin{tabular}{llcccccccc}
\toprule
 &  & \multicolumn{2}{c}{rw:insulininject} & \multicolumn{2}{c}{rw:unsafelib} & \multicolumn{2}{c}{rw:ezurio} & \multicolumn{2}{c}{std:htmldoc} \\ \cline{3-10}
\textbf{Model} & \textbf{Strategy} & Paths & Entailed & Paths & Entailed & Paths & Entailed & Paths & Entailed \\ \midrule
 \multirow{6}{*}{Opus-4.8} & Zero-Shot & \textbf{233} & \textbf{214}(91.85) & 78 & 63(80.77) & 38 & 35(92.11) & \textbf{138} & \textbf{136}(98.55) \\
  & \quad + DK & 56 & 56\textbf{(100)} & \textbf{107} & \textbf{99(92.52)} & \textbf{131} & \textbf{131(100)} & 68 & 68\textbf{(100)} \\\cline{2-10}
  & CoT & \textbf{100} & \textbf{93}(93) & \textbf{123} & \textbf{107}(86.99) & 34 & 34(100) & \textbf{169} & \textbf{169(100)} \\
  & \quad + DK & 43 & 43\textbf{(100)} & 97 & 95\textbf{(97.94)} & \textbf{84} & \textbf{84(100)} & 111 & 111(100) \\\cline{2-10}
  & Self-Refinement & \textbf{127} & \textbf{124}(97.64) & 118 & 79(66.95) & 38 & 38(100) & 117 & 114(97.44) \\
  & \quad + DK & 92 & 91\textbf{(98.91)} & \textbf{126} & \textbf{117(92.86)} & \textbf{102} & \textbf{102(100)} & \textbf{167} & \textbf{167(100)} \\
\midrule
 \multirow{6}{*}{GPT-5.5} & Zero-Shot & \textbf{129} & \textbf{114}(88.37) & \textbf{82} & 68(82.93) & \textbf{29} & 25(86.21) & \textbf{138} & \textbf{137}(99.28) \\
  & \quad + DK & 35 & 35\textbf{(100)} & 68 & 68\textbf{(100)} & 25 & 25\textbf{(100)} & 86 & 86\textbf{(100)} \\\cline{2-10}
  & CoT & \textbf{85} & \textbf{65}(76.47) & 58 & 43(74.14) & 18 & 17(94.44) & \textbf{75} & \textbf{74}(98.67) \\
  & \quad + DK & 26 & 26\textbf{(100)} & \textbf{62} & \textbf{62(100)} & \textbf{42} & \textbf{42(100)} & 66 & 66\textbf{(100)} \\\cline{2-10}
  & Self-Refinement & \textbf{135} & \textbf{97}(71.85) & \textbf{97} & 58(59.79) & 16 & 14(87.5) & \textbf{142} & \textbf{136}(95.77) \\
  & \quad + DK & 47 & 47\textbf{(100)} & 95 & \textbf{92(96.84)} & \textbf{27} & \textbf{27(100)} & 134 & 134\textbf{(100)} \\
\midrule
 \multirow{6}{*}{\shortstack[l]{Llama-4\\Scout}} & Zero-Shot & \textbf{22} & 12(54.55) & 4 & 4(100) & 29 & 29(100) & \textbf{144} & \textbf{136}(94.44) \\
  & \quad + DK & 15 & \textbf{15(100)} & \textbf{55} & \textbf{54}(98.18) & \textbf{99} & \textbf{90}(90.9) & 47 & 47\textbf{(100)} \\\cline{2-10}
  & CoT & \textbf{136} & \textbf{123}(90.44) & \textbf{62} & \textbf{48}(77.42) & \textbf{17} & \textbf{17(100)} & \textbf{130} & \textbf{130(100)} \\
  & \quad + DK & 59 & 56\textbf{(94.92)} & 40 & 40\textbf{(100)} & 15 & 15(100) & 51 & 51(100) \\\cline{2-10}
  & Self-Refinement & 33 & 33(100) & 20 & 13(65) & \textbf{18} & \textbf{18(100)} & 54 & 51(94.44) \\
  & \quad + DK & \textbf{37} & \textbf{37(100)} & \textbf{73} & \textbf{35}(47.94) & 9 & 9(100) & \textbf{73} & \textbf{73(100)} \\
\midrule
 \multirow{6}{*}{Kimi K2.5} & Zero-Shot & \textbf{296} & \textbf{184}(62.16) & \textbf{247} & \textbf{175}(70.85) & 14 & 14(100) & \textbf{288} & \textbf{272}(94.44) \\
  & \quad + DK & 6 & 6\textbf{(100)} & 51 & 44\textbf{(86.27)} & \textbf{56} & \textbf{56(100)} & 49 & 49\textbf{(100)} \\\cline{2-10}
  & CoT & \textbf{239} & \textbf{194}(81.17) & \textbf{134} & \textbf{89}(66.42) & 23 & 22(95.65) & 50 & 50(100) \\
  & \quad + DK & 52 & 50\textbf{(96.15)} & 54 & 54\textbf{(100)} & \textbf{97} & \textbf{97(100)} & \textbf{61} & \textbf{61(100)} \\\cline{2-10}
  & Self-Refinement & \textbf{277} & \textbf{93}(33.57) & 171 & 105(61.4) & 63 & 56(88.89) & 96 & 96(100) \\
  & \quad + DK & 62 & 60\textbf{(96.77)} & \textbf{225} & \textbf{215(95.56)} & \textbf{126} & \textbf{126(100)} & \textbf{104} & \textbf{104(100)} \\
\bottomrule
\end{tabular}
\end{table}
\section{The Complete Logic Program}
\label{app:rules}

The program $\program$ consists of the 42 non-ground rules listed below,
reproduced verbatim in PyReason syntax. The head annotation
\texttt{paired\_minimum\_bounds\_ann\_fn} is the annotation function
$\mu_{\mathrm{pair}}$ of the main paper: it assigns the head the componentwise
minimum of the bounds satisfied by the body clauses, so a step is never given a confidence higher than its weakest supporting evidence.

The same program is shared across all three CWEs; only the knowledge graph the reasoner reasons with changes. Throughout the experiments, we take the minimum admissible
bound $\mu_{\min}$ of the main paper's problem formulation to be $[0.25, 1]$,
matching the guard on \texttt{analystAt(CB1)} in the traversal rules: an analyst
is considered present at a block if its bound is at least $[0.25, 1]$. The bounds in the rules below are hand-set design parameters.

\subsection{Analyst-Traversal and Step-Control Rules (8)}

\begin{lstlisting}
analystAt(CB2):paired_minimum_bounds_ann_fn <-1 analystAt(CB1):[0.25,1],
    hasLabel(CB1, Lcause):[0.1,1], hasLabel(CB2, Leffect):[0.1,1],
    can_cause(Lcause, Leffect):[0.1,1], stepFrom(CB1, CB2)

analystAt(CB2):paired_minimum_bounds_ann_fn <-1 analystAt(CB1):[0.25,1],
    hasLabel(CB1, Lcontrib):[0.1,1], hasLabel(CB2, Lfault):[0.1,1],
    contributes_to(Lcontrib, Lfault):[0.1,1], stepFrom(CB1, CB2)

analystAt(CB2):paired_minimum_bounds_ann_fn <-1 analystAt(CB1):[0.25,1],
    hasLabel(CB1, Lop):[0.1,1], hasLabel(CB2, Lderived):[0.1,1],
    derives(Lop, Lderived):[0.1,1], stepFrom(CB1, CB2)

analystAt(CB2):paired_minimum_bounds_ann_fn <-1 analystAt(CB1):[0.25,1],
    hasLabel(CB1, Lunsafe):[0.1,1], hasLabel(CB2, Lsafe_concept):[0.1,1],
    unsafe_variant_of(Lunsafe, Lsafe_concept):[0.1,1], stepFrom(CB1, CB2)

analystAt(CB2):paired_minimum_bounds_ann_fn <-1 analystAt(CB1):[0.25,1],
    hasLabel(CB1, Lfault):[0.1,1], hasLabel(CB2, Lcwe):[0.1,1],
    manifestation_of(Lfault, Lcwe):[0.1,1], stepFrom(CB1, CB2)

analystAt(CB2):paired_minimum_bounds_ann_fn <-1 analystAt(CB1):[0.25,1],
    hasLabel(CB1, Lfunc):[0.1,1], hasLabel(CB2, Lop):[0.1,1],
    implements(Lfunc, Lop):[0.1,1], stepFrom(CB1, CB2)

future(Y) <-1 stepFrom(X,Y), analystAt(X):[0.01,1]
~stepFrom(X,Y) <- future(Y), ~analystAt(Y):[0.25,1], ~analystAt(X)
\end{lstlisting}

The six traversal rules advance the analyst one step per time point
(\texttt{<-1}): if the analyst occupies \texttt{CB1} with lower bound at least
$0.25$ and the trajectory contains the edge \texttt{stepFrom(CB1, CB2)}, the
step is entailed whenever some knowledge-graph relation
(\texttt{can\_cause}, \texttt{contributes\_to}, \texttt{derives},
\texttt{unsafe\_variant\_of}, \texttt{manifestation\_of}, \texttt{implements})
links a label observed on \texttt{CB1} to a label observed on \texttt{CB2}. The
clause thresholds ($[0.1,1]$ on \texttt{hasLabel} and on the relation,
$[0.25,1]$ on \texttt{analystAt}) act only as gates; the bound assigned
to the head is computed by \texttt{paired\_minimum\_bounds\_ann\_fn}, which
recovers the (cause, effect) label pairing imposed by the relation clause. For
each grounded relation edge it takes the elementwise minimum of the relation's
bound and the \texttt{hasLabel} bounds of the matching labels on the two code
blocks, so inference is never mixed across unrelated label pairs, and annotates
\texttt{analystAt(CB2)} with the pair attaining the highest lower bound. If no
pair meets the threshold, the head retains its default $[0,1]$, complete
uncertainty, and the step is not entailed.

The final two rules make that failure explicit for reporting.
\texttt{future(Y)} marks the block one step ahead on the analyst trajectory; if
the analyst never reaches there, the negation rule drives
\texttt{stepFrom(X,Y)} toward $[0,0]$, contradicting the trajectory edge
asserted at $[1,1]$. The two formulations converge. The head failing to reach
$\mu_{\min}$ and the resulting bound conflict on \texttt{stepFrom} identify the
same step, and is shown in the reasoning trace of
Figure~\ref{fig:deployment-trace}.

\subsection{Propagation Rules (28)}

\subsubsection{Label-Propagation Rules (13)}

These map raw libc labels to higher-level semantic labels through conjunctive
clauses. A single libc call is often ambiguous, but the combination is not. Head
annotations encode confidence in how strongly a combination implies the
semantic label.

\begin{lstlisting}
hasLabel(X, memory_write):[1.0,1]   <- hasLabel(X, memcpy):[0.6,1],  hasLabel(X, memmove):[0.6,1]
hasLabel(X, memory_write):[1.0,1]   <- hasLabel(X, memcpy):[0.6,1],  hasLabel(X, strcpy):[0.6,1]
hasLabel(X, memory_write):[1.0,1]   <- hasLabel(X, sprintf):[0.6,1], hasLabel(X, strcpy):[0.6,1]
hasLabel(X, memory_write):[0.778,1] <- hasLabel(X, memcpy):[0.6,1],  hasLabel(X, memset):[0.6,1]
hasLabel(X, memory_write):[0.714,1] <- hasLabel(X, memmove):[0.6,1], hasLabel(X, memset):[0.6,1]
hasLabel(X, memory_write):[0.5,1]   <- hasLabel(X, memset):[0.6,1],  hasLabel(X, snprintf):[0.6,1]
hasLabel(X, memory_write):[0.429,1] <- hasLabel(X, memset):[0.6,1],  hasLabel(X, strcpy):[0.6,1]
hasLabel(X, memory_write):[0.2,1]   <- hasLabel(X, fgets):[0.6,1],   hasLabel(X, memset):[0.6,1]

hasLabel(X, copy_operation):[1.0,1] <- hasLabel(X, memcpy):[0.6,1],  hasLabel(X, memmove):[0.6,1]
hasLabel(X, copy_operation):[1.0,1] <- hasLabel(X, memcpy):[0.6,1],  hasLabel(X, strcpy):[0.6,1]
hasLabel(X, copy_operation):[0.9,1] <- hasLabel(X, sprintf):[0.6,1], hasLabel(X, strcpy):[0.6,1]

hasLabel(X, input_operation):[1.0,1] <- hasLabel(X, fgets):[0.6,1],  hasLabel(X, sscanf):[0.6,1]
hasLabel(X, size):[1.0,1] <- hasLabel(X, destination_size_validation):[0.6,1],
    hasLabel(X, length_calculation):[0.6,1]
\end{lstlisting}

\subsubsection{Relation-Propagation Rules (15)}

These connect a code block's observed labels to the domain concepts of the DKG
through the graph's typed relations. Most relations are instantiated in both
directions with different bounds, so that strong and weak evidence propagate
asymmetrically.

\begin{lstlisting}
hasLabel(CB, Lsemantic):[0.84,1] <-0 hasLabel(CB, Lbinary_feature):[0.7,1],
    evidence_of(Lbinary_feature, Lsemantic)
hasLabel(CB, Lbinary_feature):[0.36,1] <-0 hasLabel(CB, Lsemantic):[0.95,1],
    evidence_of(Lbinary_feature, Lsemantic)

hasLabel(CB, Lparent):[1,1]  <-0 hasLabel(CB, Lchild):[0.7,1],   is_a(Lchild, Lparent)
hasLabel(CB, Lchild):[0.1,1] <-0 hasLabel(CB, Lparent):[0.95,1], is_a(Lchild, Lparent)

hasLabel(CB, Lop):[0.87,1] <-0 hasLabel(CB, Lpart):[0.7,1],
    required_component_of(Lpart, Lop)
hasLabel(CB, Lop):[0.25,1] <-0 hasLabel(CB, Lpart):[0.95,1],
    required_component_of(Lpart, Lop)

hasLabel(CB, Lop):[0.32,1]   <-0 hasLabel(CB, Lpart):[0.95,1],
    informative_component_of(Lpart, Lop)
hasLabel(CB, Lpart):[0.29,1] <-0 hasLabel(CB, Lop):[0.95,1],
    informative_component_of(Lpart, Lop)

hasLabel(CB, Lop):[0.15,1]  <-0 hasLabel(CB, Lpart):[0.95,1],
    incidental_component_of(Lpart, Lop)
hasLabel(CB, Lpart):[0.2,1] <-0 hasLabel(CB, Lop):[0.95,1],
    incidental_component_of(Lpart, Lop)

hasLabel(CB, Lrealization):[0.9,1] <-0 hasLabel(CB, Lop):[0.7,1],
    possible_realization(Lrealization, Lop)
hasLabel(CB, Lop):[0.4,1] <-0 hasLabel(CB, Lrealization):[0.9,1],
    possible_realization(Lrealization, Lop)

hasLabel(CB, Lconcept):[0.41,1] <-0 hasLabel(CB, Lop):[0.9,1], involves(Lop, Lconcept)

hasLabel(CB, Lused):[0.3,1] <-0 hasLabel(CB, Lop):[0.95,1],  may_use(Lop, Lused)
hasLabel(CB, Lop):[0.18,1]  <-0 hasLabel(CB, Lused):[0.95,1], may_use(Lop, Lused)
\end{lstlisting}

\subsection{Transitive-Closure Rules (6)}

These make a DKG relation transitive where its semantics warrant it, letting the
reasoner chain multi-hop causal and compositional relationships that no single
edge states explicitly.

\begin{lstlisting}
can_cause(E1, E3)             <- can_cause(E1, E2),             can_cause(E2, E3)
contributes_to(E1, E3)        <- contributes_to(E1, E2),        contributes_to(E2, E3)
derives(E1, E3)               <- derives(E1, E2),               derives(E2, E3)
involves(E1, E3)              <- involves(E1, E2),              involves(E2, E3)
is_a(E1, E3)                  <- is_a(E1, E2),                  is_a(E2, E3)
required_component_of(E1, E3) <- required_component_of(E1, E2),
    required_component_of(E2, E3)
\end{lstlisting}
\section{Deployment: Reasoning Trace Example}
\label{app:trace}

\begin{figure}[!htbp]
    \centering
    \includegraphics[width=0.98\columnwidth]{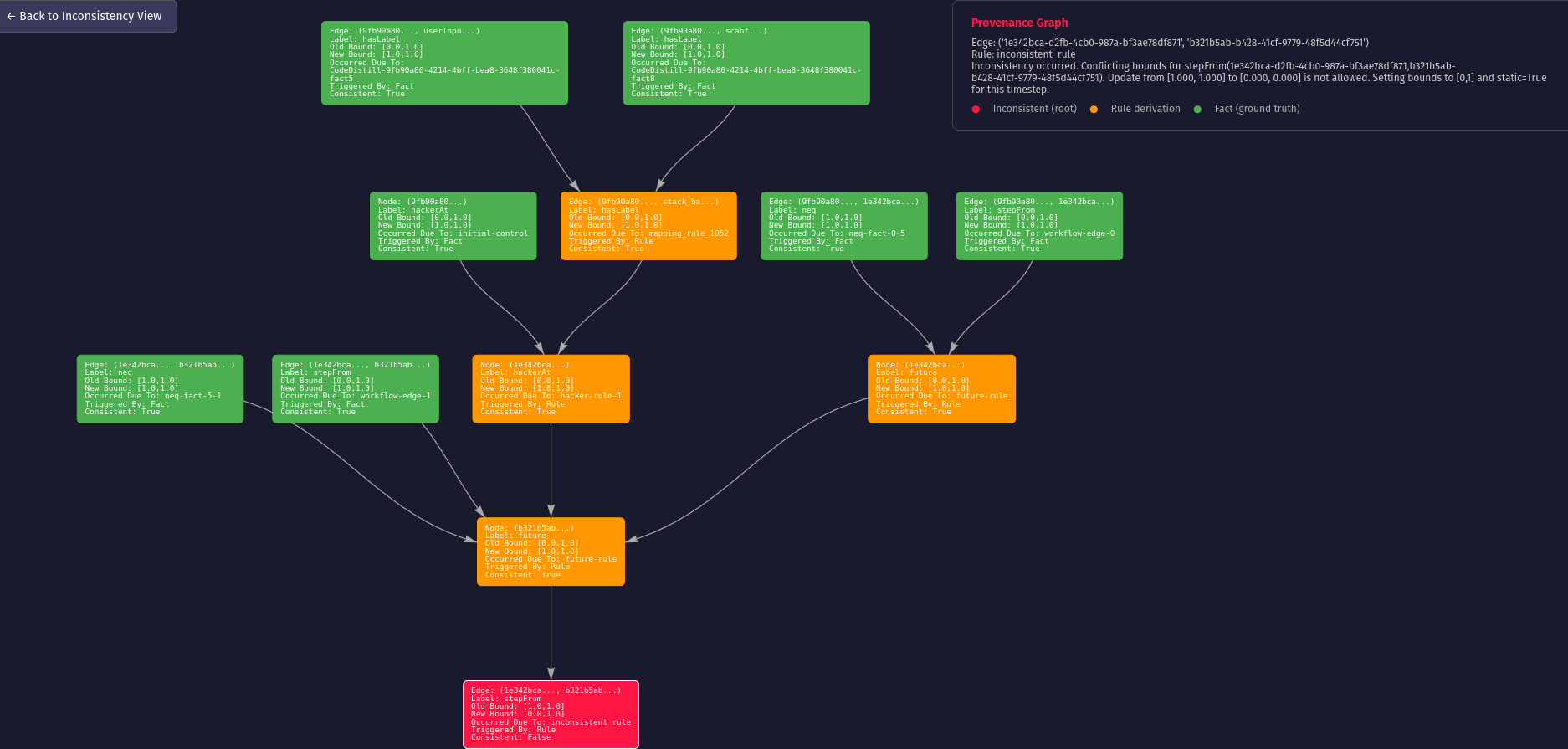}
    \caption{The reasoning trace behind a non-entailment inference, for the
    analyst path of the running example in the main paper. Green leaves are
     facts (libc label observations and the trajectory's own
    \texttt{stepFrom} edges); orange nodes are rule derivations, each annotated
    with the bound it assigns; the red root is the first step no domain-knowledge
    relation entails. Each node records the fact or rule responsible for its
    bound update. Node labels show \texttt{hackerAt}, the name the deployed end-to-end system uses for the \texttt{analystAt} predicate of the main paper.}
    \label{fig:deployment-trace}
\end{figure}

A reasoning trace accompanies every non-entailment inference. A
tree recording which observations and which rule firings produced each bound,
rooted at the first step the domain knowledge does not entail.
Figure~\ref{fig:deployment-trace} shows the trace for the running example of the
main paper, in which an agent assigns \texttt{memset} to the memory-management
class and then assumes it can perform an integer size calculation.

The leaves are facts of two kinds: libc \texttt{hasLabel} observations on
the code blocks along the path (here \texttt{userInput} and \texttt{scanf}), and
the \texttt{stepFrom} edges of the proposed trajectory itself, annotated
$[1,1]$. Interior nodes are rule firings. A relation-propagation rule fires on
the conjunction of observed labels to derive the higher-level
\texttt{stack\_\allowbreak based\_\allowbreak buffer\_\allowbreak overflow\_\allowbreak risk} label, which is what bridges
code-level evidence to the \texttt{CWE\_121} domain-knowledge graph; traversal
and step-control rules then advance the analyst along the prefix that the domain
knowledge does entail.

At the root, no relation in the domain-knowledge graph connects the labels at
the two blocks, so no traversal rule fires and \texttt{analystAt} at that block
never reaches $\mu_{\min}$. The step-control rule of
Section~\ref{app:rules} then makes the corresponding \texttt{stepFrom} bound
 $[0,0]$, contradicting the $[1,1]$ set from the trajectory, and the
program reports the conflict as a non-entailment at that step. Because every
node records both its bound update and the fact or rule that caused it, an
analyst can audit exactly which observations and which rule firings reject the
step, rather than receiving an unexplained conclusion.

\section{Prompt Templates}
\label{app:prompts}

These prompts are the LLM component of the agents (Step~1 of the main paper). All three prompting techniques share an identical \emph{base prompt}: the system
role, the CWE information, the Ghidra label vocabulary, and the constraint and
output-format blocks. The techniques differ only in a final technique-specific
block and, for self-refinement, in a multi-call structure. We therefore give the
base prompt once (Section~\ref{app:prompt-base}) and then the per-technique
additions (Sections~\ref{app:prompt-zs}--\ref{app:prompt-sr}).

The transcript below is the CWE-416 prompt for \texttt{gpac}; the main paper
shows the abridged CWE-121 form. Three fields are substituted per target: the
CWE information, the label vocabulary, and \texttt{MIN\_PATH\_EDGES}, which is
$4$ here because \texttt{gpac}'s call graph is too shallow for six-edge paths
and $6$ for every other binary (Section~\ref{app:hyperparams}). Under
domain-knowledge enrichment (+DK) the only change is that each label group in
the vocabulary is augmented with reasoner-derived DK labels
(Section~\ref{app:prompt-dk}); every other block is byte-identical. Prompts are
reproduced verbatim.

\subsection{Shared Base Prompt}
\label{app:prompt-base}

Sent in full on every call, including each of the three self-refinement calls.
Label groups beyond the three shown are omitted for space; a binary typically
yields tens of groups.

\begin{lstlisting}
=== SYSTEM PROMPT ===
You are a security analyst specializing in defensive binary code review. You
reason about which sequences of function behaviors in a binary are relevant to
a given software weakness class (CWE), so that code regions worth reviewing can
be prioritized for detection and remediation. Your task is to generate realistic
code-review paths - ordered sequences of semantic function behaviors an analyst
would examine when assessing a binary for a specific weakness type. These paths
describe which code a reviewer would inspect during analysis, not necessarily
what they would find.

=== USER PROMPT ===
CWE Information:
CWE ID: CWE-416
CWE Name: Use After Free
CWE Description: Use After Free - a program continues to use a pointer after the
memory it references has been freed, allowing an attacker to control the freed
region and influence program behavior or execution.

Binary Static Analysis Labels:
The following are groups of function labels detected via Ghidra static analysis
of the target binary. Each group represents functions that co-occur in the same
code region. Use these as your label vocabulary when constructing path stages:

Group 1: ['__stack_chk_fail', 'abort', 'clock', 'exit', 'exp', 'fwrite',
'gmtime', 'log', 'realloc', 'remove', 'scanf', 'sin', 'sprintf', 'strcat',
'strcpy', 'strtol', 'vsnprintf', 'vsprintf']
Group 2: ['abort', 'fseek', 'ftell', 'fwrite', 'vfprintf']
Group 3: ['__stack_chk_fail', 'clock', 'exit', 'fputs', 'fread', 'fseek',
'ftell', 'fwrite', 'log', 'sprintf', 'strcat', 'strcpy', 'strrchr', 'strtol']

Path Diversity:
The generated exploration paths should be meaningfully different from each other.
Each path should represent a different way an analyst might examine the code
while assessing it for the mentioned weakness.

Exploration Path Framing:
Each path represents the exploration journey of a human expert or security
researcher navigating through the binary, looking for potential vulnerabilities.
Not every stage in the path needs to directly relate to CWE-416 - a human expert
would follow interesting code paths, investigate suspicious functions, explore
memory management patterns, and sometimes backtrack from dead ends. The path
reflects how someone would realistically explore the code, not just the shortest
route to a vulnerability.

Label Vocabulary Constraint:
Every label used in every stage subset must come from the provided label
vocabularies. Do not use any label that does not appear in the provided
vocabularies.

Label Group Selection Constraint:
At each stage of the path, select one label group from the provided groups and
use only those labels from that group that makes sense for that stage. All labels
within a single stage must come from the same group.

Minimum Path Length:
Each generated path must contain at least 4 edges (transitions between stages).
Paths with fewer than 4 edges are invalid.

Per Path Reasoning:
For each exploration path, provide a brief explanation under the key 'explanation'
describing the analyst's reasoning - how this sequence of labels represents a
plausible region of code to examine when assessing the target weakness.

Exploration Likelihood Score:
For each path, provide a score between 0.0 and 1.0 under the key
'exploration_score' reflecting how likely an analyst would examine this trajectory
when reviewing the binary. A score of 1.0 means an analyst would very likely
follow this path, 0.0 means unlikely.

Number of Paths:
Generate maximum 50 exploration paths. Each path must be unique and independently
valid. You can also decide to generate fewer than 50 paths if you determine that
there are not that many high-quality, distinct paths to generate. Do not generate
more than 50 paths.

Free Label Subset Path Output:
Return your response as a valid JSON object with the following structure:
{
  "paths": [
    {
      "path_id": <integer>,
      "cluster_sequence": [
        ["label1", "label2", ...],
        ["label1", ...],
        ...
      ],
      "vulnerability_score": <float between 0.0 and 1.0>,
      "exploration_score": <float between 0.0 and 1.0>,
      "explanation": "<2 sentence explanation of why an analyst would examine
                      this path when assessing the target weakness>"
    }
  ]
}
Each entry in cluster_sequence is a subset of one or more labels chosen freely
from any of the provided label sets. Labels can be mixed from different sets
within the same stage. Do not include any text outside the JSON object. Do not
use markdown formatting or code blocks.
\end{lstlisting}

\subsection{Zero-Shot}
\label{app:prompt-zs}

Appended to the base prompt; a single call.

\begin{lstlisting}
Zero-Shot:
Using the information provided above, generate exploration paths for the given
CWE. Do not ask for clarification. Generate the paths directly.
\end{lstlisting}

\subsection{Chain-of-Thought}
\label{app:prompt-cot}

Appended to the base prompt; a single call returning one JSON object in which a
triage phase precedes generation.

\begin{lstlisting}
Chain-of-Thought (CoT):
Reason step by step before answering, and return everything as a single JSON
object that extends the output structure specified above.

Phase 1 - Label analysis. Before generating any paths, examine the label groups
provided above. Produce a top-level "label_analysis" object that, for each group
you consider relevant, classifies its labels into one of three categories for the
target CWE: "immediately_suspicious", "worth_investigating", or
"likely_irrelevant", each with a one-line justification. Use this triage to decide
which labels to choose for path stages.

Phase 2 - Path generation. Then generate the exploration paths. Keep every field
required above unchanged (path_id, cluster_sequence, vulnerability_score,
exploration_score, explanation). Additionally, for each path include a
"stage_reasoning" field: an array with one entry per stage in cluster_sequence,
where entry i explains, stage by stage, what the analyst infers from the
combination of labels present at stage i - what seeing those labels together in
the same code region suggests, and why that leads to the next stage.

The final JSON object must have this shape:
{
  "label_analysis": { ... },
  "paths": [
    { <all fields required above>, "stage_reasoning": ["stage 0 inference", ...] }
  ]
}
Do not ask for clarification. Do not include any text outside the JSON object. Do
not use markdown formatting or code blocks. Generate the analysis and paths
directly.
\end{lstlisting}

\subsection{Self-Refinement}
\label{app:prompt-sr}

A three-call loop. Every call re-sends the full base prompt; only the additions
are shown. Call~1 is identical to zero-shot, so the loop's cost is three
generations per configuration.

\begin{lstlisting}
--- Call 1: Generate (identical to zero-shot) ---
Iterative Self-Refinement:
Using the information provided above, generate exploration paths for the given
CWE. Do not ask for clarification. Generate the paths directly.

--- Call 2: Critique ---
[base prompt + Call-1 instruction], then:
   ---
   Here are the exploration paths you generated:
   [ <JSON path set from Call 1> ]

   Critically review these paths for the target CWE. What is wrong with them? Are
   they diverse enough? Are important labels being ignored? Are the vulnerability
   and exploration scores well calibrated? Respond with concise verbal feedback in
   plain text only - do not output JSON, code blocks, or paths.

--- Call 3: Refine ---
[base prompt + Call-1 instruction], then:
   ---
   Here are your original exploration paths:
   [ <JSON path set from Call 1> ]

   Here is feedback on them:
   [ <plain-text critique from Call 2> ]

   Now generate improved exploration paths that address this feedback. Return the
   refined paths using exactly the same JSON output format specified above. Do not
   include any text outside the JSON object. Do not use markdown formatting or
   code blocks.
\end{lstlisting}

\subsection{Domain-Knowledge Enrichment}
\label{app:prompt-dk}

Under +DK the label vocabulary is the only block that changes. Before
generation we run the domain knowledge's mapping rules over the libc labels
already present on each code block and append the derived labels to that block's
group. The agent's LLM therefore receives DK's vocabulary but no entailment
verdict, and the reasoner has not yet seen any trajectory at this point. A group
that reads

\begin{lstlisting}
Group 1: ['gets', 'sprintf', 'strspn']
\end{lstlisting}

\noindent in the baseline condition becomes

\begin{lstlisting}
Group 1: ['gets', 'sprintf', 'strspn', 'missing_bounds_check',
'unchecked_memory_write', 'stack_pointer_overwrite', ...]
\end{lstlisting}

\noindent under +DK. The added terms are DKG concepts rather than libc symbols,
so the model can express a stage in the vocabulary the reasoner will later check
against, instead of only in terms of the raw calls Ghidra recovered.
\section{Domain Knowledge: Ontology and Graphs}
\label{app:kg}

The domain knowledge is a CWE-focused, code-centric ontology together with a set
of instantiated knowledge graphs, one per weakness class. The ontology defines a
layered class hierarchy---from concrete Beacons, CodeEntities, CodeOperations
and CodePatterns, through FaultConditions and OutcomeLevelFaults, to CWE-level
VulnerabilityClasses, connected by a fixed set of typed relations
(e.g.\ \texttt{is\_a}, \texttt{unsafe\_variant\_of}, \texttt{can\_cause},
\texttt{mitigates}).

A central modelling choice is to represent faults as unsafe variants of
otherwise-neutral operations: an\\ \texttt{out\_of\_bounds\_write} is an
\texttt{unsafe\_variant\_of} a generic \texttt{memory\_write}. Safe and unsafe
behaviour therefore coexist in a single graph, and the reasoner can express how
missing or incorrect validation turns an ordinary operation into a
memory-corruption event. The layering supports reasoning in both directions:
forward, from observable beacons up through the class hierarchy to a CWE-level
determination, and backward, from a hypothesised classification down to the
code-level evidence that would confirm or refute it. The relation-propagation
rules of Section~\ref{app:rules} implement exactly this, which is why most
relations are instantiated in both directions with asymmetric bounds.

Each graph is populated initially by automated extraction from technical corpora
(reverse-engineering manuals, security advisories, man pages), then refined by
automated schema validation with targeted semantic review. We treat the
resulting graphs as fixed and assumed correct throughout; the extraction and
validation of domain-knowledge graphs is a separate problem that this work does
not address.

\subsection{Graph Statistics}
\label{app:kg-stats}

\begin{table}[!htbp]
\centering
\caption{The three per-CWE knowledge graphs. All graphs are directed.
\emph{Classes} counts ontology classes instantiated in the graph, \emph{Types}
the distinct relation types used, and \emph{Rel./ent.} the mean number of
relations per entity.}
\label{tab:kg}
\begin{tabular}{@{}lrrrrr@{}}
\toprule
& \multicolumn{2}{c}{\textbf{Entities}} & \multicolumn{2}{c}{\textbf{Relations}} & \\
\cmidrule(lr){2-3}\cmidrule(lr){4-5}
\textbf{DK graph} & \textbf{Classes} & \textbf{Total} & \textbf{Types} & \textbf{Total}
& \textbf{Rel./ent.} \\
\midrule
CWE-121 (stack buffer overflow) & 9 & 188 & 16 & 403 & 2.14 \\
CWE-415 (double free)           & 8 & 105 & 14 & 318 & 3.03 \\
CWE-416 (use-after-free)        & 8 & 111 & 14 & 232 & 2.09 \\
\bottomrule
\end{tabular}
\end{table}

The 16 relation types, taken as the union across graphs, are
\texttt{contributes\_to}, \texttt{can\_cause}, \texttt{manifestation\_of},
\texttt{unsafe\_variant\_of}, \texttt{involves}, \texttt{implements},
\texttt{validates}, \texttt{may\_use}, \texttt{derives},
\texttt{possible\_realization}, \texttt{is\_a}, \texttt{evidence\_of},
\texttt{mitigates}, \texttt{required\_component\_of},
\texttt{informative\_component\_of} and \\ \texttt{incidental\_component\_of}.
CWE-121 uses all 16; CWE-415 and CWE-416 use 14 each, CWE-415 omitting
\texttt{derives} and \texttt{involves} and CWE-416 omitting \texttt{derives} and
\texttt{unsafe\_variant\_of}. Six of the 16 are made transitive by the closure
rules of Section~\ref{app:rules}, and six adjudicate analyst steps in the
traversal rules.
\section{Extended Related Work}
\label{app:related}

\subsection{LLMs for Vulnerability Detection}

LLMs are increasingly applied to software vulnerability detection across source
code, repair, and binaries; recent surveys chart the rapid growth of this area
since 2023 and catalogue the models, datasets, and evaluation setups now in
use~\cite{Sheng2025}. A recurring finding, however, is that current models are
not yet dependable on their own: a comprehensive benchmarking study by Ullah et
al.~\cite{ullah2024llms} concludes that LLMs cannot yet reliably identify or
reason about security vulnerabilities. Work that does report gains typically
pairs the LLM with structure rather than trusting it in isolation---LLMxCPG, for
example, uses code-property-graph slices to focus an LLM on
vulnerability-relevant context~\cite{li2024llmxcpg}. This unreliability is the
gap our work targets: rather than treating LLM output as an answer, we treat
each proposed analyst path as a hypothesis to be checked against domain
knowledge, and we operate on decompiled binaries because source is unavailable
for the medical-device firmware we analyse.

\subsection{LLMs for Binary Reverse Engineering}

A fast-growing line of work applies LLMs directly to stripped or decompiled
binaries: LLM4Decompile trains models to recover source-like code from
assembly~\cite{tan2024llm4decompile}; ReSym recovers variable and
data-structure symbols from stripped binaries~\cite{xie2024resym}; and DeGPT
uses an LLM to refine and explain decompiler output~\cite{degpt2024}. Most
relevant to our setting, an empirical study of human--LLM teaming in reverse
engineering shows that LLMs can accelerate analyst comprehension but also
mislead when their suggestions go unchecked~\cite{basque2026decompiling}. These
efforts improve the readability or recovered structure of a
binary; they do not verify that an analyst's reasoning over that binary is
consistent with what is known about the vulnerability class. Our contribution is
complementary and sits downstream: we take the recovered call-graph structure
and libc labels as input and add a logical verification layer over the analyst
paths reasoned on top of them.

\subsection{Logic and Program Analysis for Binary Vulnerability Discovery}

Classical binary vulnerability discovery relies on static analysis and taint
tracking. Karonte propagates taint across the multiple binaries of a firmware
image to surface insecure interactions, and its authors note that
whole-firmware analysis without such cross-binary reasoning produces
overwhelming alert volumes~\cite{Redini2020}. More recently, LATTE couples an
LLM with binary taint analysis, automating the taint-propagation and inspection
rules that previously required manual expert customisation and reporting new
CVEs in real firmware~\cite{liu2025latte}. These systems share our binary-level,
source-free setting and our use of domain rules, but their goal is to
emit vulnerability alerts via taint propagation. Ours is a different
problem: we verify whether an analyst's exploration path is logically
entailed by a domain-knowledge graph under a temporal annotated logic, retaining
the exploratory structure that alert-only pipelines discard while still steering
it toward vulnerabilities. Where LATTE uses the LLM to produce the
analysis, we use logic to check the LLM output.

\subsection{Constraining and Verifying LLM Reasoning with Structure}

Because LLM hallucinations are often confident and cannot be caught by
probability-based checks, a growing body of work pairs LLMs with symbolic or
structured components. Logic-LM translates a problem into symbolic form, solves
it with a deterministic solver, and self-refines from solver
errors~\cite{pan-etal-2023-logic}; knowledge-graph-guided methods such as
Reasoning-on-Graphs~\cite{luo2024rog} and Graph-Constrained
Reasoning~\cite{luo2025gcr} constrain LLM reasoning to paths grounded in a KG to
reduce hallucination on question-answering benchmarks.

We share the premise that an independent structured layer must gate LLM output,
but differ in kind rather than degree. These methods align reasoning to a walk
within a single knowledge graph and are evaluated on natural-language QA.
In our setting the object being verified is a trajectory through the binary's
call graph, while the constraints it must satisfy live in a
separate domain-knowledge graph, so verification is an alignment
between two graphs rather than a walk within one. The temporal,
open-world semantics also admit constraints those methods do not model: analyst
position is time-indexed, so a step is licensed only at the time point it is
proposed, and negation over derived state lets the program assert what the
domain knowledge fails to establish rather than only what it establishes
(Section~\ref{app:rules}). The target is likewise different---provable,
explainable accept/reject inferences in a safety-critical binary domain, rather
than answer faithfulness on general benchmarks.

\subsection{Vulnerabilities in Medical-Device Software}

The application domain motivates the entire effort. Connected medical devices
---infusion pumps, insulin pumps, pacemakers---run long-lived embedded software
that is hard to patch and frequently analysable only as deployed binaries.
Large-scale analyses underscore the scale of the problem: a study of over
200{,}000 infusion pumps found that roughly three-quarters carried known
security weaknesses, many tied to years-old, unpatched
CVEs~\cite{unit42infusion2022}, and reviews of FDA safety communications
document repeated device vulnerabilities with direct patient-safety
consequences~\cite{menon2026cybersecurity}. This is precisely the regime our
system is built for: source-free analysis of the deployed artifact, with a
logical guarantee that retained analyst paths are consistent with encoded
domain knowledge.

\subsection{Summary}

Unlike prior LLM-for-security work, which treats the model's output as a
prediction to be trusted or scored, we treat each trajectory realized from an LLM template as a hypothesis and pose its validation as a formal
entailment-checking problem against domain-knowledge. To our
knowledge, this is the first framework to verify LLM-generated exploration
paths---as opposed to final answers or natural-language reasoning---using
temporal annotated logic.

\bibliography{master}